\documentclass{aa}
\usepackage{txfonts}
\usepackage{natbib}
\usepackage{graphicx}
\usepackage{xcolor}

\usepackage{mathrsfs} 
\usepackage[normalem]{ulem}

\def\ba{\begin{eqnarray}}
\def\ea{\end{eqnarray}}

\begin{document}

\title{Planet formation in stellar binaries: Global simulations of planetesimal growth }

\author{Kedron Silsbee\inst{1} \and Roman R. Rafikov\inst{2, 3}}
\institute{Max-Planck-Institut f\"ur Extraterrestrische Physik, 85748 Garching, Germany \email{ksilsbee@mpe.mpg.de},\and Department of Applied Mathematics and Theoretical Physics, University of Cambridge, Cambridge CB3 0WA, UK \and Institute for Advanced Study, Einstein Drive, Princeton, NJ 08540, USA; John N. Bahcall Fellow} 


\abstract{Planet formation around one component of a tight, eccentric binary system such as $\gamma$ Cephei (with semimajor axis around 20 AU) is theoretically challenging because of destructive high-velocity collisions between planetesimals. Despite this fragmentation barrier, planets are known to exist in such (so-called S-type) orbital configurations. Here we present a novel numerical framework for carrying out multi-annulus coagulation-fragmentation calculations of planetesimal growth, which fully accounts for the specifics of planetesimal dynamics in binaries, details of planetesimal collision outcomes, and the radial transport of solids in the disk due to the gas drag-driven inspiral. Our dynamical inputs properly incorporate the gravitational effects of both the eccentric stellar companion and the massive non-axisymmetric protoplanetary disk in which planetesimals reside, as well as gas drag. We identify a set of disk parameters that lead to successful planetesimal growth in systems such as $\gamma$ Cephei or $\alpha$ Centauri starting from $1-10$ km size objects. We identify the apsidal alignment of a protoplanetary disk with the binary orbit as one of the critical conditions for successful planetesimal growth: It naturally leads to the emergence of a dynamically quiet location in the disk (as long as the disk eccentricity is of order several percent), where favorable conditions for planetesimal growth exist. Accounting for the gravitational effect of a protoplanetary disk plays a key role in arriving at this conclusion, in agreement with our previous results. These findings lend support to the streaming instability as the mechanism of planetesimal formation. They provide important insights for theories of planet formation around both binary and single stars, as well as for the hydrodynamic simulations of protoplanetary disks in binaries (for which we identify a set of key diagnostics to verify).}

 
 \titlerunning{Global simulations of planetesimal growth in stellar binaries}
\authorrunning{Kedron Silsbee \& Roman R. Rafikov}
\maketitle



\section{Introduction}
 

Planets have been discovered in a wide variety of stellar systems, including stellar binaries and systems of higher multiplicity \citep{Marzari2019}. Dynamically active environments typical of multiple stellar systems are believed to be hostile for planetary assembly, positioning such planet-hosting systems as important test beds of planet formation theory.  

In particular, tight eccentric binary systems that host planets orbiting one of the stellar companions (so-called S-type systems in the classification of \citealt{Dvorak1982}), such as $\gamma$ Cephei \citep{Hatzes03}, present important challenges to theories of planet formation via core accretion, which involves a steady agglomeration of planetary cores through mutual collisions of numerous small planetesimals. Indeed, gravitational perturbations from the eccentric companion in such binaries with semimajor axes $\lesssim 20-30$ AU are expected to drive planetesimal eccentricities to high values. This naturally results in the destruction, rather than the growth, of planetesimals in mutual collisions.  A simple calculation \citep{Heppenheimer78}, including only the dominant secular effects of the stellar companion on planetesimal orbits, yields planetesimal collision velocities of a few km s$^{-1}$ at the current location of the planet in the $\gamma$ Cephei system (around 2 AU from the stellar primary). This is enough to destroy even planetesimals of hundreds of kilometers in size in catastrophic collisions, presenting an important barrier to planet formation in binaries.  

A number of ideas have been advanced over the years to overcome this "fragmentation barrier." In particular, \citet{Marzari00} noticed that aerodynamic drag due to the gaseous component of the protoplanetary disk in which planetesimals move eventually apsidally aligns their orbits. While their orbits would still have eccentricities similar to those in the \citet{Heppenheimer78} model, their relative eccentricities (which determine the magnitude of the mutual collision velocities) would be dramatically reduced by apsidal alignment.  However, it was later recognized by \citet{Thebault2006} that, even with apsidal alignment, only planetesimals of similar size would have small relative velocities as the strength of the frictional coupling to the gas disk is directly determined by the planetesimal size.  As a result, planetesimals with different sizes are not well aligned with one another and therefore experience high collision velocities \citep{Thebault08,Thebault2009,Thebault2011}. 

On the other hand, \citet{R13a,R13} noted that the gas disk should couple to planetesimals not only through gas drag, but also gravitationally.  Neglecting the gas drag in the first instance, he showed that a massive axisymmetric disk should induce a rapid apsidal precession of planetesimal orbits, severely suppressing the eccentricity excitation due to the stellar companion. However, it is not at all clear that the protoplanetary disk orbiting one of the components of the binary should be axisymmetric, given the strong gravitational perturbation due to the external stellar companion \citep{Paardekooper08, Marzari09, Regaly11}.  For this reason, \citet{SR15A} generalized the work of \citet{R13a} and calculated the gravitational effect of a massive eccentric disk on the orbits of planetesimals moving inside such a disk (again, neglecting gas drag), finding that collision outcomes depend heavily on the details of the disk structure. 

Subsequently, \citet{RS15a} extended the model of \citet{SR15A} by fully incorporating the effects of gas drag on planetesimal motion in eccentric protoplanetary disks and derived collision velocities of planetesimals as a function of their sizes. These results on planetesimal dynamics, coupled with simple prescriptions for collisional outcomes based on \citet{SL09}, were then used in \citet{RS15b} to determine the conditions under which planetesimals could grow to sizes large enough to withstand collisional fragmentation and erosion (thus providing a pathway to planetary core assembly by coagulation).  

Despite this work, some uncertainty remains regarding the collisional environments that would allow planetesimal growth. In particular, \citet{RS15b} used simple heuristics when deciding whether planetesimal growth is possible, such as assuming it to occur if there are no catastrophically disruptive collisions. However, this condition is neither necessary nor sufficient for unimpeded growth, since evolution of the planetesimal mass spectrum is affected both by coagulation at low relative velocities and by erosion and catastrophic disruption at high relative velocities. The size-dependent alignment of planetesimal orbits \citep{Thebault08,RS15a} gives rise to a dynamical environment with a mixture of low- and high-velocity collisions, making it challenging, even qualitatively, to determine the evolution of the size distribution. 

The problem is further complicated by the radial inspiral of planetesimals due to interaction with the gas.  The gas disk is slightly pressure-supported and orbits $\sim 30$ m s$^{-1}$ slower than the Keplerian speed \citep{Weidenschilling77}. Large bodies with stopping times longer than an orbital time move at the Keplerian speed and therefore experience a headwind, which causes them to spiral into the star.  The inspiral is more rapid in the case of a body having a nonzero forced eccentricity relative to the gas \citep{Adachi76}.  It was suggested in \citet{RS15b} that the dependence of the inspiral rate on planetesimal eccentricity could concentrate solid bodies in the regions of the disk with low relative particle-gas eccentricity where radial migration slows down, thus accelerating coagulation. Inspiral may also help to alleviate the detrimental effect of the erosion of growing planetesimals by small bodies by rapidly removing such bodies from the system.

In this paper we present the first realistic global model of planetesimal growth in S-type stellar binaries, which explicitly includes the aforementioned physical ingredients missing in previous studies. A specific question we address is whether the fragmentation barrier can somehow be overcome such that planetesimals can grow from some small size (that we determine as part of our analysis) to sizes of several hundred kilometers, at which point fragmentation is no longer a threat to their growth. Our modeling framework employs a multi-annulus coagulation-fragmentation code for accurately following the evolution of the planetesimal size spectrum at every radius in the disk. In addition to that, our model directly follows the exchange of mass in solid objects between the different radial locations in the disk. These components of the model use the collisional velocities and radial inspiral rates calculated in \citet{RS15a,RS15b}, which fully account for the secular gravitational effects of both the stellar companion and the massive  eccentric disk in which planetesimals orbit. Outcomes of planetesimal collisions are determined using the prescriptions derived in \citet{SL09}. With this physical model, we are able to determine whether the coagulation process as a whole is successful for a given environment (i.e., a disk model), rather than determining if it is successful based on the outcomes of individual collisions. Armed with this powerful tool, we then perform an extensive exploration of the parameter space of the disk models to determine the conditions under which planet formation is possible in compact eccentric binaries such as $\gamma$ Cephei or $\alpha$ Centauri.   

This paper is organized as follows. After describing our model setup in Sect. \ref{MSPD}, we provide an overview of the important aspects of planetesimal dynamics in S-type binaries in Sect. \ref{sect:ingredients}.  We briefly outline the functions of the coagulation code in Sect.\ \ref{codeOverview}; a detailed description of the code and its tests can be found in Appendices \ref{firstAppendix} and \ref{analTests}, respectively. In 
Sect. \ref{sect:fiducial} we describe simulations for our fiducial disk model, both in single- (Sect. \ref{singleAnnulus}) and multi-annulus (Sect. \ref{fullCalc}) setups. We then present the results of the model parameter space exploration in 
Sect. \ref{sect:par-var}. In Sect. \ref{disc} we provide a discussion of our findings with implications for planet formation in binaries (Sect. \ref{sect:implications}), planetesimal formation (Sect. \ref{sect:implications-planetesimals}), and hydrodynamic simulations of protoplanetary disks in S-type binaries (Sect. \ref{sect:implications-sims}), among other things.  We summarize our main conclusions in  Sect. \ref{conclusions}.


\section{Model setup}
\label{MSPD}


To illustrate our calculations, we consider a model S-type binary with parameters similar to that of the $\gamma$ Cephei system \citep{Hatzes03,Chauvin2011}. This binary consists of $M_p=1.6M_\odot$ (primary) and $M_s=0.4M_\odot$ (secondary) stars in an orbit with a semimajor axis $a_b=20$ AU and eccentricity $e_b=0.4$. 

A planet with mass $M_{\rm pl}=1.6M_{\rm J}$ orbits the primary star of the $\gamma$ Cephei system with the semimajor axis $a_{\rm pl}=2$ AU and eccentricity $e_{\rm pl}=0.12$, which we use to motivate our subsequent calculations. There are a number of other binaries with $a_b\lesssim 20$ AU that have similar characteristics \citep{Chauvin2011,Marzari2019}.

We assume the primary component of the binary to host a massive protoplanetary disk, coplanar with the binary orbit, with properties similar to the disk described in \citet{SR15A}: it is an eccentric disk with a power law radial dependence of the surface density $\Sigma(a)$ and eccentricity $e(a)$ on the semimajor axis $a$ of the eccentric fluid trajectories. The stated dependence on $a$ strictly applies to the surface density and eccentricity of the gas streamlines at their periastra:
\begin{equation}
\label{SigmaAndEccEquations}
\Sigma(a) = \Sigma_0 \left(\frac{a_0}{a}\right), \quad e_g(a) = e_0 \left(\frac{a_0}{a}\right)^{-1}, \quad a<a_{\rm out},
\end{equation}
where $a_0$ is a reference semimajor axis. We assume the disk to be sharply truncated by the torque due to the secondary star at the outer boundary $a_{\rm out}=5$ AU (i.e., $\Sigma(a)=0$ for $a>a_{\rm out}$).  The apsidal angle of the gas streamlines in the disk with respect to the binary orbit is denoted $\varpi_d$ and, for simplicity, is assumed to be independent of $a$, that is, the disk as a whole is apsidally aligned.  There is no particular reason for choosing $e_g(a)$ in the form (\ref{SigmaAndEccEquations}), except that it scales as the forced eccentricity due to the secondary. Development of a more accurate, better motivated model for the $e_g(a)$ behavior is beyond the scope of this study.  Throughout this paper, we assume a solid-to-gas ratio of 0.01, and the disk to be slightly flared with the $h/r$ profile given by Eq. (14) from \citet{RS15b} and varying between 0.03 and 0.05.

As a consequence of the generally nonzero disk eccentricity, the surface density distribution in the disk ends up being non-axisymmetric, as described in \citet{Ogilvie2001}, \citet{Statler01}, and \citet{SR15A}. Since in this work we fully account for the gravitational effect of the disk on the motion of the planetesimals embedded in the disk, the non-axisymmetry of the disk generally leads to nonzero torque affecting planetesimal dynamics, in addition to the torque exerted by the secondary star. We cover the roles played by these different dynamical agents next.


\section{Physical ingredients of the model}
\label{sect:ingredients}


Numerical calculations self-consistently following growth of planetesimals in protoplanetary disks from small sizes, and often all the way into the planetary regime, have been routinely used to understand planet formation around single stars. A standard approach is to follow the evolution of the size (mass) spectrum of colliding objects by solving the so-called Smoluchowski \citep{Smoluchowski16} or coagulation equation \citep{Safronov1972,Wetherill1989,Kenyon1998}, often with account being given to the possibility of collisional fragmentation \citep{Wetherill1993,Kenyon1999}. Global coagulation simulations, incorporating radial drift of mass in the disk driven by gas drag \citep{Spaute1991,Inaba2001,KenB2002,Kobayashi2010}, represent more sophisticated versions of such calculations. Some studies have also included direct N-body integration of a modest number ($\sim 10^3$) of planetary embryos formed as a result of planetesimal coagulation \citep{Bromley2011}, and accounted for the most recent results on the outcomes of planetesimal collisions \citep{SL09}. However, probably the most important ingredient of such coagulation-fragmentation calculations is the proper description of planetesimal dynamics \citep{Stewart1988,SI2000,R2003b,R2003d,R2003a,R2004}, which determines a particular mode of planetary growth \citep{Safronov1972,Ida1993,Kokubo1998,R2003c}.

While these developments significantly advanced our ideas about the origin of planets around single stars, a proper understanding of planet formation in stellar binaries requires a number of physical ingredients that are unique to these systems to be accounted for.  These include: (i) excitation of planetesimal eccentricities by gravitational perturbations due to both the stellar companion and the massive, non-axisymmetric protoplanetary disk (Sect.\ \ref{sect:perturb}); (ii) a unique size-dependent dynamical state (Sect.\ \ref{sect:dynamics}) in which planetesimals are a result of coupling between the aforementioned excitation and gas drag, strongly varying across the disk and featuring resonant locations; (iii) a nonstandard distribution of planetesimal collisional velocities (Sect. \ref{secCollVel}), different from that in disks around single stars; (iv) an important role played by the destructive collisional outcomes (erosion and catastrophic disruption) in determining the evolution of the planetesimal size distribution (Sect.\ \ref{sect:coll_outcomes}); and (v) nonuniform radial drift of planetesimals across the disk (Sect. \ref{inspiralEquations}) strongly affected by their nontrivial dynamics, which may lead to their trapping at certain locations. 
    
In the rest of this section we expand on how we calculate these effects in our models.


\subsection{Gravitational perturbations in S-type binaries}
\label{sect:perturb}


The motion of planetesimals around the primary in S-type binaries is affected by the gravity of both the secondary and the protoplanetary disk. Since the seminal work of \citet{Heppenheimer78}, the direct effect of the secondary's gravity has been accounted for using the secular approximation, which averages the (time-dependent) potential of the secondary over its orbit. However, since the gaseous protoplanetary disks in S-type binaries are themselves prone to developing eccentricity under the perturbation due to the secondary \citep{Paardekooper08,Regaly11}, one needs to also account for the gravitational potential of a non-axisymmetric, eccentric disk. \citet{SR15A} computed the potential due to an apsidally aligned disk with the surface density and eccentricity profiles given by Eq. (\ref{SigmaAndEccEquations}); their calculation was subsequently generalized for a disk with arbitrary profiles of $\Sigma(a)$ and $e_g(a)$ by \citet{Dav2018}.

Including disk potential, the secular disturbing function characterizing the perturbation of planetesimal motion by the external potential becomes \citep{SR15A}
\ba  
{\cal R}=a^2 n\left[\frac{1}{2}Ae^2 + B_b e\cos\varpi + B_d e\cos(\varpi-\varpi_d)\right],
\label{eq:R}
\ea  
where $a$, $n$, and $\varpi$ are the semimajor axis, mean motion, and apsidal angle of planetesimal orbit (both $\varpi$ and the disk apsidal angle, $\varpi_d$, are measured relative to the apsidal line of the binary), $B_b$ and $B_d$ describe the torques due to the non-axisymmetric components of the potentials of the secondary and the disk, and $A=A_b+A_d$ is the combined free precession rate due to the secondary ($A_b$) and the disk ($A_d$). The explicit expressions for $A_d$, $B_d$, $A_b$ and $B_b$ can be found in \citet{SR15A} (see their Eqs. [5]-[6] and [9]-[10], respectively). We note that while $A_d$ depends only on the $\Sigma(a)$ profile, $B_d$ is determined by both $\Sigma(a)$ and $e_g(a)$.

Of particular importance for planet formation in S-type binaries is the fact that the secondary companion always drives prograde apsidal precession of the planetesimal orbit ($A_b>0$), while the precession due to disk gravity is usually retrograde ($A_d<0$).  Since $A_b$ and $A_d$ vary with $a$ in different ways, one often finds that $A=A_b+A_d\to 0$ at a particular semimajor axis in the disk, if the disk is sufficiently massive ($\gtrsim 10^{-3}M_\odot$, \citealt{RS15b}) but not too massive (see below). At this radial location a secular resonance emerges, at which the eccentricities of planetesimals are driven to very high values (in the absence of damping effects). Such resonances due to disk gravity were previously uncovered in the context of planet formation in binaries in \citet{R13a}, \citet{Meschiari2014}, \citet{SR15A,SR15B}, and \citet{RS15a} and for massive disks with planets in \citet{Heppenheimer80}, \citet{Ward1981}, and \citet{Sefilian2020}.
\par
In addition to the secular perturbations we have considered, the companion star also produces short-period perturbations which act on the orbital period of the planetesimals \citep{Murray99}. The magnitude of the short-period perturbations is quite small for bodies which are close together.  If we consider two bodies with relative eccentricity, $e_r$, which will collide in the next orbit, then their distance is of order $e_r a_p$.  Their relative acceleration due to the tidal perturbation of the secondary is $GM_se_ra_p/R_s^3$.  If this acceleration acts over a time $\sim n_p^{-1}$ it results in a relative velocity $e_r v_K (a_p/R_s)^3$, which is suppressed by a factor $(a_p/R_s)^3$ relative to their typical relative velocity $e_r v_K$ in the absence of such short-period perturbations.  Over longer timescales such short-period perturbations average out to zero. 
\par
Because the disk we consider is small compared to the binary orbit, the resonant perturbations are also not important. Indeed, the only resonances present in the disk are of very high order, and therefore very narrow and unlikely to significantly affect the dynamics.

\par
The torque due to the secondary is expected to sharply (although not as discontinuously as we assume in Eq. (\ref{SigmaAndEccEquations})) truncate disk density at the semimajor axis $a_{\rm out}$.  We take this truncation into account when computing disk gravity.  The disk gravity-related coefficients $A_d$ and $B_d$ diverge logarithmically when a sharp disk edge is approached \citep{SR15A,Dav2018}.  For this reason, we leave a buffer of 1 AU between the outer edge of the disk and the region in which we model planetesimal growth.  The inner boundary of the disk is ignored in this calculation: As shown in Fig. 10 of \citet{SR15A}, provided that the inner boundary is located at no more than half the semimajor axis of interest, it makes no significant contribution to the disk disturbing function ${\cal R}$.


\subsection{Overview of planetesimal dynamics in S-type binaries}
\label{sect:dynamics}


The model for planetesimal dynamics employed in this work builds upon the understanding of the planetesimal disturbing function described in Sect. \ref{sect:perturb}, while additionally incorporating the dissipative effect of gas drag caused by the relative motion of planetesimals with respect to the noncircular streamlines of the eccentric protoplanetary disk. It has been developed in \citet{RS15a} for the realistic case of the quadratic gas drag \citep{Adachi76}, and here we briefly summarize its main features relevant for the present work.

\citet{RS15a} found that planetesimal dynamics admits two distinct regimes, depending on the degree of coupling between the planetesimal and the gas. Strongly coupled (small) planetesimals follow orbits nearly aligned with the local eccentric gas streamline. Weakly coupled (large) planetesimals have orbits nearly aligned with the forced eccentricity determined by the gravitational perturbations due to both the companion and the disk. Separating these two regimes is the characteristic planetesimal size $d_c$ that is defined as the planetesimal size at which the damping timescale\footnote{The damping timescale is defined such that the decay of planetesimal eccentricity vector ${\bf e}$ due to drag is described by $\dot {\bf e}_{\rm drag}=-({\bf e}-{\bf e}_g)/\tau_d$; it should be noted that $\tau_d\propto |{\bf e}-{\bf e}_g|^{-1}$ for the quadratic gas drag law.} $\tau_d$ of the planetesimal eccentricity (relative to the eccentricity vector of the local gas streamline ${\bf e}_g = e_g(\cos\varpi_d,\sin\varpi_d)$) is of order $A^{-1}$, where $A$ is the precession rate of free eccentricity in the combined potential of the disk and perturbing star (see Eq. (\ref{eq:R})). More specifically, \citet{RS15a} show that for a planetesimal of radius $d_p$
\begin{equation}
|A\tau_d| = \frac{d_p}{d_c} \left[\frac{1}{2} + \sqrt{\frac{1}{4} + \left (\frac{d_c}{d_p}\right)^2}\right]^{1/2}
\label{ATD}
\end{equation}
(see their Eq. [32]), so that $|A\tau_d|\approx 1.27$ for $d_p=d_c$.

\citet{RS15a} also show that the relative planetesimal-gas eccentricity $e_r=|{\bf e}-{\bf e}_g|$ is given by (see their Eq. [28])
\begin{equation}
e_r =  \frac{v_c}{v_K} \frac{A \tau_d}{\sqrt{1 + (A \tau_d)^2}}, 
\label{er}
\end{equation}
where $v_c = e_c v_K$ is the characteristic planetesimal-gas velocity for planetesimals with $d_p \geq d_c$: $e_c$ is the magnitude of the relative eccentricity between the gas streamlines and the forced eccentricity from the gravitational potential. An explicit expression for $e_c$ in terms of the disk and binary parameters only is provided by Eq. [29] of \citet{RS15a}. Knowing $e_c$ one can then directly compute $d_c$ through Eq. [31] of \citet{RS15a}. 

\begin{figure}
\centering
\includegraphics[width=.5\textwidth]{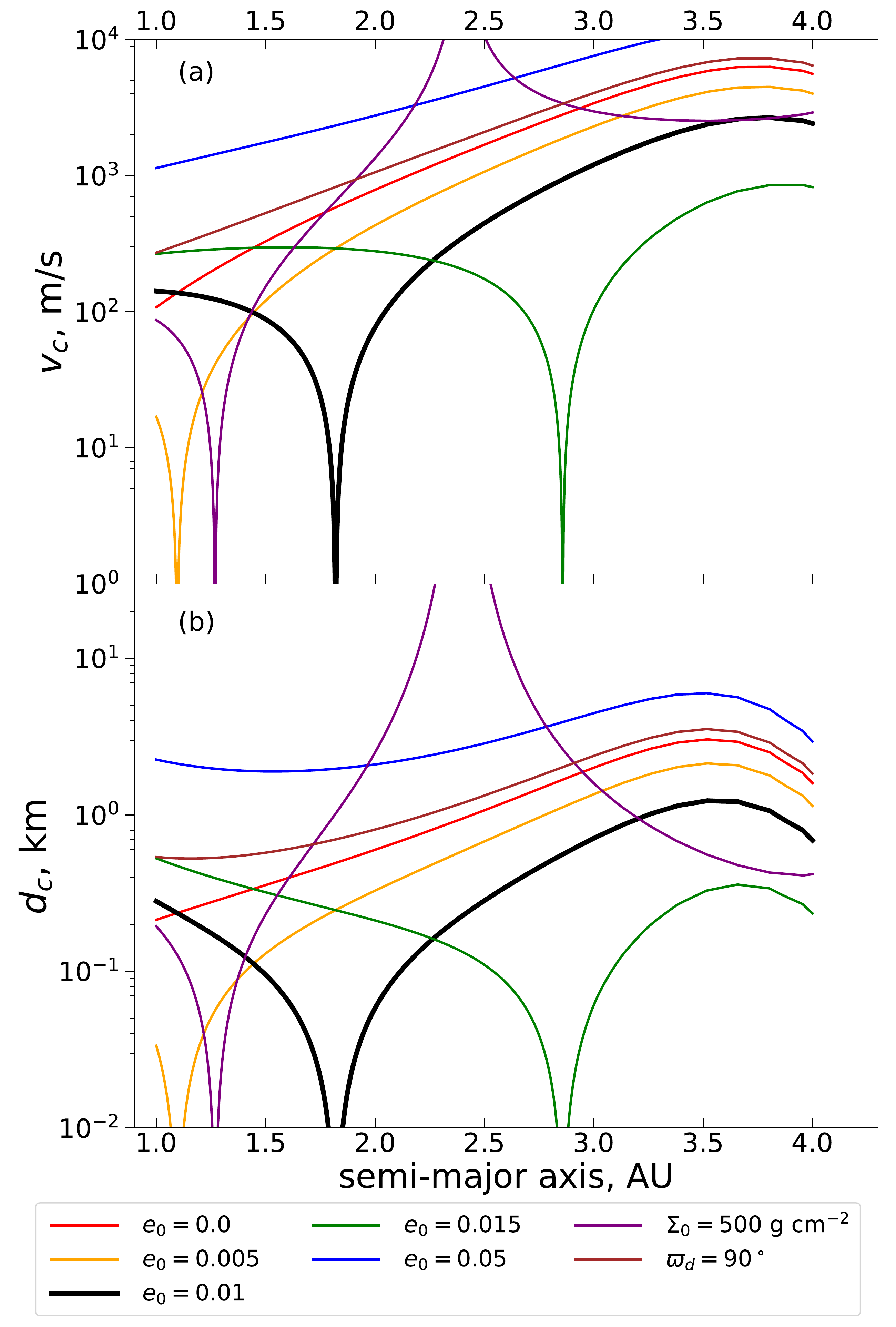}
\caption{Radial profiles of characteristic velocity $v_c$ (a) and planetesimal size $d_c$  (b) for different disk models, in which we vary disk eccentricity $e_0$, surface density normalization $\Sigma_0$, and apsidal angle $\varpi_d$ (one parameter at a time relative to the fiducial model described in Sect.\ \ref{sect:fiducial} and shown with the thick black curve), as indicated in the legend. See the text for more details on various features of these curves. }
\label{fig:ecdc}
\vspace{-.05cm}
\end{figure} 

Both $d_c$ and $v_c$ (or $e_c$) are the key metrics of planetesimal dynamics in S-type systems and will frequently appear in this work. Using these two parameters, strong coupling corresponds to small bodies, $d_p\ll d_c$, when $|A\tau_d|\approx (d_p/d_c)^{1/2}\ll 1$ and $e_r\approx e_c(d_p/d_c)^{1/2}\ll e_c$. Weak coupling is realized in the opposite limit of $d_p\ll d_c$, when $|A\tau_d|\approx d_p/d_c\gg 1$ and $e_r\approx e_c$. This suggests another intuitive definition of $v_c$ (or $e_c$) as the characteristic relative velocity (or eccentricity) between the weakly and strongly coupled (or large and small) objects. 

The parameters $v_c$ and $d_c$ vary significantly throughout the disk, and with different disk models, which is illustrated in Fig. \ref{fig:ecdc}, where we vary (one at a time) the values of the disk eccentricity $e_0$, surface density normalization $\Sigma_0$ and disk orientation $\varpi_d$. One can see that the behavior of $v_c$ and $d_c$ directly reflects many peculiar features of planetesimal dynamics that emerge when both secondary and disk gravity are important, as described in Sect. \ref{sect:perturb}. In particular, (i) $d_c$ and $v_c$ diverge at 2.5 AU in a model with $\Sigma_0=500$ g cm$^{-2}$ because of a secular resonance, which is not present in higher-mass models as it gets pushed out of the disk (i.e., disk gravity dominates planetesimal dynamics all the way out to $a_{\rm out}$); (ii) many models, in which the disk is apsidally aligned with the binary, feature a dynamically quiet location where $d_c,v_c\to 0$ (with the semimajor axis varying with $\Sigma_0$ and $e_0$); (iii) dynamically quiet locations do not appear for all values of $e_0$ and $\Sigma_0$ or for apsidally misaligned disks (e.g., for $\varpi_d=90^\circ$). These observations will greatly help us in interpreting the results of our detailed calculations in Sects. \ref{sect:fiducial} and \ref{sect:par-var}.

The concept of a dynamically quiet location in the disk will be very important for understanding the results of this work. \citet{RS15a} have demonstrated (see their Eq. [50]) that in an apsidally aligned disk, in the presence of gas drag, $d_c,e_c\to 0$ at the semimajor axis $a_{\rm dq}$ where
\ba  
A(a_{\rm dq})e_g(a_{\rm dq})+B_d(a_{\rm dq})+B_b(a_{\rm dq})=0.
\label{eq:dq_condition}
\ea  
At this location, the forced eccentricity is equal to the local gas eccentricity, so weakly and strongly coupled planetesimals are driven to identical orbits.

It should also be noted that the description of planetesimal dynamics provided in \citet{RS15a} and used in this work assumes that planetesimal eccentricities are at their steady state values, given by Eqs. [22]-[27] in \citet{RS15a}, to which they converge on a timescale $\sim \tau_d$ due to gas drag. This is a reasonable assumption since for small objects $\tau_d$ is short, $\lesssim 10^3$ yr for $d_p=1$ km objects, while for bigger bodies (for which $\tau_d\propto d_p$ is longer) the timescale for collisions with comparable objects (capable of significantly perturbing eccentricity) is long.


\subsection{Collision velocities}
\label{secCollVel}


An important feature of planetesimal dynamics in S-type binaries is that not only the planetesimal eccentricity but also the apsidal angle takes on a unique, size-dependent value at each semimajor axis as a result of a competition between gravitational perturbations and gas drag. This is different from disks around single stars, in which a balance between gas drag and dynamical excitation due to numerous embedded objects -- other planetesimals and planetary embryos -- results in a size-dependent planetesimal eccentricity distribution, whereas the apsidal angles of the planetesimals are randomly distributed. 

This difference becomes important when calculating the distribution of collision velocities of planetesimals, a procedure which is described in detail in  Sect.\ \ref{sect:collVelocity}. Around single stars, the relative velocities of planetesimals follow a 3D-Gaussian (or Schwarzschild) distribution \citep{SI2000,R2003a,R2003b}. But in S-type binaries, neglecting dynamical excitation by embedded objects, \citet{RS15a} found a very different shape for the distribution of relative velocities, given by Eq. (\ref{eq:vr}), with a velocity scale set by the relative eccentricity of the two colliding objects (see Eq. [62] of that paper). 

In this study we also account for random planetesimal motions due to dynamical excitation by embedded objects, in addition to those arising from the secondary and disk eccentricity. In particular, we assume that the two components of each planetesimal's inclination vector are drawn independently from a Gaussian distribution with standard deviation $\sigma_i$, which, for simplicity, is assumed to be independent of the size of the planetesimal.  We then assume each planetesimal's eccentricity vector to consist of the equilibrium component calculated in \citet{RS15a} as described in Sect. \ref{sect:dynamics}, plus a random component. We take this random component of the eccentricity vector to have standard deviation $\sigma_e$ which is twice $\sigma_i$ \citep{SI2000}, but allow horizontal and vertical motions to be independent of one another. The degree of random motion can therefore be fully parameterized by $\sigma_i$.
\par
If random motions result from viscous stirring, one would typically expect them to be on the order of the escape velocity from the planetesimals.  However, in the present perturbed system, because the relative motions between planetesimals are much larger than their random motions, the excitation from two-body encounters is lowered and the damping due to gas drag is enhanced.  On the other hand, if present, disk turbulence could excite larger random motions.  A study of the random planetesimal motions in a binary system would be subject to many uncertainties, and we do not attempt it here.  We note that our typical random motions correspond to the escape velocity from roughly 10 km planetesimals.

Varying $\sigma_i$ has two effects.  Let $e_{ij}$ be the relative eccentricity between two bodies in the absence of any random motions.  For collisions between bodies where $e_{ij} \gg \sigma_i$, varying $\sigma_i$ has little effect on the collision speed, but the frequency of such collisions varies inversely with $\sigma_i$, as $\sigma_i$ sets the thickness of the planetesimal disk and therefore the planetesimal number density.  In the limit that $e_{ij} \ll \sigma_i$, the collision velocity is proportional to $\sigma_i$ and the collision rate becomes independent of $\sigma_i$ provided that $\sigma_i v_K \gg v_{\rm esc}$ (where $v_{\rm esc}^2=2G(m_1+m_2)/(d_1+d_2)$ for colliding objects with sizes $d_1, d_2$ and masses $m_1, m_2$), that is, gravitational focusing is negligible; the collision rate scales as $\sigma_i^{-2}$ in the opposite limit $\sigma_i v_K \ll v_{\rm esc}$ when gravitational focusing enhances the collision rate.  The effect of random velocities on the collision velocities and rates and their implementation in our numerical framework are discussed in detail in Sects. \ref{sect:collVelocity} and \ref{sect:CollisionRate}.

\begin{figure}
\centering
\includegraphics[width=.45\textwidth]{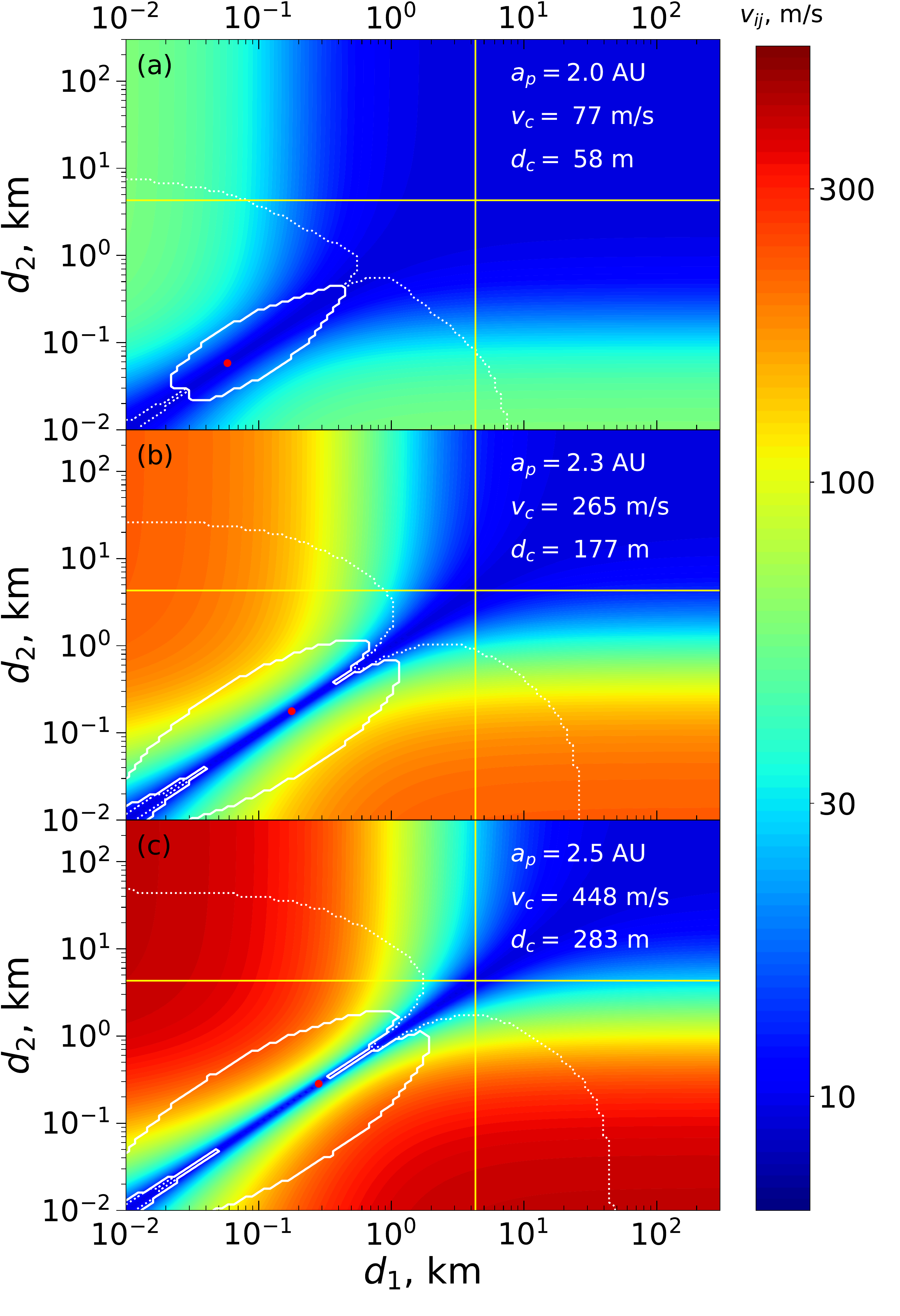}
\caption{Collision velocities between two planetesimals as a function of their sizes $d_1$ and $d_2$ in a disk with the fiducial set of parameters given in Sect. \ref{sect:fiducial} and $\sigma_i = 10^{-4}$.  Each panel corresponds to a different location in the disk, which leads to the varying values of $v_c$ and $d_c$ labeled on the panels.  Regions of erosive collisions are outlined in dotted white lines, and regions of catastrophic disruption are enclosed by solid white lines.  The horizontal and vertical yellow lines correspond to the initial planetesimal size $d_{\rm init}=4.3$ km used in the simulation described in Sect. \ref{singleAnnulus}.  The point $(d_c, d_c)$ is shown as a red dot. }
\label{env001}
\vspace{-.05cm}
\end{figure}

Figure \ref{env001} illustrates the collision velocities $v_{\rm coll}$ between planetesimals as a function of their sizes.  These are given by Eq. \eqref{combinedVcoll}, with $\lambda$ and $e_{\rm rand}$ assigned their mean values of 0.81 and $2 \sqrt{\pi} \sigma_i$, respectively.  This figure is made for the fiducial system described in Sect.\ \ref{sect:fiducial} with rather small $\sigma_i = 10^{-4}$, so that relative velocities are typically dominated by the secondary and disk forcing (i.e., by $e_{ij}$). The three panels differ in the assumed location $a_p$ of the colliding planetesimals in the disk (which determines their $v_c$ and $d_c$), as labeled on the figure. As mentioned in Sect.\ \ref{sect:dynamics}, $v_{\rm coll} \sim v_c$ for pairs of collision partners in which one object has size $d_1\ll d_c$ and another has size $d_2\gg d_c$.  If the collision partners have similar size, or both are much larger (or both much smaller) than $d_c$, then the collision velocities are determined predominantly by random motions and become independent of the sizes of the collision partners (since we take $\sigma_i$ to be size independent in this work). 
\par
As mentioned above, Fig. \ref{env001} would look very different if it were drawn assuming planetesimal dynamics typical for disks around single stars (i.e., without apsidal alignment): in that case the "valley" at $d_1\approx d_2$ would disappear and $v_{\rm coll}$ would monotonically increase along this line (since smaller planetesimals have lower random velocities as a result of gas drag, \citealt{R2003b,R2004}). These differences clearly demonstrate the importance of apsidal alignment naturally emerging in disks around binaries for setting the pattern of planetesimal collision velocities.


\subsection{Collision outcomes}
\label{sect:coll_outcomes}


Armed with the understanding of the distribution of relative velocities of planetesimals (Sect.\ \ref{secCollVel}), we determine the outcome of their physical collision using the recipes provided in the work of \citet{SL09}. This study assumes, quite generally, that as a result of a collision one is left with a large remnant body and a spectrum of small fragments. The size of the largest remnant is the main ingredient in their description of the collision outcome (see our Eq. \eqref{Mlr}). We provide the details of our implementation of this prescription in Sect.\ \ref{sect:fragmentMassDistribution}, but will briefly highlight the differences with the single-star case here.

Following \citet{RS15b}, we divide the collision outcomes into three classes.  We say that a collision leads to catastrophic disruption if the largest remnant contains less than half the combined mass of the two incoming planetesimals, erosion if the largest remnant is smaller than the larger of the two incoming bodies, and growth if the largest remnant is larger than either of the incoming bodies. We can then use the recipe of \citet{SL09} to delineate in Fig. \ref{env001} the domains in $d_1-d_2$ phase space, corresponding to each to type of the outcome: regions of catastrophic disruption lie within the solid white lines, regions of erosive collisions are outlined in dotted white, and regions leading to growth lie outside both white lines.  We use the ``strong rock" planetesimal composition of \citet{SL09} in this calculation.  One can see that, unless $v_c$ is very high, catastrophic disruption occurs only between collision partners with similar (but not equal) sizes. It is mainly relevant for objects with sizes around $d_c$, as that is where even objects of comparable size experience high-velocity collisions (although the regions of catastrophic disruption are not exactly centered on $d_c$ because the strength of planetesimals is size-dependent, with a minimum around 0.1 km \citep{SL09}.  In contrast to catastrophic disruption, erosion is possible even when the sizes of the colliding bodies are very unequal.

Due to the complexity of secular dynamics in binaries outlined in Sect. \ref{sect:dynamics}, there is a significant variation in the collisional environment across the disk, as different panels of Fig. \ref{env001} demonstrate. The regions corresponding to erosion and catastrophic disruption grow larger as the characteristic velocity $v_c$ increases. The weakest planetesimals with sizes around 0.1 km may be destroyed even in the absence of any secularly excited eccentricities, in collisions arising just from random motions, which is almost what is happening in Fig. \ref{env001}a. But it is still clear from panels (b) and (c) of that figure that secular pumping of planetesimal eccentricities in binaries endows these systems with much harsher collisional environments than in disks around single stars.

The heuristic picture of collision outcomes based on the relative velocity maps has been used in \citet{RS15a} and \citet{SR15B} to understand the possibility of planetary growth, which was assumed to take place when neither catastrophic disruption nor substantial erosion were taking place at certain locations in the disk. Obviously, such a simplistic approach cannot provide a perfect characterization of the conditions necessary for sustained planetesimal growth. Indeed, persistent planetesimal accretion may be possible even if some collisions are destructive; however, one does not know a priori how harsh of a collisional environment can be tolerated.  \citet{Thebault08,Thebault2009}, \citet{Thebault2011}, and \citet{RS15b} all had to make certain assumptions when drawing their conclusions on the likelihood of planet formation in binaries (see Sect. \ref{sect:heuristics} for an assessment of their realism). For this reason, in this work we follow the evolution of the planetesimal size distribution in full, using detailed  coagulation-fragmentation calculations with realistic physical inputs as described in Sect. \ref{codeOverview} and Appendix A.


\subsection{Radial inspiral}
\label{inspiralEquations}


In this work we also explicitly include the radial inspiral of planetesimals, which we expect to impact planetesimal growth in two different ways. First, inspiral drains the disk of solid material.  If this happens more rapidly than coagulation, then large bodies will not form due to lack of solid material needed for their growth.  Second, since small bodies inspiral more rapidly than large ones, one might expect that, once small bodies are flushed out, the effect of erosive collisions would be reduced, thus favoring the growth of large planetesimals.  

Using the results of \citet{Adachi76}, \citet{RS15b} found that the inspiral rate $\dot a_p$ is given by (their Eq. 11):
\begin{equation}
\label{dotap}
\dot a_p = -\pi \frac{a_p}{e_r \tau_d} \left(\frac{5}{8}e_r^2 + \eta^2\right)^{1/2}\left[\left(\frac{\alpha}{4} + \frac{5}{16}\right)e_r^2 + \eta \right],
\end{equation}
where $e_r$ and $\tau_d$ are given by Eqs. (\ref{er}) and  (\ref{ATD}), respectively, $\eta \sim (h/a_p)^2\sim 0.003$ ($h$ is the disk scale height) is a measure of the particle-gas velocity differential due to the pressure support in the gas disk, and $\alpha\sim 1$ is a constant that depends on the disk model as described in \citet{RS15b}. From Eqs. \eqref{ATD} and \eqref{er}, one can show \citep{RS15b} that $e_r \tau_d\propto d_p$. As a result, one finds that $\dot a_p \propto d_p^{-1}$ if either (i) $d_p \gg d_c$, when $e_r$ becomes constant, or (ii) $e_r \ll \eta$, so that the last term in brackets in Eq. (\ref{dotap}) dominates. 

\begin{figure}
\centering
\includegraphics[width=.52\textwidth]{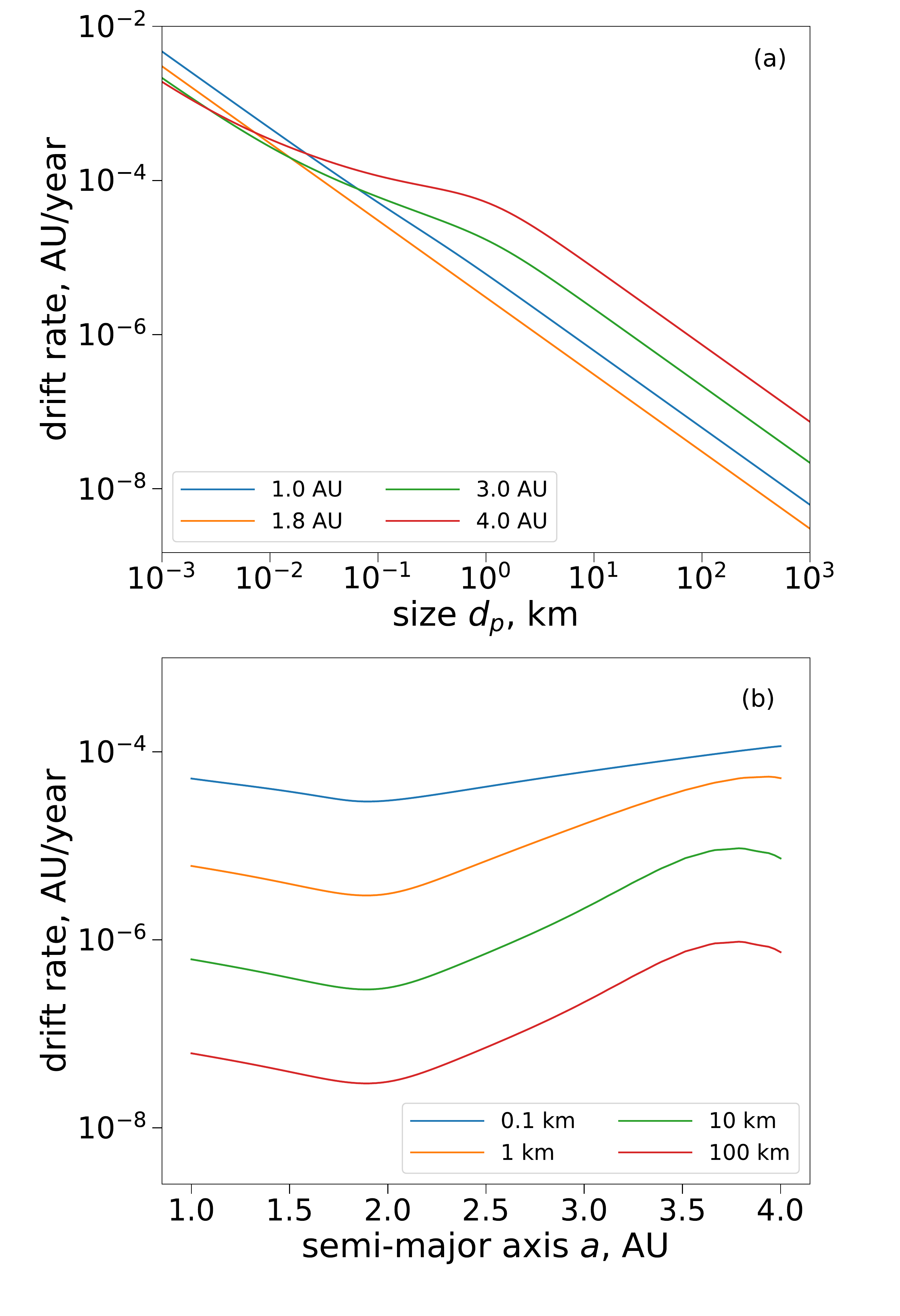}
\caption{Inspiral speed as a function of planetesimal size at different locations in the disk (a) (see legend),   and as a function of location in the disk  for different planetesimal sizes (b) (see legend).  The figure was made assuming the disk model described in Sect. \ref{sect:fiducial}, with $e_0 = 0.01$, $\Sigma_0 = 1000$ g cm$^{-2}$, and $\varpi_d = 0$. }
\label{inspiralSpeeds}
\vspace{-.05cm}
\end{figure}

Figure \ref{inspiralSpeeds} shows the speed of radial drift for different planetesimal sizes and locations in the disk in the $\gamma$ Cephei system, assuming fiducial disk parameters as in Sect.\ \ref{sect:fiducial}. We see that the curves in the upper panel match the limiting cases discussed in the previous paragraph; the regions where $\dot a_p \propto d_p^{-1}$ are joined by an intermediate region where the inspiral velocity depends less strongly on $d_p$.  In all cases, inspiral speed is a monotonically decreasing function of $d_p$, meaning that small bodies are preferentially flushed out of the system, thus reducing the erosion suffered by larger objects. 

In the lower panel we see that for moderately large planetesimals ($\gtrsim 0.1$ km), there is a noticeable dip in the inspiral rate in the broad region around 1.8 AU.  This is because for this disk model $v_c$ vanishes at 1.8 AU, so the only contribution to the inspiral comes from the sub-Keplerian rotation of the gas and not from any relative eccentricity between the planetesimals and gas, significantly reducing $\dot a_p$.  

These considerations highlight important differences in the radial drift behavior between the binary and single stars. First, high planetesimal eccentricities driven by the secular effect of the disk and the secondary result in faster radial drift in disks in binaries compared to single systems (see Eq. (\ref{dotap})).  Second, because higher $e_r$ increases the inspiral rate, planetesimals will be naturally concentrated in regions where $e_r$ is low, and thus where they can grow more easily.  This effect does not exist for planetesimals around single stars (unless one includes additional physics).


\section{Numerical ingredients of the model}
\label{codeOverview}


Our calculations of planetesimal growth in binaries employ a multi-annulus (or multi-zone) coagulation-fragmentation code specifically designed for this task. Here we briefly summarize its main features and provide references to relevant sections in Appendices \ref{firstAppendix} and \ref{analTests}. 


\subsection{Basics of the code structure}
\label{codeStruct}


Our code models the evolution of the planetesimal size distribution in $N_{\rm ann}$ discrete spatial annuli placed at different radii from the primary.  These should really be thought of as bins in the space of the semimajor axis, rather than radius,  but we do not make that distinction in the following discussion.  Within each annulus, the planetesimal size distribution is represented using $N_{\rm bins}$ logarithmically spaced mass bins.  At any point in time, in a given annulus, all of the information about the size distribution is encoded in a single vector $\vec n$, such that $n_i$ is the number of planetesimals in mass bin $i$. To follow the evolution of $n_i$ in each annulus the code performs two basic functions.  

First, in each of the annuli, it calculates the evolution of the size distribution due to planetesimal-planetesimal collisions. This is done in a standard fashion, effectively by solving a discretized version of the Smoluchowski equation \citep{Smoluchowski16} accounting for the possibility of fragmentation (Sects. \ref{collisions}, \ref{sect:fragmentMassDistribution}, and \ref{sect:coag_timestep}). Inclusion of fragmentation generally makes this an $O(N_{\rm bins}^3)$ calculation at each time step, which is numerically expensive. To overcome this problem, our code employs the new fragmentation algorithm developed in \citet{Rafikov20}, for which the numerical cost goes only as $O(N_{\rm bins}^2)$, as long as the size distribution of fragments formed in a collision of two objects is self-similar (which is a standard and physically motivated assumption anyway). The coagulation-fragmentation component of the code has been extensively tested against the known analytical solutions as described in Sects. \ref{coagTests} and \ref{fragTests}.

To compute the number of collisions between different mass bins we use collision rates calculated using the relative velocities from \citet{RS15a} and accounting for both the forced eccentricity and random velocities (see Sects. \ref{sect:dynamics} and \ref{secCollVel}). Our collision rates include gravitational focusing and smoothly interpolate between the shear- and dispersion-dominated velocity regimes (Sect. \ref{sect:CollisionRate}). The full distribution of collision velocities is also used to model collision outcomes following the recipes in \citet{SL09} (see Sects. \ref{sect:coll_outcomes} and \ref{sect:fragmentMassDistribution}). 

Our implementation of many code components has certain elements of stochasticity in it, which is quite important. It has been previously found \citep{Windmark12, Garaud13} that including the distribution of collision velocities can qualitatively change the outcome of the coagulation process.  In particular, in a study of dust growth, \citet{Windmark12} found that using a Maxwellian collision velocity distribution instead of a delta function at the rms velocity of the Maxwellian can cause a few particles to experience a series of lucky low-velocity collisions and grow to larger sizes; the larger bodies formed via lucky collisions are more resistant to destruction in subsequent collisions and would continue to grow to arbitrarily large sizes.  In our case, the strong variation in collision velocity with collision partner size provides the dominant source of such randomness.  In addition, even for bodies with given sizes, there is a range in their relative collision velocity (see Sect. \ref{sect:collVelocity}).  To account for such a possibility, we also draw collision velocity between two bodies from a physically motivated distribution (\citealt{RS15a}; see Sect. \ref{secCollVel}), as described in Sect. \ref{sect:collVelocity}. 

The second operation performed by the code is the radial redistribution of mass between different annuli caused by the size-dependent inward migration of planetesimals due to gas drag (Sect.\ \ref{inspiralEquations}). The implementation of this procedure and its tests are described in Sects. \ref{sect:radialInspiral}, \ref{sect:drift_timestep}, and \ref{inspiralTests}, respectively. 
\par
Our model implicitly assumes that other than redistribution caused by radial drift, planetesimals interact only with other planetesimals within their semimajor axis bin. In practice this is not strictly true: Planetesimals in neighboring bins can also collide. However, because the growth we are modeling occurs in regions where the relative eccentricity between planetesimals of different sizes is very small (typically less than 1\%), we expect the coupling of planetesimals with those in adjacent semimajor axis bins to be a minor effect.


\subsection{Parameters of the numerical model}
\label{sect:modelpars}


Multiple modules comprising our code employ a number of parameters, both numerical and physical. All our simulations use a standard set of these parameters described here and listed in Table \ref{tbl:num-pars}.

Our grid in mass space employs $N_{\rm bins}=100$ bins; the smallest size (corresponding to the lowest bin) below which mass is supposed to be flushed out from the system due to gas drag is $d_p=10$ m. The largest mass bin in our simulations corresponds to an object with $d_p=300$ km, since beyond this point destructive collisions are no longer able to prevent growth to larger sizes. This results in a mass ratio between (logarithmically uniformly spaced) mass bins of $M_{i+1}/M_i = 1.37$. To represent our global simulation domain extending from $1$ AU to $4$ AU we use $N_{\rm ann}=60$ annuli, spaced in semimajor axis as described in Sect. \ref{sect:radialInspiral}. In choosing this radial range we are motivated by the planetary semimajor axes in several S-type systems: $a_{\rm pl}=2$ AU in $\gamma$ Cephei, $a_{\rm pl}=2.6$ AU in HD 196885, and $a_{\rm pl}=1.6$ AU in HD 41004 \citep{Chauvin2011}.

Some additional parameters used in our simulations are as follows: the slope of the fragment mass spectrum $\xi=-1$ (Sect. \ref{sect:fragmentMassDistribution}); parameter $b=10^{-2}$ used to decide whether the largest fragment is distinct from the continuous spectrum of fragments (Sect. \ref{sect:fragmentMassDistribution}); tolerance parameters $\epsilon_1=0.05$ and $\epsilon_2=10^{-6}$ used in time step determination  (Sect. \ref{sect:timestep}). 

We studied the sensitivity of the code to changes of these parameters and found only slight variations in the outcomes (see Appendix \ref{modelParams}).  


\section{Results: Fiducial model}
\label{sect:fiducial}


We start presentation of our results by describing in detail a particular (fiducial) simulation, which will also illustrate the metric we employ to determine the success of planet formation in a given disk model (Sect. \ref{sect:outcomeDiagnostic}). Our fiducial disk model uses the orbital parameters of the $\gamma$ Cephei system and disk characteristics as described in Sect. \ref{MSPD}. For the disk and planetesimal parameters it adopts $e_0 = 0.01$, $\Sigma_0 = 1000$ g cm$^{-2}$ at $a_0 = 1$AU, $\varpi_d = 0^\circ$, $\sigma_i=10^{-4}$, and a solid to gas ratio of 0.01.  The total disk mass is $3.7M_\mathrm{J}$.  We present the results for other disk models in Sect. \ref{sect:par-var}.

To assist in interpreting the outcomes of our calculations, we first describe the results of several one-zone simulations of planetesimal growth at different semimajor axes in Sect. \ref{singleAnnulus}. We then move on to the global, multi-zone calculation of planet formation in the fiducial model in Sect. \ref{fullCalc}. This approach not only allows us to separate local and global factors affecting planet formation but also highlights the role played by the gas drag-driven radial inspiral (naturally absent in the former case).


\subsection{Coagulation in a single annulus}
\label{singleAnnulus}


\begin{figure}
\centering
\includegraphics[width=.5\textwidth]{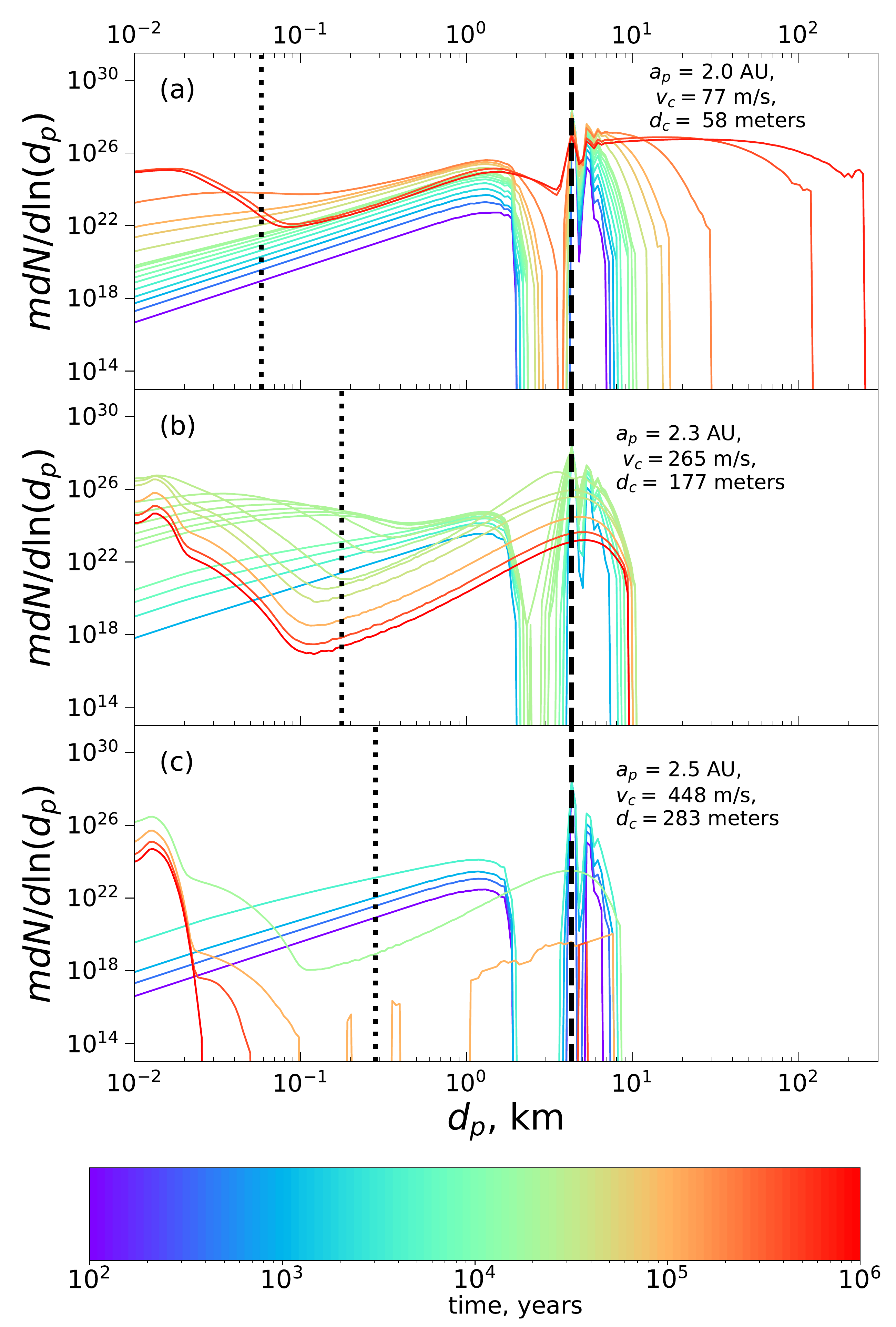}
\caption{Evolution of the size distribution as a function of time in a set of single-annulus simulations, described in Sect. \ref{singleAnnulus}, which should be consulted for details. The three panels correspond to environments with the different values of the critical velocity $v_c$ and critical size $d_c$, previously shown in Fig. \ref{env001}, at different locations (as labeled) in the disk around the primary star in the $\gamma$ Cephei system.  The vertical dotted lines correspond to the critical size $d_c$, and the vertical dashed lines correspond to the initial size of the planetesimals $d_{\rm init}$. }
\label{f001}
\vspace{-.05cm}
\end{figure} 

We ran our single-annulus calculations of planetesimal growth at three semimajor axes -- $a_p=2$ AU, $2.3$ AU, and $2.5$ AU -- in our fiducial disk model. These are the same locations for which Fig. \ref{env001} illustrates the distribution of collision velocities, which helps us in understanding the role of planetesimal dynamics on their growth. The simulations started with all planetesimals having the same size $d_{\rm init}=4.3$ km; for the adopted bulk density $\rho_p = 3$ g cm$^{-3}$ this corresponds to planetesimal mass of $m = 10^{18}$ g. 

Figure \ref{f001} illustrates how the size distribution in our model system changes as a function of time. What is actually shown is $mdN/d\ln d_p$, a variable equal to the mass per unit log planetesimal size.  Different curves correspond to different times since the beginning of the simulation, as indicated by the color bar. The simulation for panel (a) was stopped when the largest body went beyond the largest mass bin in the simulation ($d_p=300$ km). In panels (b) and (c), the simulation was stopped after $10^6$ yr. 

In all plots, the curves corresponding to times $\lesssim 10^4$ years display features due to the initial conditions and finite spacing of the mass bins.  Initially, all of the mass is distributed in just two mass bins surrounding the initial (seed) size $d_{\rm init}$, which is marked by the dashed vertical black line.  The wiggles just to the right of $d_{\rm init}$ are related to the finite spacing of the mass bins. The gap just to the left of $d_{\rm init}$ exists because debris of that size is not created in collisions of objects with size $d_{\rm init}$ --- they produce only smaller fragments to the left of the gap. The size spectrum of objects to the left from the gap is universal at early times ($\sim 10^3-10^4$ yr), as it simply reflects the size distribution of fragments formed in collisions of roughly equal mass planetesimals of size $d_{\rm init}$. The gap region is only later filled in by coagulation of the debris particles produced early on and erosion of the seed bodies by small debris. 

Over time, memory of the initial conditions is erased, the gaps near $d_{\rm init}$ get filled, and the size distribution becomes smooth. There are some other notable features that develop in the size distribution at late times. First, after a few times $10^4$ years $mdN/d\ln d_p$ starts featuring a dip around $d_p\sim 0.1$ km, especially pronounced in panel (b). Its location corresponds to the planetesimal size which is most easily destroyed in collisions (see \citealt{RS15b}). In each panel, the vertical dotted line corresponds to $d_c$ for that simulation, which varies by a factor of several across the different environments, but stays pretty close to $d_p\sim 0.1$ km. Particles with sizes around $d_c$ experience high-velocity collisions with almost all other bodies, and are therefore preferentially destroyed, which likely affects the depth of the dip in different panels. 

Second, one can see (especially in panel (b)) $mdN/d\ln d_p$ developing power law behavior to the left and right of the $d_p\sim 0.1$ km line, with different slopes. We believe this behavior to be real but will refrain from offering an explanation for the slope of these segments. This cannot be done based on standard results regarding fragmentation \citep{OBrien03, Tanaka96} since they assume a power-law dependence of the collision rate on colliding partner size.

Third, panel (c) exhibits an accumulation of particles of $10-20$ m in size; a similar accumulation is also present in panel (b) as a small bump in the same size range. This feature is artificial and arises because in our calculations we do not follow the fate of debris smaller than 10 m in size, simply removing it from the system. As a result, 10-20 m objects do not have smaller bodies to erode them, and their own relative velocities are too low to result in catastrophic disruption: because of their comparable size their collisional velocities are small, thanks to the peculiarity of planetesimal dynamics in binaries (see Fig. \ref{env001}). Nevertheless, the emergence of this feature near the bottom of our mass grid does not affect the outcome of planetesimal evolution for $d_p\gtrsim d_{\rm init}$.  

Careful examination of our simulation outputs shows that the largest objects grow primarily in collisions with objects of similar size, certainly for $d_p> d_{\rm init}$. This is because most solid mass is contained in such objects, but also because growth is most favorable in collisions of such objects.

\begin{figure}
\centering
\includegraphics[width=.5\textwidth]{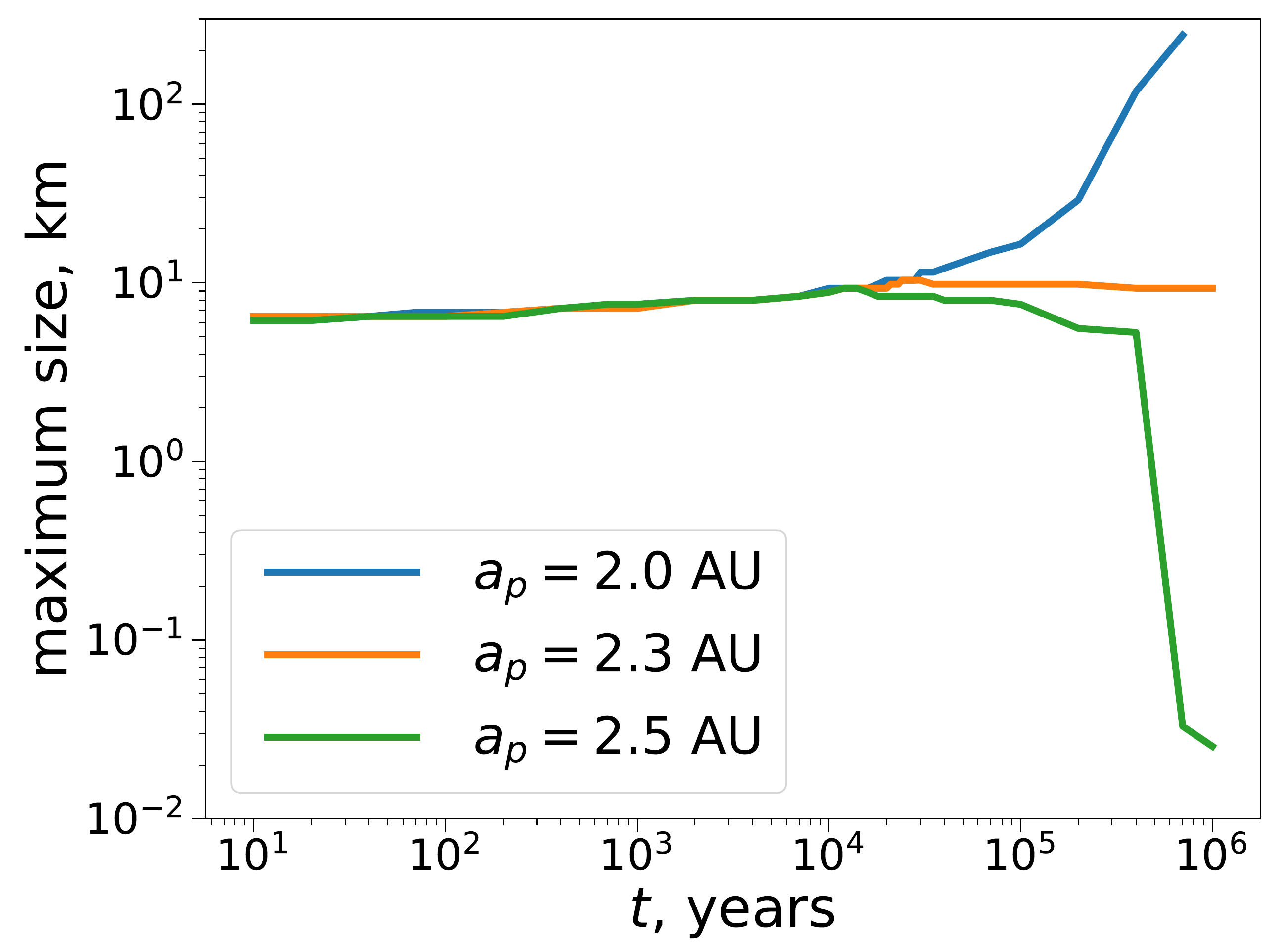}
\caption{Time evolution of the size of the biggest object in the system, shown for the three simulations depicted in Fig. \ref{f001} and labeled according to their semimajor axis. }
\label{fig:biggest-single}
\vspace{-.05cm}
\end{figure} 

We observe, not surprisingly, that higher $v_c$ suppresses growth.  In panel (a), with $v_c = 77$ m/s, growth to hundreds of km occurs within several $\times 10^5$ years. This can also be seen in Fig. \ref{fig:biggest-single} where we show the size of the largest body as a function of time for the three local simulations discussed here. Were we to simulate larger bodies, beyond 300 km, growth would continue in panel (a) until the mass reservoir is fully  depleted. Because of low $v_c$, even when a significant amount of solid mass in the system has been converted to debris (at late times), collisions of the largest objects with debris particles are not energetic enough to significantly impede their growth. 

In panel (b) with $v_c = 265$ m/s, planetesimals can only grow to a few times the initial size. After that growth of the largest objects stalls (see also Fig. \ref{fig:biggest-single}), while the population of objects with $d_p\sim d_{\rm init}$ gets gradually eroded by smaller objects produced in previous collisions (this can be seen in the decay of amplitude of $dN/d\ln d_p$ around that size). If we were to run this simulation beyond $1$ Myr, large objects with $d_p\sim d_{\rm init}$ would eventually be eroded, followed by smaller bodies also grinding themselves down. 

Finally, in panel (c) of Fig. \ref{f001}, collision velocities are so high that there is barely any growth --- the largest objects present in the system at early time have grown in size beyond $d_{\rm init}$ only by $\sim 2$. But after a population of small debris develops in the system (already by $10^3$ yr), Fig. \ref{fig:biggest-single} shows that large bodies get eroded away by $10^5$ yr, leaving only a small population of debris near the lower end of the simulation mass range (which is an artifact of our calculation, as we mentioned earlier).

Figure \ref{fig:biggest-single} shows that in all three environments, the largest bodies initially evolve in a very similar fashion, steadily growing in size. This universality is caused by the fact that initial growth occurs by mergers of objects with $d_p=d_{\rm init}$, which have low relative velocities. The differences emerge only after $\sim 10^4$ yr when a sufficient amount of small debris capable of efficiently eroding big bodies in high-$v_c$ environments accumulates in the system.

Here we again consider Fig. \ref{env001}, which illustrates, in particular, collision outcomes as a function of the sizes of the collision partners, for the same environments that the simulations shown in Fig. \ref{f001} were run for. Interestingly, we see that to halt planetesimal growth it is not necessary to have catastrophic disruption of objects at $d_p\sim d_{\rm init}$. Indeed, although Fig. \ref{f001} shows that growth is halted in panels (B) and (C), Fig. \ref{env001} shows no catastrophic disruption of planetesimals at the initial size: the point $(d_{\rm init},d_{\rm init})$ is outside the solid white contours for all dynamic  environments, as $d_{\rm init}=4.3$ km considerably exceeds $d_c$ in all panels. 

It may seem strange that in all three panels in Fig. \ref{env001} the point $(d_{\rm init},d_{\rm init})$ also lies outside the dotted contours, bordering the region of erosive encounters --- it would seem natural that one should then find growth starting with a population of $d_p=d_{\rm init}$ objects in all three environments. The resolution of this apparent paradox involves two factors. First, according to our definition of erosion in Fig. \ref{env001} dotted contours correspond to collisions, in which the mass of the largest remnant is equal to the mass of the bigger object (target) involved in the collision; the mass equal to the mass of the smaller object (projectile) is reduced to debris. Similarly, even though the point $(d_{\rm init},d_{\rm init})$ sits outside the dotted contour, some amount of debris will still be produced in collisions of two objects of initial size $d_{\rm init}$; the amount of created debris is larger the closer this point is to the dotted contour (i.e., in panels (b) and (c) relative to (a)). And once some debris particles are created, Fig. \ref{env001} shows that their collisions with seed bodies ($d_p=d_{\rm init}$) are very erosive, especially in panel (c), but only barely so in panel (a). 

Second, relative velocity maps in Fig. \ref{env001} assume a particular (mean) value of encounter velocity, whereas in practice $v_{\rm coll}$ has a reasonably broad distribution around this mean (see Sect.\ \ref{sect:collVelocity}). As a result, some collisions occur at higher velocities and are more destructive --- a fact that is fully accounted for in our coagulation-fragmentation simulations depicted in Fig. \ref{f001}. This is another clear illustration of the importance of considering the distribution of planetesimal velocities when studying their growth. The combination of the two aforementioned factors leads to erosion eventually becoming the dominant player in the collisional evolution in panels (b) and (c) of Fig. \ref{f001}.


\subsection{Full global calculation with inspiral}
\label{fullCalc}


We now proceed to describe a fully global, multi-annulus simulation of planetesimal growth, fully accounting for the effects of size-dependent radial inspiral (see Sect. \ref{inspiralEquations}) and using the fiducial disk parameters. This simulation is also used to motivate our choice of a particular metric determining whether planetesimal growth in the system successfully overcomes the fragmentation barrier (see next section).


\subsubsection{Outcome diagnostic}
\label{sect:outcomeDiagnostic}


In any reasonable environment, planetesimal growth would proceed to very large sizes, eventually leading to planet formation, provided that the initial planetesimal size $d_{\rm init}$ is large enough. Indeed, objects with sizes $d_p\gg d_c$ do not experience catastrophic disruption and get eroded only by much smaller planetesimals, which cannot compete with the addition of mass in collisions with larger bodies. Thus, given large enough $d_{\rm init}$, planet formation is guaranteed to be successful (e.g., \citealt{Thebault2011}) even in dynamically harsh environments of the binary stars.

However, in many environments, if the initial planetesimals are too small, then they will be eroded or destroyed and large bodies will not form (like in panels (b) and (c) of Fig. \ref{f001}). For this reason, the appropriate question to ask is not whether planetesimal growth can occur, as the answer would depend on the initial condition -- the size of the seed planetesimals $d_{\rm init}$. Instead, one should be trying to figure out from what initial planetesimal size $d_{\rm init}$ can sustained planetesimal growth occur.  In a given dynamical environment, there will be a minimum size $d_{\rm min}$ such that if the initial bodies are smaller than $d_{\rm min}$ then planetesimal growth in the system will eventually be halted, as in Fig. \ref{f001}b,c, whereas if seed objects have $d_{\rm init}>d_{\rm min}$, then planetesimal coagulation will eventually form large bodies, thus overcoming the fragmentation barrier. 

This question, in principle, can be asked locally, at a given semimajor axis in the disk. In this case Fig. \ref{f001} makes it clear that, for a fiducial disk model, $d_{\rm min}<4.3$ km at $a_p=2$ AU, whereas $d_{\rm min}>4.3$ km at $a_p=2.3$ and $2.5$ AU. However, the calculations presented in Sect.\ \ref{singleAnnulus} do not account for radial mass transport due to gas drag and may thus be inaccurate. For this reason, we will be more interested in a general question of what $d_{\rm min}$ is needed for the fragmentation barrier to be overcome and for a planet to form at some location in the whole disk, given its parameters. Once $d_{\rm min}$ is determined, some other information, for example, the location where planetesimal growth is fastest, would be the outcome of our global calculations, giving them certain predictive power. Of course, $d_{\rm init}$ may (and probably should) vary across the disk. Moreover, at each location, a whole spectrum of initial sizes is likely to be present. However, in this work, to reduce the number of degrees of freedom, we assume that seed planetesimals have the same size $d_{\rm init}$ across the whole disk and determine a single value of $d_{\rm min}$ for a given disk model based on this assumption. 

To determine $d_{\rm min}$ defined in this manner, we employed the following iterative algorithm. We chose a value of $d_{\rm min}$ and used that as our starting size $d_{\rm init}$. The simulation was run until one of three conditions were met: (i) 1 Myr has passed, (ii) a 300 km body is formed, or (iii) there is no longer sufficient solid mass remaining in the simulation to produce a 300 km body. If a 300 km body forms, then we know $d_{\rm min} < d_{\rm init}$, whereas if it does not, then $d_{\rm min} > d_{\rm init}$.  By trying several values of $d_{\rm init}$ and running our global simulations for each of them, we were able to first bracket $d_{\rm min}$ within some interval, and then to converge to its value with high accuracy. This procedure is demonstrated next (allowing us to also illustrate the sensitivity of global planetesimal growth to $d_{\rm init}$).

\begin{figure}
\centering
\includegraphics[width=.5\textwidth]{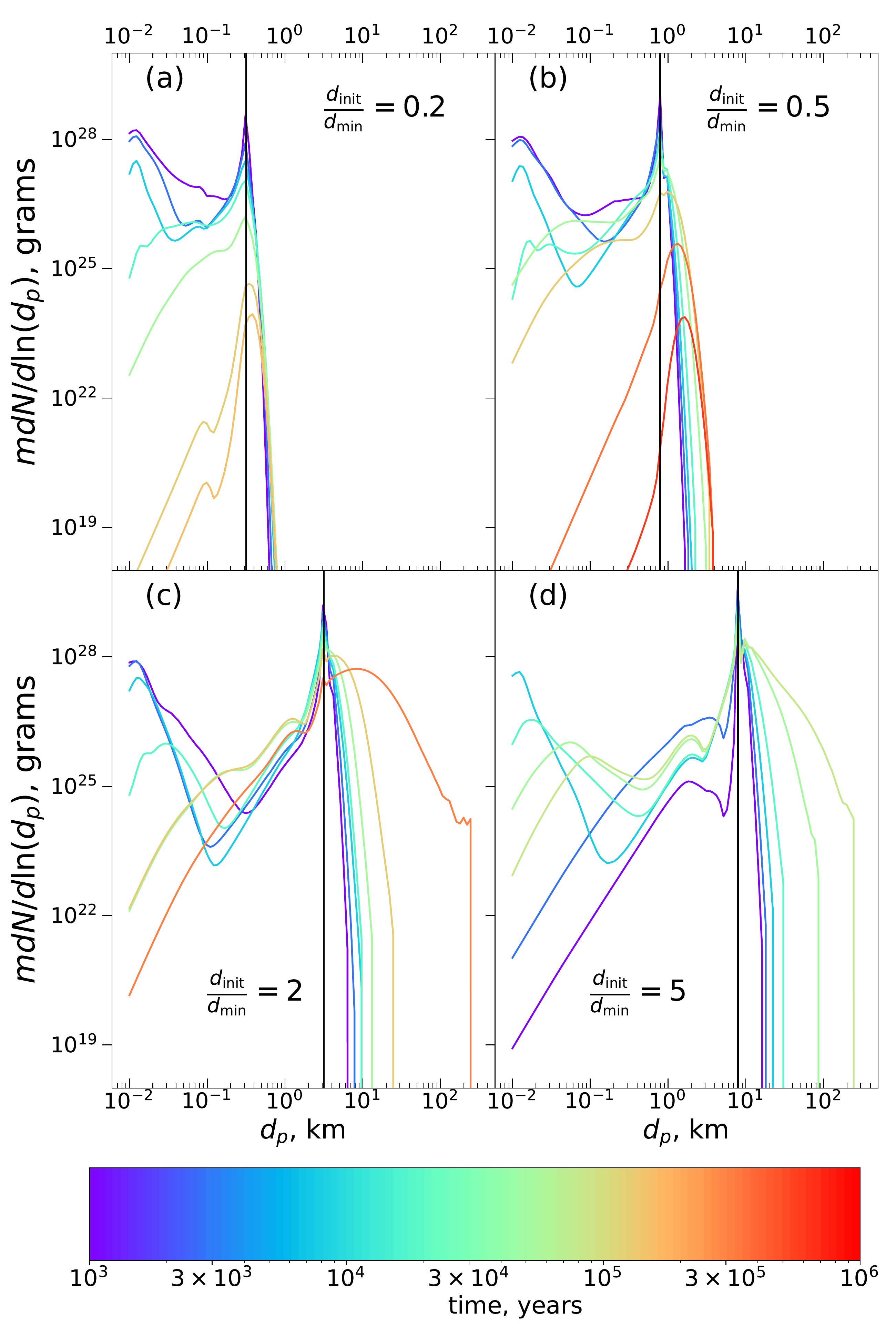}
\caption{Evolution of the mass distribution in the entire disk in our fiducial multi-zone model (Simulation 1).  Different panels are produced for different initial planetesimal sizes $d_{\rm init}$ in terms of $d_{\rm min}$ (as labeled on the plot), which is $d_{\rm min}=1.6$ km for this simulation. }
\label{totalMass}
\vspace{-.05cm}
\end{figure} 

Since the planetesimal composition is uncertain, in addition to our default simulations that use the material properties of ``strong rock" bodies from \citet{SL09},  we also ran simulations using the strength parameters for their ``rubble pile" planetesimals.  We denote the minimum size assuming rubble pile planetesimals as $d_{\rm min}^{\rm rubble}$.  


\subsubsection{Evolution of the system as a whole} 


Figure \ref{totalMass} shows the evolution of the planetesimal size distribution in the entire disk for several runs of our fiducial simulation with different initial sizes $d_{\rm init}$ expressed in terms of $d_{\rm min} = 1.6$ km, as determined for this disk model. Curves of different color correspond to different times since the simulation was started. Because collisional processes proceed at different rates across the disk, reflecting the complexity of planetesimal dynamics in binaries (see Fig. \ref{fig:ecdc}), the evolution of the planetesimal mass spectrum looks somewhat different from that in a single annulus (see Fig. \ref{f001}; e.g., the initial gaps to the left of $d_{\rm init}$ are barely there, and there is no pile-up at small sizes). This is to be expected since the global size spectrum in Fig. \ref{totalMass} is an average of multiple single-zone size distributions (such as the ones shown in Fig. \ref{f001}) with the added complication of radial inspiral of solids. Nevertheless, some key features of the size distribution evolution (e.g., a dip in $mdN/d\ln d_p$ near $d_p\sim 0.1$ km, mass flow toward small sizes, and so on) are still preserved even in the full disk calculations. 

We can see that in panels (a) and (b), essentially no growth occurs beyond the initial size.  The behavior shown in these panels differs only in how rapidly the bodies are ground down.  The lack of any appreciable growth in panel (b) with initial size only a factor of two less than $d_{\rm min}$ shows that there is a sharp transition over less than a factor of two in $d_{\rm init}$ from growth to large sizes to essentially no growth whatsoever. 

In panels (c) and (d), coagulation to 300 km does occur. In both cases, though, less than half of the initial solid mass in the simulation remains at the end.  The time required to produce a 300 km body is almost four times less for the simulation with $d_{\rm init}/d_{\rm min} = 5$ than for the one with $d_{\rm init}/d_{\rm min} = 2$ ($8 \times 10^{4}$ years vs. $3 \times 10^5$  years), and in the former case more than twice as much solid material remains (47\% of the original $7M_\oplus$ in panel (d) vs. 21\% in panel (c)). In the simulation with $d_{\rm init}/d_{\rm min} = 2$ (5), 79\% (88\%) of the mass is lost to the creation of rubble smaller than the smallest body tracked in the simulation, with the remaining amount lost to inspiral into the central star. 

\begin{figure}
\centering
\includegraphics[width=.5\textwidth]{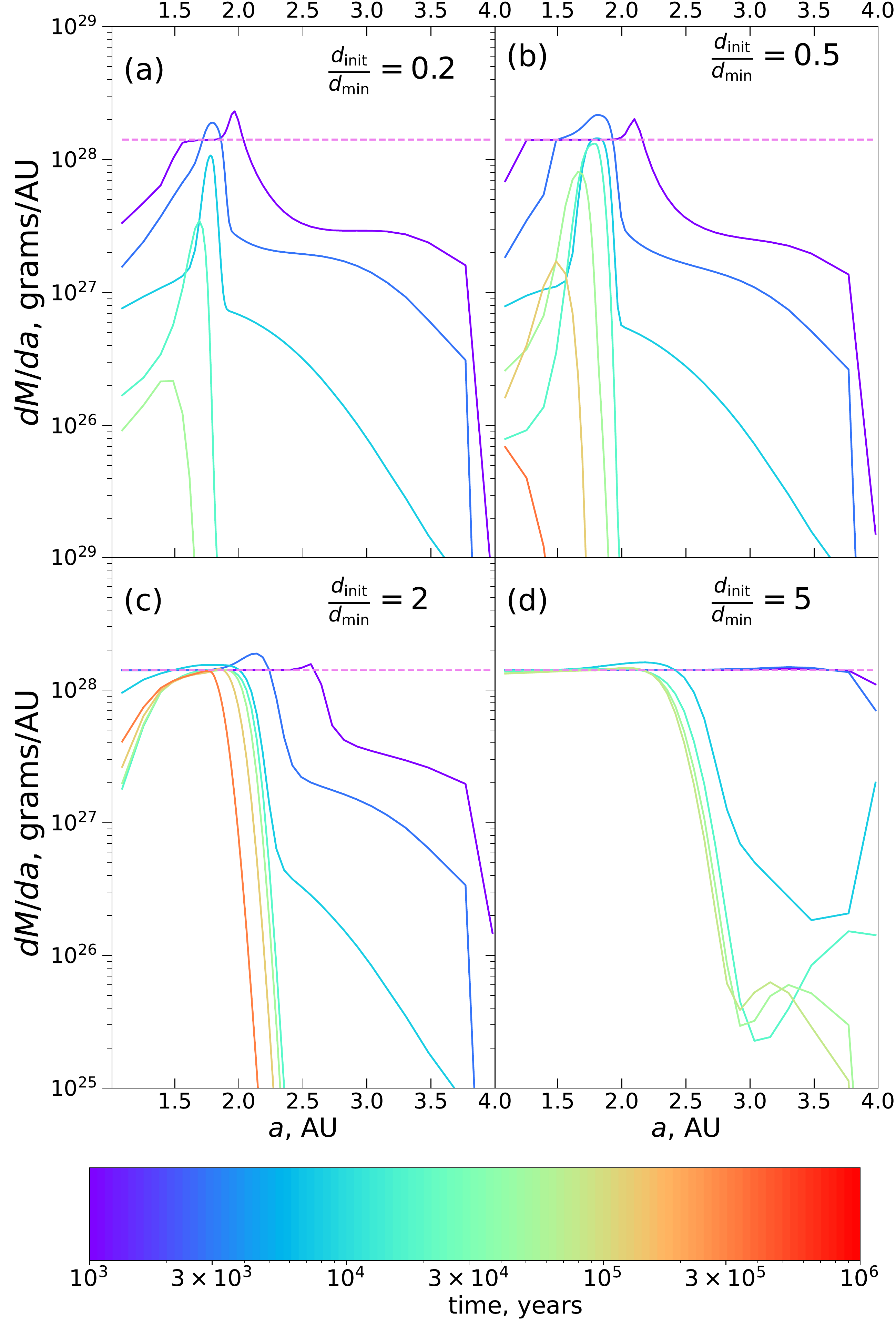}
\caption{Mass in planetesimals of all sizes (tracked in our mass grid) in the disk per unit semimajor axis $dM/da=2\pi a\Sigma(a)$ as a function of $a$ in our fiducial model at different moments of time (see color bar). The different panels correspond to the same initial planetesimal sizes $d_{\rm init}$ as in Fig. \ref{totalMass}.  The dashed violet horizontal line corresponds to the mass distribution at $t = 0$.   }
\label{MofR}
\vspace{-.05cm}
\end{figure} 

\begin{figure}
\centering
\includegraphics[width=.5\textwidth]{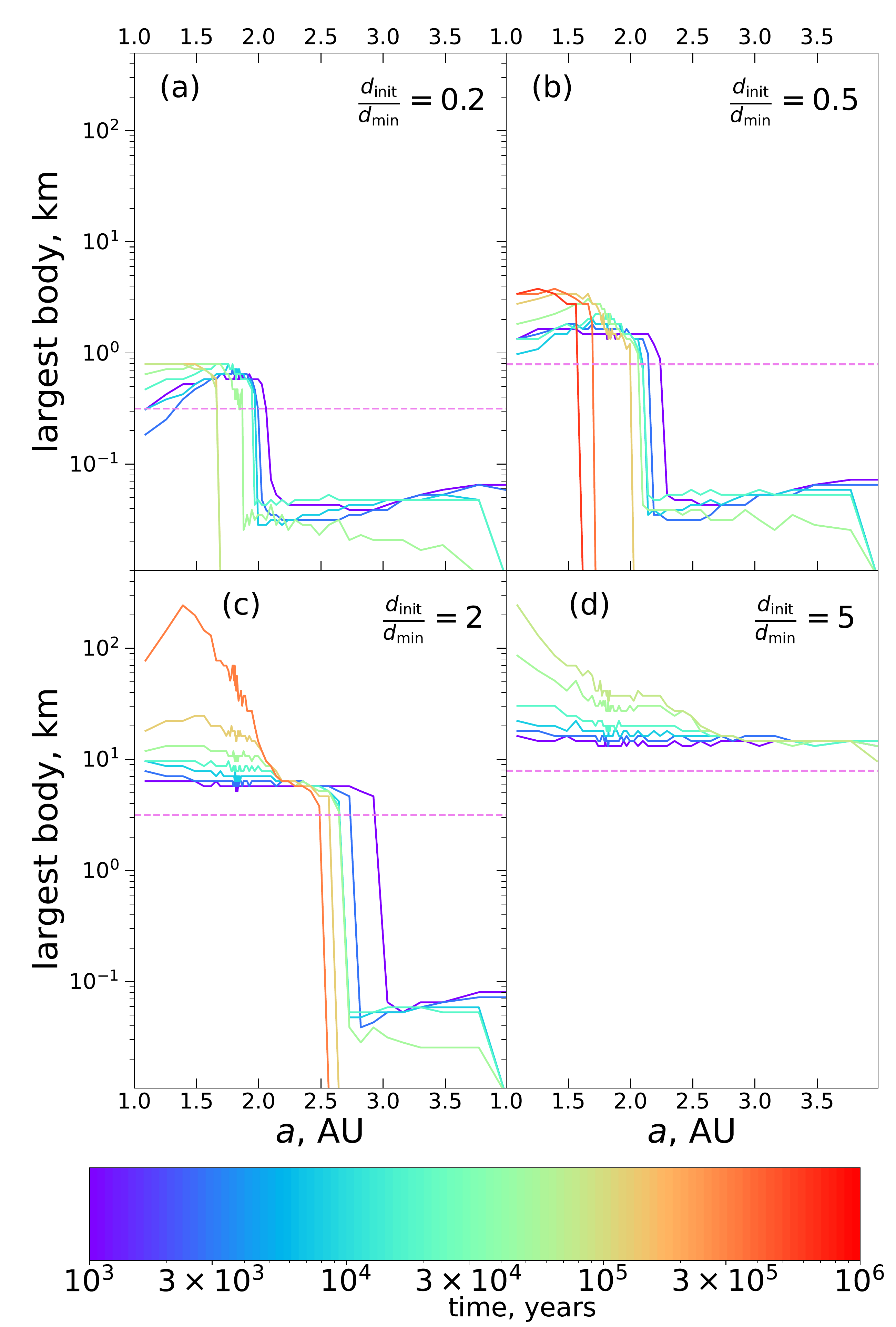}
\caption{Similar to Fig. \ref{MofR} but showing the size of the largest body as a function of the semimajor axis. The dashed violet horizontal line now shows the size of seed planetesimals $d_{\rm init}$.}
\label{fig:biggest-of-r}
\vspace{-.05cm}
\end{figure} 

Figures \ref{MofR} and \ref{fig:biggest-of-r} provide details on how the coagulation-fragmentation process proceeds as a function of semimajor axis in the disk, for each global simulation shown in Fig. \ref{totalMass}. Figure \ref{MofR} shows $dM/da$ --- the total solid mass (in planetesimals of all sizes) per unit $a$, whereas Fig. \ref{fig:biggest-of-r} depicts the size of the largest body (in a given annulus), both as functions of $a$ and time. In Fig. \ref{MofR}, the horizontal dashed violet line represents the initial mass distribution, which is flat because of our chosen $\Sigma$ profile (see Eq. \eqref{SigmaAndEccEquations}).  In Fig.  \ref{fig:biggest-of-r} that line corresponds to the size of seed objects $d_{\rm init}$, which is constant across the disk according to our assumption (see Sect.\ \ref{sect:outcomeDiagnostic}). 

\begin{figure}
\centering
\includegraphics[width=.5\textwidth]{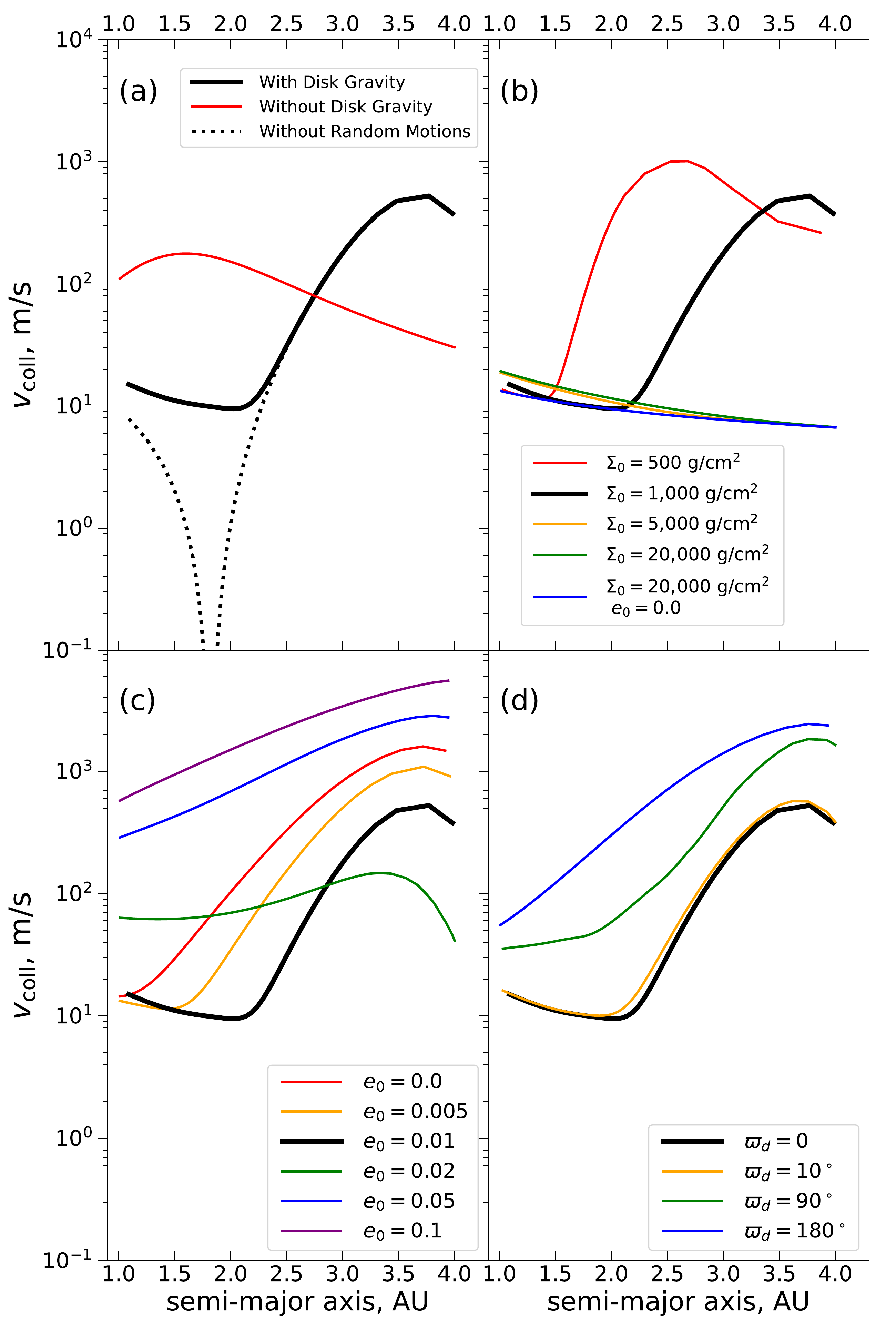}
\caption{Collision velocity between $d_1=2$ and $d_2=5$ km planetesimals as a function of the semimajor axis $a$ for the different values of model parameters varied one at a time (with respect to the fiducial set of parameters) and shown in the subplot legends.  The fiducial value of the varied parameter in each subplot is shown with the solid black curve.  The dotted line in panel (a) shows the collision velocities in the limit that $\sigma_i = 0$. }
\label{ecdcRun}
\vspace{-.05cm}
\end{figure}

To illustrate how the dynamics of planetesimals affect their size evolution, we also show in Fig. \ref{ecdcRun} the run of the collisional velocity $v_{\rm coll}$ between $d_1=2$ and $d_2=5$ km planetesimals as a function of semimajor axis (for many of the different simulations carried out in this work, see Sect.\ \ref{sect:par-var}). It accounts for both the secular and random velocity contributions and is calculated using Eq. \eqref{combinedVcoll} with parameters $\lambda$ and $e_{\rm rand}$ taking their mean values. The black solid line corresponds to the fiducial disk model, whereas the black dotted line in panel (a) shows $v_{\rm coll}$ computed in the absence of random motions. One can easily recognize the "valley" forming around $1.8$ AU as the dynamically quiet location where $v_c,d_c\to 0$ in Fig. \ref{fig:ecdc}. Nonzero random velocities of planetesimals "fill" this valley, not allowing $v_{\rm coll}$ to drop to zero in their presence. Nevertheless, $v_{\rm coll}$ is still relatively low around this region. 

A large increase in $v_{\rm coll}$ in the outer disk starting around 2 AU arises because for $\Sigma_0=10^3$ g cm$^{-2}$ the free precession rate $A$ becomes very small there, while the excitation by the companion grows large. As a result, both $v_c$ and $v_{\rm coll}$ become very large, of order km s$^{-1}$, even though a true secular resonance does not appear in this disk model. We note that $v_{\rm coll}$ is often several times smaller than $v_c$, because $d_c$ is often less than $d_1=2$ km.

Figure \ref{MofR} shows that, in the cases with $d_{\rm init}/d_{\rm min} < 1$, most of the solid mass is removed from the outer and inner portions of the disk on timescales of $10^4$ years, leaving the majority of the remaining solids in a narrow band around 1.8 AU. This band coincides with the dynamically quiet location in the fiducial disk model (see Fig. \ref{ecdcRun}), where (i) the velocity of inward drift of planetesimals goes down (see Fig. \ref{inspiralSpeeds}) and (ii) $v_{\rm coll}$ is relatively low. These factors naturally cause planetesimal accumulation ($dM/da$ temporarily exceeds its initial value due to arrival of mass from larger radii) and promote growth at this location (see Fig. \ref{fig:biggest-of-r}). But over time, the mass concentration around 1.8 AU is eroded and moves inward due to gas drag.  

In the outer disk, outside 2 AU, both the amount of mass and the size of the largest object drops precipitously early on, primarily because of very high $v_{\rm coll}\sim 1$ km s$^{-1}$ in this region (see the black curve in Fig. \ref{ecdcRun}) resulting in rapid planetesimal erosion down to the smallest size (10 m) tracked in our simulations\footnote{The saturation of the size of the largest body at the level of several tens of meters in the outer disk is an artifact of our calculations (see Sect.\ \ref{singleAnnulus}).}. Some of this debris gets swept into the dynamically quiet zone through inspiral before it is fully ground down. In the inner disk, below 1.8 AU, $v_{\rm coll}$ is much lower (enabling some temporary growth for the largest objects) and mass is removed predominantly by gas drag. 

In the simulations in which coagulation is successful ($d_{\rm init}/d_{\rm min} > 1$), the mass is again quickly removed from the outer part of the disk. But in the inner disk, particularly around 1.8 AU, the mass density (per unit semimajor axis) remains roughly constant as large objects are able to efficiently accrete smaller ones, preventing them from getting lost via inspiral or grinding down. In the run with $d_{\rm init}/d_{\rm min}=5$ we find that in the outer disk, the largest bodies have size $(2-3)d_{\rm init}$ for the duration of the simulation. This illustrates that larger seed objects find it much easier to survive than smaller ones even in the dynamically harsh environments.

\begin{figure*}
\centering
\includegraphics[width=1.0\textwidth]{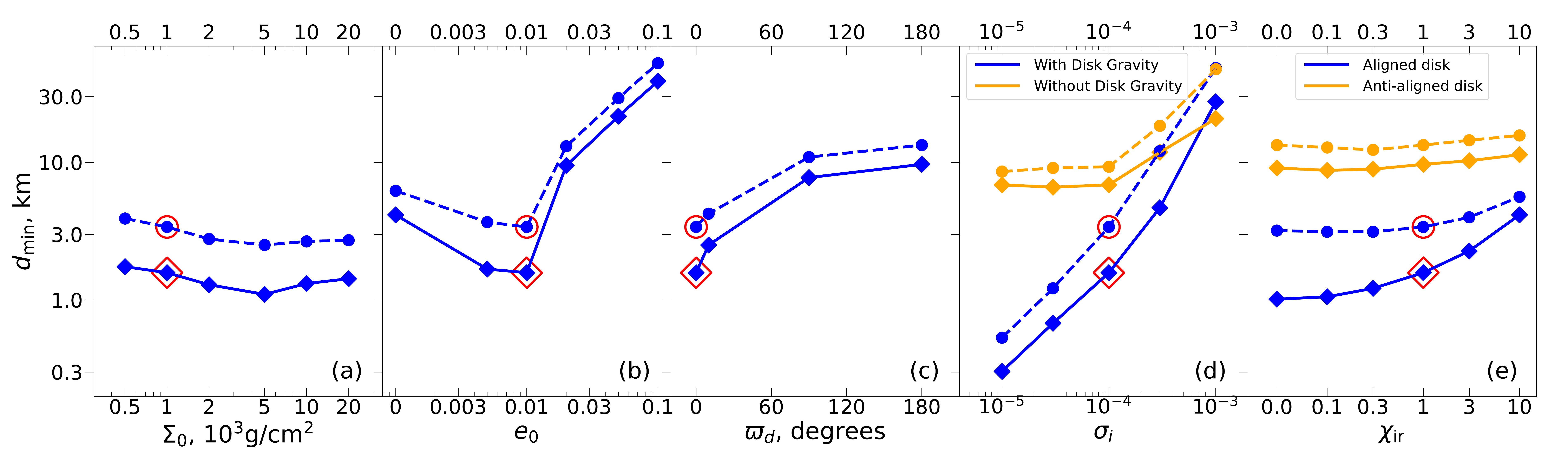}
\caption{Dependence of our key metric $d_{\rm min}$ -- a minimum size of seed planetesimals that results in their eventual growth beyond 300 km -- on some parameters of our simulations: (a) surface density normalization $\Sigma_0$, (b) disk eccentricity normalization $e_0$, (c) apsidal angle of the disk $\varpi_d$, (d) amplitude of the random velocity of planetesimals $\sigma_i$, and (e) parameter $\chi_{\rm ir}$ (artificially) regulating the magnitude of planetesimal inspiral.  Solid lines and diamond points correspond to $d_{\rm min}$, whereas dashed lines and circles are for $d_{\rm min}^{\rm rubble}$ -- the minimum size for collisionally weak planetesimals.  The causes of the trends shown here are discussed in Sects. \ref{sdVariation}-\ref{inspRate}.  In each panel, the fiducial simulation is highlighted in red. }
\label{dminPlot}
\vspace{-.05cm}
\end{figure*} 


\section{Results: Variation of system parameters}
\label{sect:par-var}


Having discussed in detail the evolution of the global planetesimal size distribution for a particular disk model (Sect.\ \ref{sect:fiducial}), we proceed to explore the changes brought in by varying disk model parameters relative to their fiducial values. We vary several key physical parameters -- surface density normalization $\Sigma_0$, disk eccentricity normalization $e_0$, disk orientation $\varpi_d$, amplitude of random velocity $\sigma_i$ -- which control planetesimal dynamics. We also run calculations in which we artificially turn off disk gravity (but not gas drag), and vary the inspiral rate, to explore the role played by these physical processes in determining the outcome of planetesimal evolution. In carrying out this parameter exploration we use the minimum planetesimal size $d_{\rm min}$ (at which growth beyond 300 km becomes possible) as our metric to characterize the success of planet formation in a given model. 

Table \ref{physicalParameters} lists the parameters of the different models that have been explored, as well as their $d_{\rm min}$ and $d_{\rm min}^{\rm rubble}$. These two simulation metrics are also shown in Fig. \ref{dminPlot} and are discussed in the rest of this section. Simulation 1 is our fiducial model, and all other simulations vary one or two of the parameters while leaving the rest fixed.  In each row, the parameters that differ from Simulation 1 are highlighted in bold. We also vary a number of numerical parameters, and show that these choices make little difference to the outcome (see Appendix \ref{modelParams} and Table \ref{tbl:num-pars} for details).


\begin{table*}[htpb]
\caption{{\large \hspace{4cm} \textbf{Parameters of our model runs.}}}
\vspace{-1mm}
\centering
\begin{tabular}{c c c c c c c c c c c c c c c}
\hline\hline
Sim \# &  $\Sigma_0$$^1$ (g cm$^{-2}$)& $e_0$ $^2$ & $\varpi_d$ $^3$  & $\sigma_i$ $^{4}$ &  Disk gravity $^{5}$ & $\chi_{\rm ir}$ $^6$ &$d_{\rm min}$ (km) $^7$ & $d_{\rm min}^{\rm rubble}$ (km) $^8$ &    \\ [0.5ex] 
\hline 
\hline
\hline
1  & 1,000 & 0.01& $0^\circ$   &  $10^{-4}$& yes & 1 & 1.6 & 3.4   \\[1ex]
\hline
\hline
\hline
2  &    {\bf 500} & 0.01& $0^\circ$    &  $10^{-4}$& yes & 1& 1.7 & 3.9   \\[1ex]
3  &{\bf 2,000} & 0.01& $0^\circ$    &  $10^{-4}$& yes & 1& 1.3 & 2.8   \\[1ex]
4  &{\bf 5,000} & 0.01& $0^\circ$    &  $10^{-4}$& yes & 1& 1.1 & 2.5   \\[1ex]
5  &{\bf 10,000} & 0.01& $0^\circ$    &  $10^{-4}$& yes & 1& 1.3 &  2.7 \\[1ex]
6  &{\bf 20,000} & 0.01& $0^\circ$    &  $10^{-4}$& yes & 1& 1.4 & 2.7   \\[1ex]
7  &{\bf 20,000} & {\bf 0.0} & $0^\circ$    &  $10^{-4}$& yes & 1& 0.7 & 2.1  \\[1ex]

\hline
8  &1,000 & {\bf 0.0}& $0^\circ$   &  $10^{-4}$ & yes & 1& 4.2 &  6.2   \\[1ex]
9  &1,000 & {\bf 0.005}& $0^\circ$   &  $10^{-4}$ & yes & 1& 1.7 & 3.7   \\[1ex]
10  & 1,000 & {\bf 0.02} & $0^\circ$   &  $10^{-4}$& yes & 1& 9.5 & 13.1    \\[1ex]
11  &1,000 & {\bf 0.05} & $0^\circ$  &  $10^{-4}$& yes & 1& 21.7 & 29.3   \\[1ex]
12  &1,000 & {\bf 0.1} & $0^\circ$  &  $10^{-4}$& yes & 1& 38.9 & 52.6   \\[1ex]
\hline
13  &1,000 & 0.01& ${\bf 10^\circ}$  &  $10^{-4}$& yes & 1& 2.5  & 4.2 \\[1ex]
14  &1,000 & 0.01& ${\bf 90^\circ}$  &  $10^{-4}$& yes & 1& 7.8 & 10.9 \\[1ex]
15 &1,000 & 0.01& ${\bf 180^\circ}$   &  $10^{-4}$& yes & 1& 9.7 & 13.4     \\[1ex]
\hline
16   &1,000 & 0.01 & $0^\circ$   &  ${\bf10^{-5}}$&  yes & 1& 0.3 & 0.5    \\[1ex]
17  &1,000 & 0.01 & $0^\circ$  &  ${\bf3\cdot10^{-5}}$& yes  & 1& 0.7 & 1.2   \\[1ex]
18   &1,000 & 0.01 & $0^\circ$    &  ${\bf 3\cdot10^{-4}}$&  yes  & 1& 4.7 & 12.1  \\[1ex]
19  &1,000 & 0.01 & $0^\circ$   &   ${\bf 10^{-3}}$&  yes & 1& 27.6 & 48.5  \\[1ex]
\hline
20   &1,000 & 0.01 & $0^\circ$  &  ${\bf 10^{-5}}$& {\bf no} & 1& 6.9 & 8.6  \\[1ex]
21   & 1,000 & 0.01 & $0^\circ$   &   ${\bf 3\cdot 10^{-5}}$& {\bf no}  & 1& 6.6 & 9.1  \\[1ex]
22   &1,000 & 0.01 & $0^\circ$    &  $\bf 10^{-4}$& {\bf no} & 1& 6.9 & 9.3   \\[1ex]
23   & 1,000 & 0.01 & $0^\circ$     &  ${\bf 3\cdot10^{-4}}$  & {\bf no} & 1& 11.8 & 18.5   \\[1ex]
24   &1,000 & 0.01 & $0^\circ$    &  ${\bf10^{-3}}$& {\bf no} & 1& 20.8 & 47.5  \\[1ex]
\hline
25  & 1,000 & 0.01& $0^\circ$   &  $10^{-4}$& yes & {\bf 0.0}& 1.0 & 3.2    \\[1ex]
26  & 1,000 & 0.01& $0^\circ$   &  $10^{-4}$& yes & {\bf 0.1}& 1.1 & 3.1   \\[1ex]
27  & 1,000 & 0.01& $0^\circ$   &  $10^{-4}$& yes & {\bf 0.3}& 1.2 & 3.1  \\[1ex]
28  & 1,000 & 0.01& $0^\circ$   &  $10^{-4}$& yes & {\bf 3.0}& 2.3 & 4.0 \\[1ex]
29  & 1,000 & 0.01& $0^\circ$   &  $10^{-4}$& yes & {\bf 10.0}& 4.2 & 5.6    \\[1ex]
30  & 1,000 & 0.01& $0^\circ$   &  $10^{-4}$& yes & {\bf 30.0} & 5.6 & 7.8   \\[1ex]
31  & 1,000 & 0.01& $0^\circ$   &  $10^{-4}$& yes & {\bf 100.0}& 8.6 & 9.5  \\[1ex]
\hline
32 &1,000 & 0.01& ${\bf 180^\circ}$   &  $10^{-4}$& yes & {\bf 0.0} & 9.1 &  13.4  \\[1ex]
33 &1,000 & 0.01& ${\bf 180^\circ}$   &  $10^{-4}$& yes & {\bf 0.1} & 8.8 &   12.8  \\[1ex]
34 &1,000 & 0.01& ${\bf 180^\circ}$   &  $10^{-4}$& yes & {\bf 0.3} & 8.9 &  12.3  \\[1ex]
35 &1,000 & 0.01& ${\bf 180^\circ}$   &  $10^{-4}$& yes & {\bf 3.0} & 10.3 & 14.5    \\[1ex]
36 &1,000 & 0.01& ${\bf 180^\circ}$   &  $10^{-4}$& yes & {\bf 10.0} & 11.4 & 15.7  \\[1ex]
\end{tabular}
\begin{flushleft}
1. Disk surface density at $a_0=1$ AU \\
2. Disk eccentricity at $a_0=1$ AU \\
3. Angle between the apsidal line of the disk and that of the binary.\\
4.  Dispersion of the distribution for each component of the planetesimal inclination vector. \\
5.  If ``yes", then disk gravity is included when calculating collision rates and outcomes, if ``no", then it is ignored.  \\
6.  The inspiral rate is artificially multiplied by the inspiral rate multiplier $\chi_{\rm ir}$.\\
7. Smallest initial planetesimals size which results in a 300 km body forming within 1 Myr.\\
8. Smallest initial planetesimal size which results in a 300 km body forming within 1 Myr, assuming rubble-pile planetesimals.\\
\end{flushleft}
\vspace{80mm}
\label{physicalParameters}
\end{table*}

\subsection{Variation of surface density:  Simulations 2-7}
\label{sdVariation}


Somewhat unexpectedly, we find that variation in the surface density normalization $\Sigma_0$ within a broad range relative to the fiducial value of $10^3$ g cm$^{-2}$ has almost no effect on $d_{\rm min}$, which ranges between 1.1 and 1.7 km (see Fig. \ref{dminPlot}a and Table \ref{physicalParameters}). Even though both the dynamically quiet location at which $v_c$ and $d_c$ vanish and the location (and existence) of the secular resonance change with $\Sigma_0$ (see Fig. \ref{fig:ecdc}), this does not dramatically affect $d_{\rm min}$.

Figure \ref{dminPlot}a does show a weak decrease in $d_{\rm min}$ as $\Sigma_0$ increases from 500 to 5000 g cm$^{-2}$. This is because, as in the fiducial model, the most favorable conditions for planetesimal growth occur at the dynamically quiet location, where $v_c$ and $d_c$ vanish, leaving only random velocity to produce relative velocity between the colliding planetesimals (i.e.,  $v_{\rm coll}\sim \sigma_i v_{\rm K}$). As $\Sigma_0$ increases, the dynamically quiet location moves out in the disk.  Since the Keplerian speed drops with distance, planetesimals collide at lower velocities at the more distant dynamically quiet locations (i.e., for higher $\Sigma_0$). As a result, within the range of $\Sigma_0$ where the dynamically quiet location exists, collisions are less destructive and planetesimal growth becomes possible starting at lower $d_{\rm init}$ for larger $\Sigma_0$.

We note that in the $\Sigma_0=500$ g cm$^2$ model a secular resonance \citep{SR15A} appears in the disk around 2.4 AU (see Fig. \ref{fig:ecdc}), driving $v_{\rm coll}$ to very high values\footnote{Unlike $v_c\propto A^{-1}$, neither $e_r$ nor collisional velocity diverge at the resonance (where $A\to 0$), since the product $v_cA$ in Eq. (\ref{er}) remains finite as $A \rightarrow 0$.} at moderate (2 - 2.5 AU) semimajor axes. However, $d_{\rm min}$ is only slightly larger than for other disk models (not featuring resonances), simply because most growth occurs at the dynamically quiet location, which is far from this resonance. 

The situation changes for even more massive disks, $\Sigma_0 > 5000$ g cm$^{-2}$, in which the disk gravity is dominant over gravity from the companion throughout the whole disk. As a consequence, such disks have no dynamically quiet location\footnote{We note that Fig. \ref{ecdcRun} does not reflect these changes well: The $v_{\rm coll}$ curves in its panel B look very similar to each other for $\Sigma_0 \geq 2000$ g cm$^{-2}$ (as if they were fully dominated by random motions), whereas the $d_{\rm min}$ outcomes are not exactly the same. This is because this figure is made only for a particular pair of planetesimal sizes, not revealing the full picture of the behavior of $v_{\rm coll}$ across all planetesimal sizes.} and we find a slight increase in $d_{\rm min}$ with $\Sigma_0$.

In Simulation 7 we also considered a model of a massive axisymmetric disk, with $\Sigma_0 = 20,\!000$ g cm$^{-2}$ (corresponding to a total disk mass of 0.07 $M_\odot$ interior to 5 AU) and $e_0 = 0$. In this setup, previously explored in \citet{R13}, disk gravity does not give rise to planetesimal eccentricity excitation but effectively suppresses the dynamical excitation due to the secondary's torque. As a result, in this model collision velocities are reduced even further (compared to, e.g., Simulation 6), and we find a significantly lower $d_{\rm min} = 0.7$ km (see Table \ref{physicalParameters}). 


\subsection{Variation of $e_0$: Simulations 8-12}
\label{sect:e_0-var}


Figure \ref{dminPlot}b shows that disk eccentricity has a dramatic effect on planetesimal growth: $d_{\rm min}$ varies by 1.4 dex as the disk eccentricity normalization $e_0$ changes from $0$ to $0.1$. This variation is non-monotonic. 

As $e_0$ increases from zero to the fiducial value $e_0=0.01$, we find $d_{\rm min}$ decreasing for reasons similar to the ones mentioned in Sect.\ \ref{sdVariation}: Higher $e_0$ means larger $|B_d|\propto e_0\Sigma_0$ \citep{SR15A,RS15a}, shifting the dynamically quiet location further from the primary (here we again assume $\varpi_d=0^\circ$; see Fig. \ref{fig:ecdc}). As $v_{\rm coll}\sim \sigma_i v_{\rm K}\propto a^{-1/2}$ at this location, higher $e_0$ results in lower $v_{\rm coll}$ (see Fig. \ref{ecdcRun}c) promoting planetesimal growth and leading to lower $d_{\rm min}$. In an axisymmetric disk ($e_0=0$), $B_d=0$ and cancellation of the secondary's torque does not occur, leading to a rather high $d_{\rm min}=4.2$ km. The dynamically quiet location exists in our simulation domain for values of $e_0$ between 0.004 and 0.018.  

For $e_0 > 0.018$ the torque due to disk gravity starts to dominate planetesimal eccentricity excitation, and $v_c$ is nonzero throughout the whole simulated portion of the disk; the dynamically quiet zone disappears. In this regime $v_c\propto |B_d|\propto e_0$ \citep{SR15A}, leading to higher $v_{\rm coll}$ and a less growth-friendly environment as $e_0$ increases (see Fig. \ref{ecdcRun}c); hence the rapidly increasing values of $d_{\rm min}$. Even a disk with $e_0=0.05$ requires seed planetesimals with $d_{\rm min}\approx 22$ km to ensure their growth into planetary regime.

We note that while the location of the dynamically quiet region is  sensitive to $e_0$, variation of disk eccentricity does not affect the presence (or location) of the secular resonance in the disk. The latter is determined by the disk mass only, while the former depends on both $\Sigma_0$ and $e_0$ \citep{R13,SR15A,RS15a}.


\subsection{Variation of $\varpi_d$: Simulations 13-15}
\label{sect:varpi_d-var}


We also find substantial sensitivity of $d_{\rm min}$ to $\varpi_d$. Figure \ref{dminPlot}c shows that increasing the deviation of the disk orientation from apsidal alignment with the binary orbit leads to larger $d_{\rm min}$: Keeping everything else fixed, $d_{\rm min}$ grows from 1.6 km for an aligned disk ($\varpi_d=0^\circ$) to 9.7 km for an anti-aligned disk ($\varpi_d=180^\circ$). Even a $\varpi_d=10^{\circ}$ misalignment is sufficient to increase $d_{\rm min}$ to 2.5 km.

This trend, again, owes its origin to the presence (or absence) of a dynamically quiet location in the disk. Secularly forced $v_c$ can vanish only in apsidally aligned disks (see Fig. \ref{fig:ecdc}a). A small misalignment makes $v_c$ nonzero everywhere \citep{RS15a,RS15b} but only slightly modifies the behavior of $v_{\rm coll}$ (see, e.g., the $\varpi_d=10^\circ$ curve in Fig. \ref{ecdcRun}d), because (i) $v_c$ is still dramatically reduced at the location where Eq. \eqref{eq:dq_condition} is satisfied and (ii) random velocity $\sigma_i$ is nonzero. However, for a significant misalignment ($\varpi_d\sim 1$), collisional velocities end up being much higher than in the aligned disk (see $\varpi_d=90^\circ$ and $\varpi_d=180^\circ$ curves in Fig. \ref{ecdcRun}d), requiring larger seed planetesimals to overcome the fragmentation barrier. Therefore, it is reasonable that $d_{\rm min}$ increases with the degree of misalignment, as our results show.  


\subsection{Variation of $\sigma_i$: Simulations 16-19}
\label{sect:sigma_i-var}


Figure \ref{dminPlot}d demonstrates a strong, monotonic dependence of $d_{\rm min}$ on the magnitude of random motions $\sigma_i$: For collisionally strong planetesimals $d_{\rm min}$ rises from 0.3 km at $\sigma_i = 10^{-5}$ to almost 30 km at $\sigma_i = 10^{-3}$.  This is to be expected for disk models in Simulations 16-19, which all feature a dynamically quiet zone around 1.8 AU (set by the values of $\Sigma_0$ and $e_0$). In this narrow zone $v_c$ is very low and $v_{\rm coll}$ is set entirely by random velocity $\sigma_iv_{\rm K}$ (see the black solid and dotted curves in Fig. \ref{ecdcRun}a). As a result, even at this favorable location in the disk, $d_{\rm min}$ is limited by a combination of radial inspiral and random motions that can destroy small planetesimals if $\sigma_i$ gets large, requiring larger seed bodies to ensure steady growth.  We anticipate that $\sigma_i$ would be less important in models which do not have a dynamically quiet region (see Sect. \ref{sect:no-grav}).


\subsection{Strong versus weak planetesimals}
\label{sect:weak-pl}


Dashed curves in Fig. \ref{dminPlot} show our results for collisionally weak, rubble-pile planetesimals as defined in \citet{SL09}. Rather unsurprisingly, we find that for such planetesimals $d_{\rm min}^{\rm rubble}$ is larger than $d_{\rm min}$ for collisionally stronger objects, typically by a factor of 1.5-3. Thus, more massive seed bodies would be necessary to enable sustained growth if they are collisionally weak.


\subsection{The case without disk gravity: Simulations 20-24 }
\label{sect:no-grav}


In order to better understand the impact of properly accounting for disk gravity and to isolate its effect on planetesimal growth, we also ran simulations where disk gravity was artificially turned off in our secular solutions for planetesimal eccentricity, that is, we set $B_d = A_d = 0$. In this case, the secular eccentricity of planetesimals is set only by the gravity of the companion and gas drag \citep{Thebault08}. We carried out this exercise while varying $\sigma_i$, with the results presented by orange curves and points in Fig. \ref{dminPlot}d. 

Turning off disk gravity eliminates the region of vanishing $v_c$ (see Fig. \ref{fig:ecdc}a) and collision velocities around 1.8 AU go up substantially (see Fig. \ref{ecdcRun}a). Therefore, as one might expect, planetesimal growth requires significantly higher $d_{\rm min}=6.9$ km, compared to 1.6 km when disk gravity is fully accounted for (assuming the fiducial $\sigma_i = 10^{-4}$). We note that without disk gravity $v_{\rm coll}$ in the outer disk is lower than with it, but this still does not help to keep $d_{\rm min}$ low.

In the absence of disk gravity, reducing $\sigma_i$ has very little effect on $d_{\rm min}$ because for $\sigma_i \lesssim 10^{-4}$, destructive collisions arise mostly due to secularly excited eccentricities and are therefore unaffected by further reducing $\sigma_i$. But at sufficiently high $\sigma_i$, the collision velocities are dominated by random motions, and $d_{\rm min}$ is no longer affected by the disk model.  We do not consider such high values of $\sigma_i$, but we do find at $\sigma_i = 10^{-3}$ that $d_{\rm min}$ is lower in the model with disk gravity than without.  This reversal occurs because for $\sigma_i = 10^{-3}$, the outer part of the disk becomes the most favorable environment for planet formation in the absence of disk gravity because the random velocities are lower there. With disk gravity the collision velocities in the outer disk are still dominated by the secular eccentricities, so planet formation must occur in the inner disk where the random velocities are higher. 

Interestingly, $d_{\rm min}$ increases very slightly as we go to very low values of $\sigma_i$, going from 6.6 km at $\sigma_i = 3 \times 10^{-5}$ to 6.9 km at $\sigma_i = 10^{-5}$.  Possibly that is because decreasing $\sigma_i$ decreases the timescale for collisional evolution (see Eq. (\ref{eq:R2})) while leaving the timescale for radial inspiral unaffected. As a result, erosion in collisions is more of an issue at low $\sigma_i$ because small bodies have less time to be removed from the system before they collide with larger ones. But quantitatively, this effect is rather small, and in fact not present for the rubble-pile planetesimals.


\subsection{The effect of artificially altering the inspiral rate: Simulations 25-36}
\label{inspRate}


\begin{figure}
\centering
\includegraphics[width=.5\textwidth]{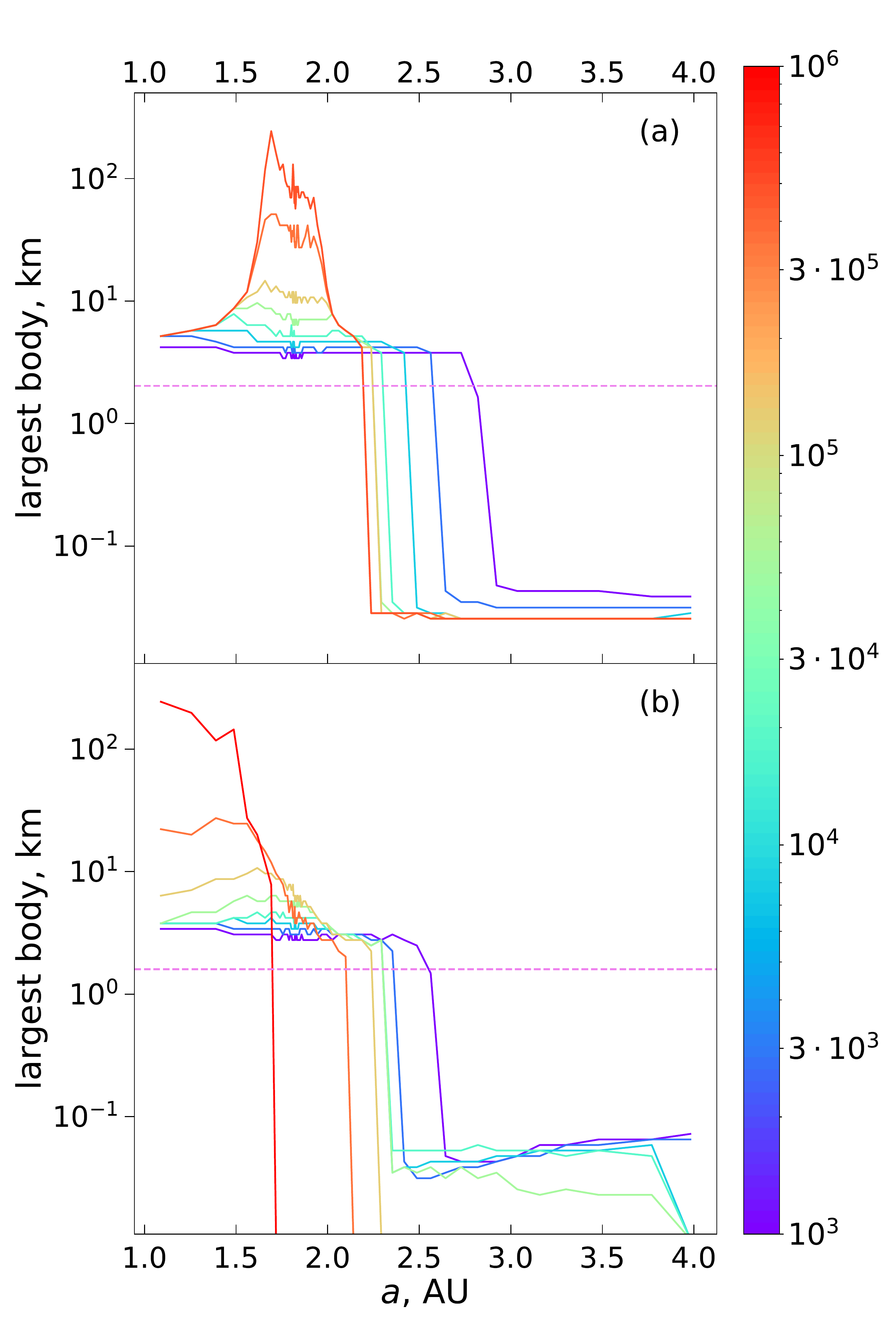}
\caption{Evolution of the size of the largest object at different locations in the disk for two models used to illustrate certain aspects of our calculations. (a) Fiducial model with gas drag turned off, $\chi_{\rm ir}=0$, computed for $d_{\rm init}/d_{\rm min}=2$, to be compared with Fig. \ref{fig:biggest-of-r}c (see Sect. \ref{inspRate} for details). (b) Fiducial model computed for $d_{\rm init}/d_{\rm min}$ just slightly exceeding unity, to be compared with Fig. \ref{fig:biggest-of-r}c,d (see Sect. \ref{sect:implications} for details). }
\label{combinedFig}
\vspace{-.05cm}
\end{figure} 

The key factor not allowing us to treat the global size evolution of planetesimals as a simple superposition of one-zone coagulation-fragmentation simulations such as the one presented in Sect. \ref{singleAnnulus}, is the mass exchange between the different annuli caused by radial inspiral due to gas drag. Thus, to determine the degree to which the growth process is affected by the mass exchange, we also carried out some simulations with the inspiral rate artificially increased or decreased compared to the value given by Eq. (\ref{dotap}) by a factor $\chi_{\rm ir}$ as indicated in Table \ref{physicalParameters}. Simulations 25-31 explore what happens in the fiducial model when the inspiral rate is varied with $\chi_{\rm ir}$ changing from 0 to 100. 

One possibility that we wanted to test through this exercise is that the removal of small planetesimals by inspiral should reduce the detrimental effect of erosion on planetesimal growth, with the expectation that high $\chi_{\rm ir}>1$ (i.e., faster removal of small fragments) would facilitate growth, while $\chi_{\rm ir}<1$ (weakened inspiral) would suppress it. However, Fig. \ref{dminPlot}e reveals a pattern of $d_{\rm min}$ behavior that runs counter to this expectation. In fact, in the no-inspiral case ($\chi_{\rm ir}=0$) we find $d_{\rm min}\approx 1$ km to be lower than in the case with full inspiral ($\chi_{\rm ir}=1$) for which $d_{\rm min}\approx 1.6$ km. And $\chi_{\rm ir}>1$ results in even larger $d_{\rm min}$.

We believe that there are two reasons for this behavior. First, in the outer disk, inspiral does not help much with cleaning out small fragments, simply because they get primarily ground down below the smallest size captured in our simulations. Second, the radial inspiral negatively impacts the growth of large planetesimals by shifting their semimajor axis away from the dynamically quiet zone where their growth is most favorable. Indeed, Fig. \ref{combinedFig}a shows the results of a no-inspiral simulation with fiducial parameters ($\chi_{\rm ir}=0$, analogous to a simple addition of a number of multi-zone simulations not interacting with each other) for $d_{\rm init}/d_{\rm min}=2$, which can be directly compared to Fig. \ref{fig:biggest-of-r}c (for which $\chi_{\rm ir}=1$). One can see that without inspiral, the largest objects form closer to $a_{\rm dq}=1.8$ AU where $v_c=0$, which certainly facilitates their growth compared to the full-inspiral calculation, in which the 300 km object emerges interior to 1.4 AU. But the general conclusion that we can draw from this exercise is that the gas drag-driven inspiral has a rather modest effect on the outcome of planetesimal growth.


\section{Discussion}
\label{disc}


The results of the previous sections highlight the important differences between planet formation around single stars and in binaries. In the binary case, erosion plays a much more important role in planetesimal size evolution, which is caused by the obvious differences in underlying dynamics: whereas in disks around single stars planetesimal motions are excited only by gravitational perturbations between numerous planetesimals (and, possibly, turbulence in the disk), planetesimals in binaries get additionally (and very strongly) excited by the gravity of the companion star and the disk. This can easily raise the collision speeds of planetesimals of comparable size into the km s$^{-1}$ regime (see Fig. \ref{ecdcRun}), much higher than in disks around single stars. As a result, growth to $10^2-10^3$-km size objects in binaries is possible only if planetesimals start already rather large, with $d_{\rm init}\sim 1-10$ km. 

This places planets in binaries such as $\gamma$ Cep in a unique category of dynamical extremophiles or hyperdynamophiles\footnote{By analogy with hyperthermophiles -- extremophilic organisms thriving in extremely hot environments.} --- objects capable of surviving and growing despite the harsh dynamical environment in which they reside. The very existence of such planets puts planet formation theories to a very stringent test.

Our parameter space exploration in Sect.\ \ref{sect:par-var} is not entirely exhaustive: because of the multi-dimensional space of physical parameters we could vary at most one or two of them at once with respect to our fiducial model, meaning that some domains of the parameter space are left unexplored. Also, our models rely on certain simplifying assumptions that were necessary to reduce the number of model inputs, for example, the initial planetesimal size $d_{\rm init}$ is the same everywhere in the disk, the amplitude of random motions $\sigma_i$ is the same regardless of location and object's size, etc. Nevertheless, our results do allow us to robustly single out the most important physical ingredients determining the outcome of planetesimal growth in binaries. 

The gravitational effect of the protoplanetary disk is the main such ingredient and should not be overlooked. The inclusion of disk gravity significantly reduces $d_{\rm min}$ in our fiducial case, by about a factor of four compared to the case in which disk gravity is ignored (see Fig. \ref{dminPlot}d). Eccentricity $e_0$ and orientation $\varpi_d$ of the protoplanetary disk are two other key parameters determining the outcome of planetesimal growth. Acting in conjunction with disk gravity, they can lead to the emergence of dynamically quiet locations in the disk, providing favorable conditions for coagulation. 

Inclusion of radial inspiral of solids is what makes our models fully global. We find that inspiral somewhat complicates planet formation by causing the migration of growing planetesimals out of the dynamically quiet zone in the disk and raising $d_{\rm min}$ compared to the case in which inspiral is ignored (see Fig. \ref{dminPlot}e). But the overall effect is not as large as one might have expected (see Sect. \ref{inspRate}). 
Next we discuss some other consequences of our results.


\subsection{Implications for planet formation in binaries}
\label{sect:implications}


About a dozen binaries are currently known to harbor planets in S-type configurations \citep{Chauvin2011,Marzari2019}, some of them with binary semimajor axis even smaller than that of $\gamma$ Cep (i.e., $a_b<20$ AU). Our results provide a pathway to understanding the origin of such planets. Our previous attempt to explain their existence \citep{RS15b} focused on explaining the formation of planets in some of these systems \citep{Chauvin2011} at their current locations. But in this work we are more interested in the overall possibility of planet formation at one to a few AU separation in a $\gamma$ Cep-like system. 

Our finding that $d_{\rm min}$ only weakly depends on $\Sigma_0$ is quite important. As we vary $\Sigma_0$ from 500 g cm$^{-2}$ to 20,000 g cm$^{-2}$ (i.e., between 0.3 to 12 times the minimum mass solar nebula density at 1 AU \citep{Hayashi81}), implying the variation of the protoplanetary disk mass from $1.8M_{\rm J}$ to $70M_{\rm J}$ (for the assumed $\Sigma$ profile (\ref{SigmaAndEccEquations}) with $a_0=1$ AU and $a_{\rm out}=5$ AU), $d_{\rm min}$ changes by less than a factor of two. What this means physically is that, even with the variation of $\Sigma_0$, the dynamically quiet region can still exist somewhere in the disk, ensuring that planetesimal growth can be achieved in the disk globally (things may get more complicated at very low $\Sigma_0$ when a secular resonance appears in the disk; see Fig. \ref{fig:ecdc}).
\par

\begin{figure}
\centering
\includegraphics[width=.45\textwidth]{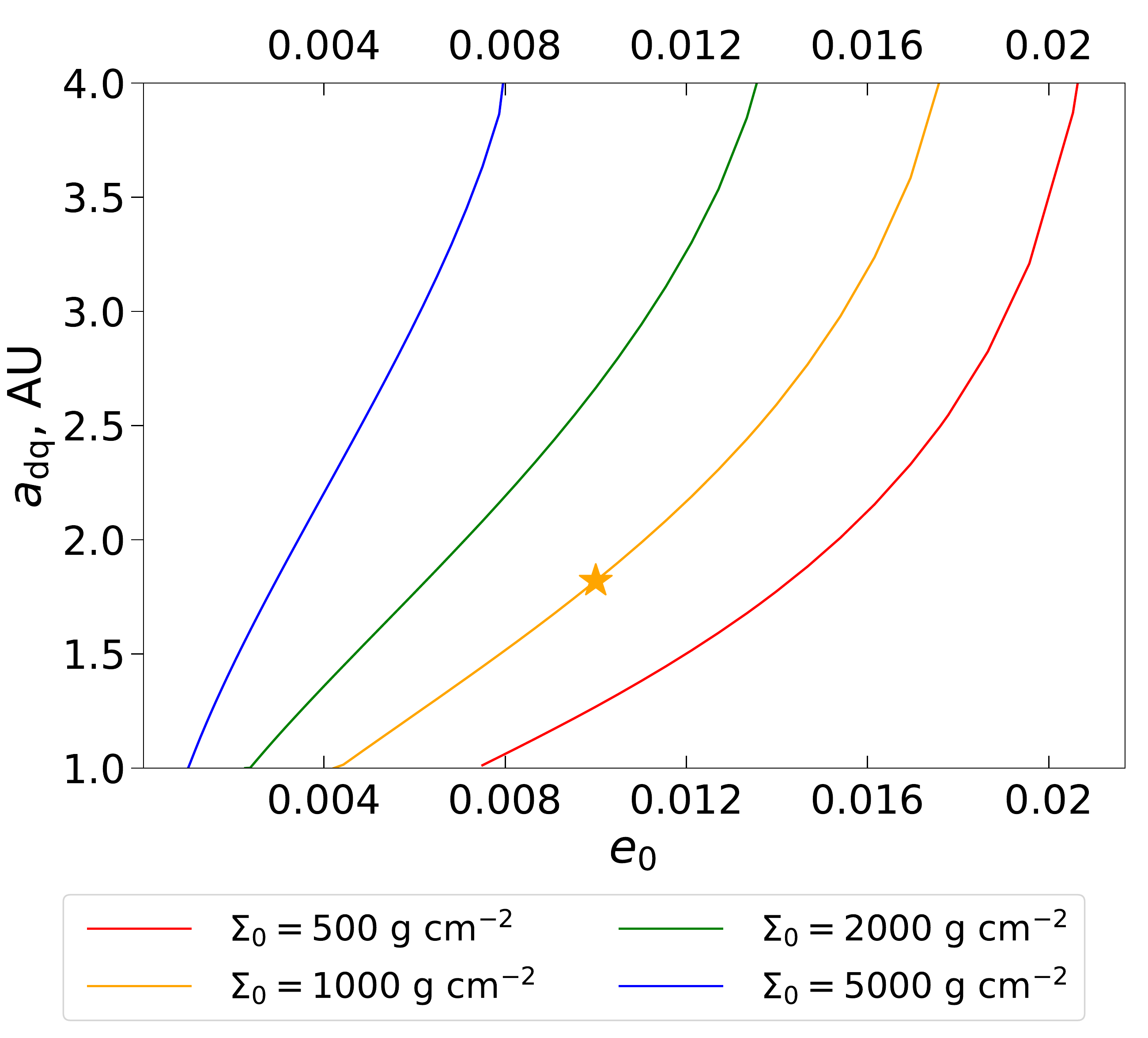}
\caption{Location $a_{\rm dq}$ of the dynamically quiet zone in an apsidally aligned disk.  We assume the fiducial disk plus binary parameters, except for $\Sigma_0$ and $e_0$ which are varied as shown on the plot. The fiducial model is marked by an orange star along the curve corresponding to $\Sigma_0 = 1000$ g cm$^{-2}$.} 
\label{Eq8Comparison}
\vspace{-.05cm}
\end{figure}
\par

Figure \ref{Eq8Comparison} shows $a_{\rm dq}$ as a function of $e_0$ for various values of $\Sigma_0$ in a system which has, other than the variation of $e_0$ and $\Sigma_0$, our fiducial parameters.  We have marked the fiducial model with an orange star along the $\Sigma_0 = 1000$ g cm$^{-2}$ curve.  This shows that over a wide range of disk masses, there is a dynamically quiet region in the disk.  For each disk mass however, there is only a relatively narrow (approximately factor of two) range in $e_0$ over which the dynamically quiet region exists in the part of the disk outside 1 AU.
\par
 At low disk masses, a secular resonance appears in the disk (see the curve for $\Sigma_0=500$ g cm$^{-2}$ in Fig. \ref{fig:ecdc}).  This resonance dramatically increases collision velocities in the outer disk as illustrated in Fig. \ref{ecdcRun}b. Collision velocities between $d_p>10$ km planetesimals are even higher than in that figure, since their orbits are less affected by gas drag. 
\par
In a static disk model such as ours, this resonance is localized, and planet formation can still proceed in other regions of the disk. In reality though, the disk evaporates over time, which changes the resonance location and causes the resonance to ``sweep" through the disk.  This effect is similar to the sweeping secular resonances involving planets  \citep{Ward1981}.  
\par 
On the other hand, the dynamically quiet location also sweeps through the disk, meaning that over the disk lifetime, planetesimal growth may be promoted in many parts of the disk.  Exactly what is the effect of this combined sweeping dynamically quiet location and sweeping resonance will depend on the details and timing of the disk dispersal.
\par
It is worth noting that Fig. \ref{Eq8Comparison} shows that even for a quite low surface density disk with $\Sigma_0 = 500$ g cm$^{-2}$ (corresponding to $6 M_\oplus$ of solid material), the dynamically quiet region will not exist if $e_0$ is greater than about 0.02, corresponding to a mass-weighted mean disk eccentricity of 0.05.  This is within the range found in numerical simulations, but on the low end of that range (see Sect. \ref{sect:implications-sims} for details).
\par
The low levels of planetesimal random motions required to drive $d_{\rm min}$ into the kilometer size range may seem problematic: $\sigma_i=10^{-4}$ featured in our fiducial model corresponds to random velocities of about $10$ m s$^{-1}$.  However, even this is above the level of the escape speed from the surface of our initial planetesimals in most cases, which is the most the two-body relaxation would pump up the random motions to.  Moreover, given the high relative velocities (up to $\sim v_c$) between the planetesimal populations of different sizes, both viscous stirring and dynamical friction are likely to be very inefficient \citep{SI2000,R2003b}, further reducing the amplitude of planetesimal random motions. 

The precise location and time at which planetesimals reach our assumed threshold size of 300 km are important outcomes of our models. They turn out to be rather strong functions of $d_{\rm init}/d_{\rm min}\ge 1$. If the size of seed planetesimals $d_{\rm init}$ is close to $d_{\rm min}$, planetesimals grow rather slowly, causing their substantial gas drag-driven inspiral, so that $d_p$ reaches 300 km somewhat closer to the star than $a_{\rm dq}$ (see Fig. \ref{combinedFig}b). As $d_{\rm init}/d_{\rm min}$ increases, planetesimals tend to grow to large sizes closer to $a_{\rm dq}$ (see Fig. \ref{fig:biggest-of-r}c for $d_{\rm init}/d_{\rm min}=2$). However, further increase in $d_{\rm init}/d_{\rm min}$ causes the location of preferential planetesimal growth to shift inward again (see Fig. \ref{fig:biggest-of-r}d), presumably because of faster coagulation there. Such non-monotonic variation with $d_{\rm init}/d_{\rm min}$, as well as the possibility of late-time planet migration (that must have certainly affected S-type systems harboring hot Jupiters), complicate the effort to constrain the initial planetesimal size $d_{\rm init}$ based on the semimajor axes of observed planets within binaries.

At the same time, the time to reach 300 km is a monotonic function of $d_{\rm init}/d_{\rm min}$, steadily falling from $\sim 1$ Myr to $<10^5$ yr as this ratio increases from 1 to 5 in our fiducial model. The expectation of short disk lifetimes in S-type binaries \citep{Harris2012,Baren2019,Zurlo2020} may then favor larger $d_{\rm init}$, for which the growth time is shorter. 
\par
As previously stated, our fiducial disk model contains less than four Jupiter masses of material.  The planet in the $\gamma$ Cephei system was determined to have around ten Jupiter masses \citep{Benedict18}.  Unless the disk is being replenished by material infalling from the circumbinary disk, this would rule out all disk models with less than $\Sigma_0 \lesssim 5000$ g cm$^{-2}$ in this particular system.


\subsection{Implications for planetesimal origin}
\label{sect:implications-planetesimals}

 
Our finding that some minimum size $d_{\rm min}\sim 1-10$ km of seed planetesimals is necessary for ensuring the success of planetary genesis in S-type binary configurations has important consequences for theories of planetesimal formation. Two main ideas for the origin of planetesimals --- steady growth by coagulation and rapid formation via streaming instability \citep{Johansen2014}--- predict very different formation pathways. 

Growth by particle sticking must necessarily cover all sizes in the range from dust grains to multi-kilometer objects (subject to substantial gravitational focusing). This formation mode is known to be problematic because of the bouncing barrier around millimeter to centimeter sizes \citep{Blum2008} and the rapid radial inspiral threatening meter-size objects \citep{Weidenschilling77}. In S-type binaries this planetesimal formation mechanism encounters another major bottleneck -- efficient growth is impossible until seed objects are 1-10 km in size, substantially larger than any of the aforementioned barriers. This additional obstacle significantly reduces chances of slow planetesimal growth by coagulation in binaries.   

On the contrary, the competing mechanism relying on the operation of streaming instability appears to naturally produce large planetesimals with sizes in the range from tens to hundreds of kilometers \citep{Simon2016,Schafer2017}. This mechanism is also rather fast (with a typical timescale of tens of local orbital periods), resulting in a rapid production of the population of seed planetesimals with $d_{\rm init}$ exceeding $d_{\rm min}$, which would enable further growth into the planetary regime as we have demonstrated here.  

Thus, results of our work provide indirect support for the idea that streaming instability is a pathway for forming planetesimals in binaries (similar reasoning can be found in \citealt{Thebault2011}). Of course, whether and how this instability would operate in an eccentric disk in presence of strong perturbations from the secondary, is an open question. But if it works in binaries, then there is no reason for it not to be the mechanism for forming planetesimals also in disks around single stars.


\subsection{Implications for simulations of gaseous disks in binaries}
\label{sect:implications-sims}


Given the strong sensitivity of planetesimal growth to disk eccentricity $e_0$ and orientation $\varpi_d$ that we identify in this work, it is imperative to better constrain these variables. Unfortunately, observations are not yet at a stage where one can reliably determine eccentricity and orientation of protoplanetary disks in binaries with small separations $a_b\lesssim 30$ AU. Instead, one has to rely on numerical simulations for guidance in that matter. 

A number of studies have explored the amplitude of disk eccentricity excited by the gravity of an eccentric companion in a coplanar S-type configuration using 2D hydrodynamic simulations. The conclusions that can be drawn from these works very strongly depend on the assumed physical inputs: equation of state, treatment of disk gravity, inner boundary conditions, and so on. 

For example, studies using a locally isothermal equation of state \citep{Kley2008,Paardekooper08,Marzari09,Zsom2011,Regaly11,Martin2020} typically find rather high (mass-weighted mean) disk eccentricities, $\langle e_d\rangle\sim 0.1-0.3$, with $e_d$ profile typically exhibiting a double-peaked structure as a function of semimajor axis and with high values of $e_d$ close to the disk center. These results may need to be viewed with caution in light of the recently identified issues with using the locally isothermal equation of state for numerical modeling of the disk-perturber tidal interaction \citep{Miranda2019}. On the contrary, studies that attempt to represent disk thermodynamics in a more realistic way \citep{Marzari2012,Gyer2014} find a dramatic reduction of disk eccentricity\footnote{Another work accounting for radiative effects by \citet{Picogna2013} is very different from the rest --- it is a 3D Smoothed Particle Hydrodynamics (SPH) study --- and found high disk eccentricity. }, down to $\langle e_d\rangle\sim 0.02-0.05$. As our results demonstrate, disk eccentricity at this level would allow planetesimal growth to proceed successfully starting from objects of a few to 10 km in size (see Fig. \ref{fig:ecdc}b).  We note that in our model, $\langle e_d \rangle = 2.5 e_0$.  

Similarly, accounting for the effect of disk self-gravity on its dynamics has also been shown to suppress $\langle e_d\rangle$ \citep{Marzari09}. Another interesting finding is that S-type disks tend to be less eccentric in more eccentric binaries \citep{Marzari09,Marzari2012,Regaly11}, presumably because of their more compact size. 

Also important for the possibility of efficient planetesimal growth is the orientation of the disk, since the dynamically quiet zone, greatly facilitating planetesimal growth, arises only in disks which are close to apsidal alignment ($\varpi_d\approx 0$) with the binary orbit (see Fig. \ref{fig:ecdc}c). Unfortunately, at the moment there is little clarity regarding this issue.  \citet{Marzari09} find the disk to be precessing, that is, there is not a well-defined value for $\varpi_d$.  In some other studies, the disk is aligned with the binary orbit \citep{Muller2012,Martin2020}. \citet{Gyer2014} find the disk to be close to alignment, with $|\varpi_d|\approx 0.6$ rad. \citet{Marzari2012} find anti-alignment ($\varpi_d\approx 180^\circ$), whereas \citet{Marzari09} find either anti-alignment or precession, depending on whether self-gravity is included.  \citet{Paardekooper08} find either anti-alignment or precession depending on the flux limiter used in their code.

To summarize, the existing numerical studies of S-type disks in binaries have not yet arrived at a consensus regarding the values of either $e_0$ or $\varpi_d$, and their dependence on various physical inputs, for example, realistic thermodynamics, disk self-gravity, viscosity, etc. Given the importance of these characteristics for understanding planet formation in S-type binaries, we encourage future numerical efforts to focus on measuring the true value of $\varpi_d$ as well as disk eccentricity $e_0$, using the most realistic physical inputs, such as disk thermodynamics and self-gravity.


\subsection{Comparison of detailed coagulation-fragmentation calculations with simple heuristics}
\label{sect:heuristics}


Full coagulation-fragmentation simulations like the ones featured in this work are the most accurate predictor of the outcome of planetesimal growth. However, they are numerically expensive. For that reason, in our previous work \citep[e.g.,][]{RS15b}, we employed simple local (i.e., neglecting radial inspiral) heuristic criteria to determine conditions under which growth would occur.   

The simplest of these is the assumption that growth occurs if and only if there are no catastrophically disruptive collisions.  However, as shown in Sect. \ref{singleAnnulus}, it is also possible that erosion can prevent growth even in the absence of catastrophic disruption. For that reason \citet{RS15b} considered an alternative criterion, in which growth happens provided that no erosive collisions occur between planetesimals of  ``similar" sizes (i.e., bodies with a mass ratio greater than $10^{-2}$).

\begin{figure}
\centering
\includegraphics[width=.45\textwidth]{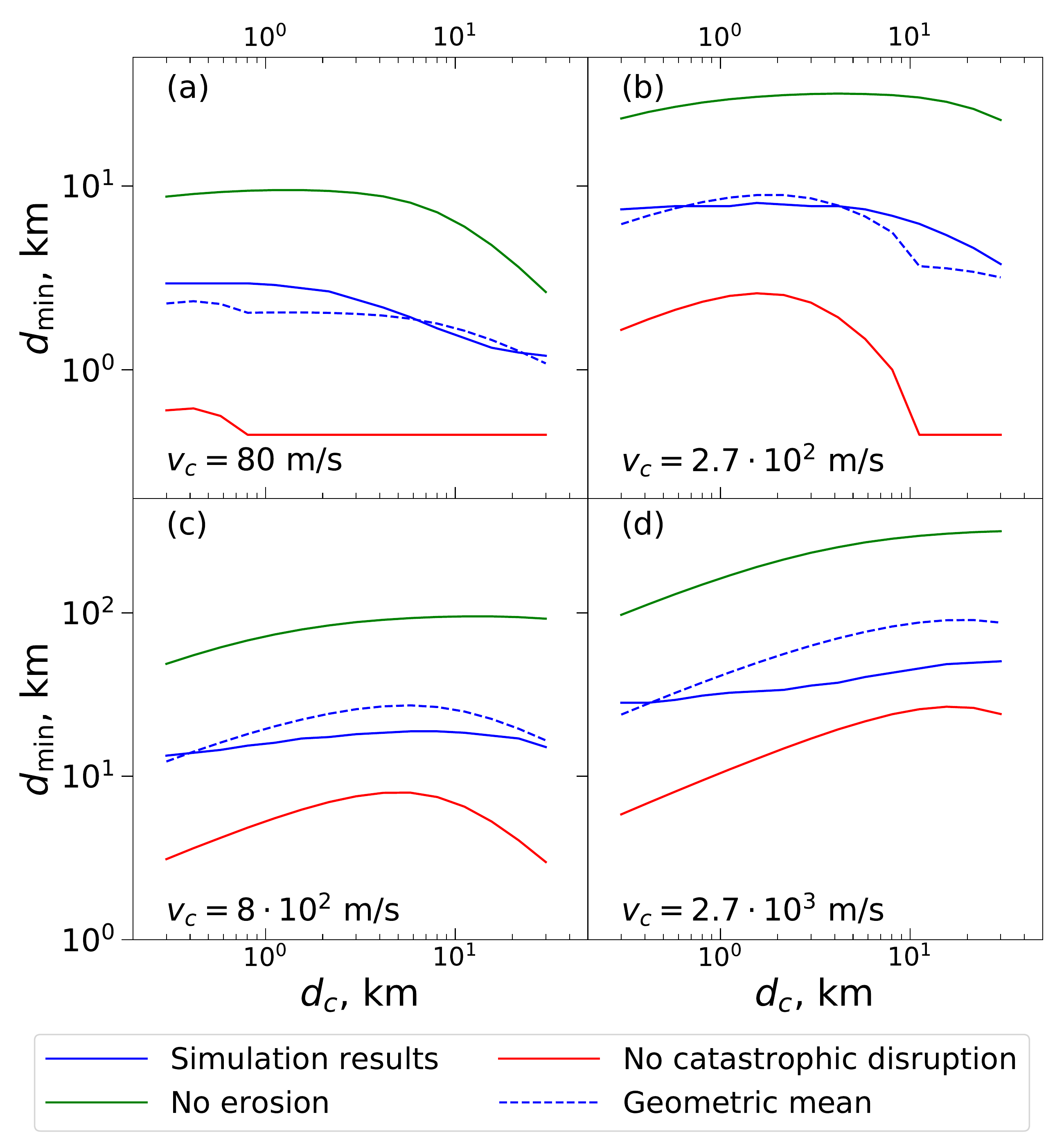}
\caption{Minimum planetesimal size $d_{\rm min}$ resulting in the formation of large objects obtained with different methods for different values of $v_c$ (indicated in panels) and $d_c$. The solid blue line shows $d_{\rm min}$ computed using our multi-annulus simulations. The green line is the analytic estimate for $d_{\rm min}$ assuming that growth can occur starting from the smallest size that does not suffer erosion by any particle greater than 1\% of its mass, according to the \citet{SL09} prescription.  The red line is the analytic estimate for $d_{\rm min}$ assuming that growth can occur as long as catastrophic disruptions are absent.  The dashed blue line shows the geometric mean of the red and green lines. }
\label{ecdcFigFinal}
\vspace{-.05cm}
\end{figure} 

Now, armed with the full coagulation-simulation code, we can test the performance of each of these growth criteria in predicting the success of planetesimal evolution. To that effect we used our simulations in a single-annulus mode to calculate $d_{\rm min}$ for a variety of models featuring different values of $d_c$ and $v_c$. As discussed in \citet{RS15a}, the collision rates and velocities are entirely determined by $d_c$ and $v_c$ (for the assumed level of random motions), regardless of what combination of disk parameters lead to the specific $d_c$ and $v_c$ (we set $v_K$ as appropriate at 2 AU in the disk around $\gamma$ Cephei).  Therefore, by exploring the range of plausible $v_c$ and $d_c$ we check the validity of these heuristics for all values of the disk parameters in our model.  

Figure \ref{ecdcFigFinal} displays the results of this calculation (each panel corresponds to a different value of $v_c$), with the solid blue line showing $d_{\rm min}$ as a function of $d_c$, computed by our code using the procedure described in Sect. \ref{sect:outcomeDiagnostic}.   The red line shows the minimum value of $d_{\rm init}$ such that no catastrophic disruption occurs for planetesimals of size $d_{\rm init}$ or larger.  The green line shows the minimum value of $d_{\rm init}$ such that no body larger than $d_{\rm init}$ suffers an erosive collision with a partner with mass ratio greater than 1\%. The same criteria were used to determine the gray and black regions, respectively, in Fig. 4 of \citet{RS15b}. 

This figure shows that the criterion of no catastrophic disruption is too optimistic, typically predicting $d_{\rm min}$ between two and five times smaller (depending on $d_c$ and $v_c$) than the one determined by the coagulation-fragmentation simulation. On the contrary, the criterion of no erosion between bodies with a mass ratio greater than 1\% is too restrictive, with the green curve exceeding the blue curve by a factor two to five. 

Interestingly, the blue curve is often roughly equally separated from both red and green ones (in log space), which motivated us to also show the dashed blue line computed as the geometric mean of the predictions from the ``no catastrophic disruption" and ``no erosion" models. As one can see, it matches the simulation-based $d_{\rm min}$  (blue curve) surprisingly well, within a factor of two or even better (although we do not claim this to have any deep meaning). Thus, the geometric mean of the predictions for $d_{\rm min}$ based on the two criteria used in \citet{RS15b} can be employed as a rough approximation for predicting the success of planetesimal growth in the absence of sophisticated coagulation-fragmentation simulations.


\subsection{Comparison with previous work}
\label{sect:prev-work}


A number of past studies of planet formation in S-type binaries have arrived at rather pessimistic conclusions regarding planetesimal growth starting from a population of 1-10 km objects  \citep{Thebault2006,Thebault08,Thebault2009,Beauge2010}. In particular, \citet{Thebault2011} found that even $d_{\rm init}\gtrsim 10^2$ km planetesimals do not ensure planetesimal growth in HD 196885 system at the location of the known planet. All these studies included the effect of gas drag on planetesimal dynamics, and \citet{Paardekooper08} and \citet{Beauge2010} even considered the possibility of the disk being eccentric. 

However, as noted already in \citet{R13} and \citet{RS15a,RS15b}, all these calculations ignored the effect of disk gravity, which dramatically changes planetesimal dynamics. For example, \citet{Beauge2010} find low planetesimal collisional velocities in the outer disk, whereas we find a very hostile dynamical environment there (see Fig. \ref{ecdcRun}). The existence of the dynamically quiet location in apsidally aligned disks substantially alleviates the fragmentation bottleneck, allowing in many cases planetesimal growth to proceed even starting at $d_{\rm init}\sim 1$ km (see Fig. \ref{dminPlot}). 

Another factor that distinguishes our study from others in terms of realism is our reliance on the global coagulation-fragmentation calculations fully accounting for the gas drag-driven inspiral of planetesimals. Whereas \citet{Thebault2006,Thebault08, Thebault2009} and \citet{RS15b} drew their conclusions based on individual collisional outcomes (see Sect. \ref{sect:heuristics}), we are able to model the evolution of the planetesimal size distribution in its entirety, fully accounting for the fine details of planetesimal dynamics at every step. 

Compared to \citet{RS15b}, we find less need for a massive ($\gtrsim 30M_{\rm J}$) protoplanetary disk: the lowest value of $\Sigma_0$ allowing growth to proceed starting at $d_{\rm init}\sim 1$ km in Fig. \ref{dminPlot}a corresponds to a disk mass of just $2M_{\rm J}$. This difference arises primarily from our current full treatment of the planetesimal size evolution, but also from focusing on planet formation not at a given location but globally, in the whole disk.  


\section{Conclusions}
\label{conclusions}

 
We present the first fully global coagulation-fragmentation framework for exploring planetesimal growth in S-type binary systems, such as $\alpha$ Cen and $\gamma$ Cep. Our framework fully accounts for the specifics of planetesimal dynamics in binaries, including perturbations not only due to the eccentric stellar companion, but also due to the gravity of the eccentric protoplanetary disk in which planetesimals orbit. We include the effects of gas drag on the eccentricity dynamics of planetesimals, as well as on their radial inspiral. Thus, this framework brings studies of planet formation in binaries to an entirely new level of realism. 

By running a suite of simulations with varied model inputs, we are able to delineate the conditions under which planetesimals can grow to large sizes (hundreds of kilometers) despite the adverse effects of fragmentation. The difficulty of planet formation for a given disk model was measured via the parameter $d_{\rm min}$ -- the smallest size of seed planetesimals that allows their growth into the planetary regime -- with the smaller $d_{\rm min}$ implying more favorable conditions for planet formation. Based on the results of these calculations, we are able to draw the following general conclusions. 

\begin{itemize}

\item For most disk parameters considered in this paper, planet formation in binaries such as $\gamma$ Cephei can successfully occur provided that the initial planetesimal size is $\gtrsim 10$ km; however, for favorable disk parameters, this minimum initial size can go down to $\lesssim 1$ km. 

\item The gravitational effect of the protoplanetary disk plays the key role in lowering the minimum initial planetesimal size necessary for sustained growth by a factor of four.  This reduction can be achieved in protoplanetary disks apsidally aligned with the binary, in which a dynamically quiet zone appears within the disk provided that the mass-weighted mean disk eccentricity $\lesssim 0.05$ (corresponding, in the context of our model, to $e_0 \lesssim 0.02$).

\item We identify disk eccentricity ($e_0$) and the apsidal misalignment angle ($\varpi_d$) as the parameters to which planetesimal growth is most sensitive. Whenever the dynamically quiet location exists in the disk, growth is also strongly dependent on the level of random motion of planetesimals.

\item For the models explored in this work, the outcome of planetesimal evolution is only weakly dependent on the disk surface density $\Sigma_0$.

\item Radial inspiral of solid material appears to make little difference to the critical planetesimal size from which collisional agglomeration can occur. 

\item The planetesimal strength, as parameterized by the two models of \citet{SL09}, makes only about a factor of two difference to the critical planetesimal size.

\item Planetesimal growth to large sizes can occur even in the face of some erosion, but in the environments with frequent collisions leading to catastrophic disruption, growth will not occur. We used our coagulation-fragmentation simulations to provide a new heuristic criterion for successful planetesimal evolution in Sect. \ref{sect:heuristics}.

\item Our results provide indirect support for models in which planetesimals are born large ($\gtrsim 10$ km), such as those based on streaming instability.

\end{itemize}

The general framework for the realistic treatment of planetesimal growth in binaries developed in this work can be applied to study a range of other problems: circumbinary planet formation in P-type systems \citep{SR15B}, collisional evolution of debris disks perturbed by planetary \citep{Nesvold2015,Sefilian2020} or stellar \citep{Thebault2020} companions, and so on.
\\

The authors acknowledge helpful discussion with Lucas Jordan.  K.S. acknowledges the support of the Max Planck Society.  R.R.R. thanks NASA grant 15-XRP15-2-0139, STFC grant ST/T00049X/1, and John N. Bahcall Fellowship for financial support. 


\bibliographystyle{apj}
\bibliography{apj-jour,finalPaper}


\appendix


\section{Detailed description of the numerical methods}
\label{firstAppendix}


Here we describe a number of technical details of the implementation of our code, including both numerical details and physical inputs. 


\subsection{Mass redistribution due to collisions}
\label{collisions}


In a given annulus, our code solves the standard coagulation-fragmentation problem, which effectively is a version of the coagulation equation of \citet{Smoluchowski16} modified to account for fragmentation (see Eq. [1] of \citealt{Rafikov20}). Our framework employs mass space discretization in the form of  logarithmically spaced mass bins with constant $M_{i+1}/M_i$.  The grid extends from the minimum mass corresponding to a 10 m object to a maximum mass corresponding to a 300 km object. Fragments with sizes below 10 m that form in planetesimal collisions are simply removed from the system (i.e., total mass is not conserved for this reason alone). This leads to certain numerical artifacts discussed in Sect.\ \ref{singleAnnulus}.

At each time step, the code calculates the mean collision rate $R_{ij}$ between planetesimals in bin $i$ and bin $j$ for all $i$ and $j$ in $[1, N_{\rm bins}]$.  This calculation is described in Appendix \ref{sect:CollisionRate}.  The number of collisions $Z_{ij}$ in a given time step of length $\Delta t$ between bodies in mass bins $i$ and $j$ is then drawn randomly from a Poisson distribution with the mean $R_{ij} \Delta t$:
\begin{equation}
Z_{ij} = Poisson(R_{ij} \Delta t).
\label{eq:Poisson}
\end{equation}
We define an operator $\hat F$ such that $F_{ijk}$ is the number of fragments in mass bin $i$ produced by a collision of a particle in bin $j$ with a particle in bin $k$ (see Sect. \ref{sect:fragmentMassDistribution}).  

The mass distribution $\vec n$ is updated as 
\begin{equation}
n_i = n_i + \Sigma_{j = 1}^{N_{\rm bins}} \Sigma_{k=1}^{N_{\rm bins}} C_{ijk},
\end{equation}
where
\begin{equation}
C_{ijk} = Z_{jk} \frac{1+\delta_{jk}}{2}\left[F_{ijk} - \delta_{ij} - \delta_{ik}\right].
\end{equation} 
This takes into account both addition of particles to bin $i$ from collisional fragments (the $F_{ijk}$ operator), as well as collisional destruction of particles in bin $i$ (the $-\delta_{ij}$ and $-\delta_{ik}$ terms).  The factor of $(1+ \delta_{jk})/2$ is to prevent over-counting when summing over both $j$ and $k$. 

At times, due to a fluctuation, a number of particles in a particular mass bin may become negative.  In this case we reset the number of particles in such a bin to zero; this effectively leads to addition of mass to the system. However, we checked several runs of our fiducial simulation, and found that relative to the total mass in the simulation, less than 1 part in $10^5$ was added in this way.


\subsection{Collisional outcomes}
\label{sect:fragmentMassDistribution}


This section describes the physics which goes into the determination of $F_{ijk}$.  The outcome of a collision between a body in mass bin $i$ and one in mass bin $j$ is dependent on the collision velocity.  High-velocity collisions will result in many small fragments, and no large remnant bodies.  Low-velocity collisions will result in one body which contains nearly all of the pre-collision mass and only very small fragments.  Calculation of collision velocity $v_{\rm coll}$ is described in Sect. \ref{sect:collVelocity}. 

Given a value of $v_{\rm coll}$, we use the recipe provided in \citet{SL09} to determine the mass of the largest remnant $M_{\rm lr}$.  We take the mass of the largest remnant to be
\begin{equation}
\frac{M_{\rm lr}}{M_{\rm tot}} = 1-0.5 \frac{Q_R}{Q^*_{\rm RD}},
\label{Mlr}
\end{equation}
where $M_{\rm tot} = m_i + m_j$ is the total mass of the two incoming bodies, $Q_R$ is the specific center-of-mass energy of the collision given by 
\begin{equation}
Q_R = 0.5 \frac{m_i m_j v_{\rm coll}^2}{M_{\rm tot}},
\end{equation}
and $Q^*_{\rm RD}$ is the critical value of $Q_R$ which leads to catastrophic disruption (which corresponds to $M_{\rm lr} < 0.5 M_{\rm tot}$); $Q^*_{\rm RD}$ is a function of the sizes and composition of the planetesimals, and $v_{\rm coll}$, which is given by Eq. 2 from \citet{SL09}.  As our default, we used the parameters appropriate for their ``strong rock" planetesimals; however we also ran simulations using their ``rubble pile" planetesimals.

When $M_{\rm lr}$ is calculated, it will in general be between $M_i$ and $M_{i+1}$ for some $i$.   We then add a fraction $P_i$ of the mass to bin $i$ and $P_{i+1}$ to bin $i+1$ given by 
\begin{equation}
P_i = \frac{M_{i+1}-M_{\rm lr}}{M_{i+1} - M_i}, \quad P_{i+1} = \frac{M_{\rm lr}- M_i}{M_{i+1} - M_i}.
\end{equation}
\par

We assume that the rest of the fragments follow a mass distribution with a power law index $\xi$.  The precise value of $\xi$ does not affect the long-term outcome of fragmentation cascade \citep{OBrien03}.  The upper cutoff mass for this distribution is given by
\begin{equation}
    M_{\rm cutoff} = \begin{cases} 
          0.5(M_{\rm tot} - M_{\rm lr}) & M_{\rm lr}/M_{\rm tot} \geq 0.5, \\
          \max(M_{\rm tot}b, 0.5 M_{\rm lr}) & M_{\rm lr}/M_{\rm tot} < 0.5. \\
       \end{cases}
       \label{cutoffMass}
  \end{equation}
Here $b$ is an adjustable parameter which determines the cutoff mass for catastrophic collisions: whenever Eq. \eqref{Mlr} yields $M_{\rm lr} < 2bM_{\rm tot}$, we put all of the mass into fragments (as described below) without leaving a single largest remnant. This situation often arises in very energetic collisions when $M_{\rm lr}$ is small. 
\par
The fragment mass is assigned to mass bins with $0<i\leq i_{\rm cutoff}$, where $i_{\rm cutoff}$ is the index of the largest bin with mass less than $M_{\rm cutoff}$. This is done such that
\begin{equation}
F_{ijk} = \Upsilon_{jk} M_i^{1+\xi},
\label{tequation}
\end{equation}
where $\Upsilon_{jk}$ is a normalization constant. The exponent of $M_i$ is $1+\xi$ rather than $\xi$ because of the logarithmic spacing of the mass bins.
\par
In order to solve for $\Upsilon_{jk}$, we note that 
\begin{equation}
    \sum_{i = -\infty}^{i_{\rm cutoff}} \Upsilon_{jk} M_i^{2+\xi} = M_{\rm tot} - M_{\rm lr},
\end{equation}
where nonpositive values of $i$ correspond to (hypothetical) mass bins with $M_i$ less than the minimum size considered in our simulations. Letting $\mathfrak{R} \equiv M_{i+1}/M_i$, we can then solve for $\Upsilon_{jk}$ and write
\begin{equation}
F_{ijk} = \frac{M_{\rm tot} - M_{\rm lr}}{M_{\rm cutoff}}\times(M_i/M_{\rm cutoff})^{1+\xi} \times\left(1-\mathfrak{R}^{-\xi-2}\right)
\label{eq:FijkExplicit}
.\end{equation}

\par

Some mass is lost from the simulation because it is contained in bodies with mass less than $M_1$.  This lost mass is equal to
\begin{equation}
M^{\rm lost}_{jk} = \Upsilon_{jk}\sum_{i = -\infty}^0 M_i^{2+\xi} = (M_{\rm tot} - M_{\rm lr}) \left[ \frac{M_0}{M_{\rm cut}} \right]^{2+ \xi}
.\end{equation}

Because of the self-similar form of the fragment mass spectrum, we were able to implement this fragmentation algorithm with numerical cost scaling as $O(N_{\rm bins}^2)$ using explicit time stepping (see \citealt{Rafikov20} for details). This considerably reduces the computational burden of our calculations.


\subsection{Collision velocity}
\label{sect:collVelocity}


We calculate the collision velocity $v_{\rm coll}$ by considering two separate dynamical regimes.  In the first, the relative motion of planetesimals is dominated by their secularly excited eccentricities. Equation (64) of RS15a then gives the relative eccentricity $e_{ij}$ between planetesimals as a function of their masses.  Given $e_{ij}$, there is a distribution of relative encounter velocities $v_{ij}$ between $v_{\rm min} = 0.5 e_{ij} v_K$ and $v_{\rm max} = e_{ij} v_K$, where $v_K$ is the local Keplerian speed.  This distribution is given by Eq. (62) of RS15a as 
\begin{equation}
\label{eq:vr}
\frac{df_{ij}}{dv_{ij}} = \frac{v_{\rm max}^{-1}}{E(\sqrt{3}/2)} \frac{v_{ij}^2}{\left[(v_{\rm max}^2 - v_{ij}^2)(v_{ij}^2 - v_{\rm min}^2)\right]}.
\end{equation}
At each time step, for each set of collision partners, a random number $\lambda$ between 0.5 and 1 is drawn based on the distribution in Eq. \eqref{eq:vr}, and the collision velocity in this regime is taken to be
\begin{equation}
v_{\rm coll} = \lambda e_{ij} v_K.
\end{equation}

In the second dynamical regime, we assume that random motions dominate over the secularly excited eccentricities.  We assume that planetesimals have eccentricity vectors $(k, h) = (k_s + k_{\rm rand}, h_s + h_{\rm rand})$ where $(k_s, h_s)$ and $(k_{\rm rand}, h_{\rm rand})$ are the secular and random components, respectively; $k_{\rm rand}$ and $h_{\rm rand}$ are assumed to be drawn from independent Gaussian distributions
\begin{eqnarray}
&\rho(k_{\rm rand}) = \frac{1}{\sigma_e \sqrt{2\pi}} \exp\left[-\frac{(k_{\rm rand})^2}{2 \sigma_e^2}\right],
\nonumber\\
&\rho(h_{\rm rand}) = \frac{1}{\sigma_e \sqrt{2\pi}} \exp\left[-\frac{(h_{\rm rand})^2}{2 \sigma_e^2}\right],
\end{eqnarray}
where we take $\sigma_e=2\sigma_i$. This leads to a distribution of the relative random eccentricity $e_{\rm rand} = \sqrt{(k_{\rm rand,1} - k_{\rm rand,2})^2 + (h_{\rm rand,1} - h_{\rm rand,2})^2}$ given by the Rayleigh distribution
\begin{equation}
\rho(e_{\rm rand}) = \frac{e_{\rm rand}}{2\sigma_e^2} \exp\left[-\frac{(e_{\rm rand})^2}{4\sigma_e^2}\right].
\label{rayleighE}
\end{equation}
In principle, for each set of collision partners, we should then take the collision velocity to be given by 
\begin{equation}
    v_{\rm coll} = v_K\sqrt{\left(\lambda e_{\rm ij}\right)^2 + e_{\rm rand}^2},
\label{combinedVcoll}
\end{equation}
where $e_{\rm rand}$ is drawn from the distribution in Eq. \eqref{rayleighE}, and $\lambda$ is drawn from the distribution in Eq. \eqref{eq:vr}.

In practice, doing this at each time-step would be too time-consuming.  Instead, we use the following algorithm. Let $F(v)$ be the true collision velocity distribution for a given set of collision partners in a given annulus. We approximately calculate $N_1$ points in $v$-space, such that there is a constant fraction of the distribution $F(v)$ between consecutive points: this means that the $i^{\rm th}$ point $v_i$ satisfies the relation $\int_0^{v_i} F(v)dv = (i+0.5)/N_1$, for $i \in [0, N_1-1]$.  We then calculate the collision rates and outcomes for each set of collision partners at the beginning of the simulation, using the $N_1$ different velocities.  During the simulation, in each timestep, and for each set of collision partners, we randomly choose one of the $N_1$ collisional outcomes and rates to use for that timestep.
\par
To calculate the $N_1$ values $v_i$, we initially create two vectors ${\bf z}_{\rm rand}$ and ${\bf z}_{\lambda}$ of length $N_2$ with a constant fraction of the distribution for $e_{\rm rand}$ between each entry of ${\bf z}_{\rm rand}$ and a constant fraction of the distribution of $\lambda$ between each entry of ${\bf z}_{\lambda}$.  This means that $\int_{\bf z_{\rm rand}^i}^{\bf z_{\rm rand}^{i+1}} \rho(e_{\rm rand}) =  1/N_2$, and a similar relation exists between ${\bf z}_{\lambda}$ and the distribution for $\lambda$ implied by Eq. \eqref{eq:vr}.  
\par
We then draw $N_3$ samples of the collision velocity using Eq. \eqref{combinedVcoll} and randomly selected entries from ${\bf z}_{\rm rand}$ and ${\bf z}_\lambda$.  We then select the $N_1$ values of $v_i$ so there are an equal number of the $N_3$ sample velocities between consecutive values of $v_i$.  We take $N_1 = 11$, $N_2 = 100$, and $N_3 = 990$.    

We ignore shear motion in calculating the collision velocity because if the shear motion is the dominant effect, then the planetesimals will completely merge in collisions, so there is no reason to consider shear motion except in how it affects the collision rate.

We note that the differential radial drift described in Sect. \ref{inspiralEquations} is not included in our calculation of $v_{\rm coll}$ as it does not typically contribute substantially to the collision velocities between bodies of the sizes that we consider in this paper.


\subsection{Collision rate}
\label{sect:CollisionRate}


Next we describe our calculation of the collision rate. In the first dynamical regime (where the relative motion of planetesimals is dominated by their secularly excited eccentricities), we use Eq. (63) from RS15a (corrected so that $e_{\rm max}$ is replaced with $v_{\rm max}/a_p$).  This gives a flux  $F_{ij}$ of planetesimals from bin $j$ past a planetesimal in bin $i$ of 
\begin{equation}
F_{ij} = \frac{E(\sqrt{3}/2)\Sigma_j v_{\rm max}}{\pi^2 a_p m_j I_{ij}}, 
\end{equation}
where $I_{ij}$ is the relative inclination between the two groups of particles.  We assume that all of the planetesimal inclination is due to random motions.  We assume all planetesimals have inclination vectors $(q, p)$ with $q$ and $p$ drawn from independent Gaussian distributions:
\begin{equation}
\rho(q) = \frac{1}{\sigma_i \sqrt{2\pi}} \exp{\left(-\frac{q^2}{2 \sigma_i^2}\right)}; \quad \rho(p) = \frac{1}{\sigma_i \sqrt{2\pi}} \exp{\left(-\frac{p^2}{2 \sigma_i^2}\right)}.
\end{equation}
The distribution of $I_{\rm ij} = \sqrt{(q_i - q_j)^2 + (p_i - p_j)^2}$ is therefore given [analogous to Eq. \eqref{rayleighE}] by 
\begin{equation}
\rho(I_{ij}) = \frac{1}{2\sigma_i^2}I_{ij} \exp{\left(-\frac{I_{ ij}^2}{4\sigma_i^2}\right)}.
\end{equation}
Then, averaged over $I_{ij}$, the average value of $F_{ij}$ is given by 
\begin{equation}
\langle F_{ij} \rangle = \int_{I_{ij} = 0}^\infty \rho(I_{ij}) F_{ ij} dI_{ij} = \frac{E(\sqrt{3}/2)\Sigma_j v_{\rm max}}{2 \pi^{3/2}\sigma_i m_j a_p}.
\label{fij}
\end{equation}
The collision rate for a given particle in bin $i$ with the population in bin $j$ is given by $R_{ij} = \langle F_{ij}\rangle \sigma_{ij}$, where $\sigma_{ij}$ is the collision cross section including gravitational focusing:
\begin{equation}
\sigma_{ij} = A_{\rm g}\left[1 + \left(\frac{v_{\rm esc}}{\lambda e_{ij} v_K}\right)^2\right],
\label{sigmaij}
\end{equation}
where $A_{\rm g} = \pi(d_i + d_j)^2$ is the geometric cross section, and $v_{\rm esc} = \sqrt{2G(m_i + m_j)/(d_i + d_j)}$.  As discussed in Sect. \ref{sect:collVelocity}, at each time-step, we assume a particular value of the collision velocity drawn from the distribution $F(v)$.  For the purposes of calculating the collision rate, we assume $\lambda$ to be in the same quantile of the distribution for $\lambda$ as $v$ is for $F(v)$.

In the second dynamical regime of dominant random motions ($\sqrt{\pi} \sigma_e > e_{ij}$) we use the two-body accretion formula from Eq. 17 of \citet{Greenzweig92}:
\begin{equation}
R_{ij} = \frac{\Sigma_j A_{\rm g}}{2\pi^2 m_j}\left[\mathscr{F}(I_*) + \frac{v_{\rm esc}^2}{v_K^2} \frac{\mathscr{G}(I_*)}{e_*^2}\right].
\label{GLrate}
\end{equation}
Here, $\mathscr{F}$ and $\mathscr{G}$ are functions of $I_* = \sigma_i/\sigma_e$.  We assume that $I_* = 1/2$, in which case $\mathscr{F}(1/2) = 17.3$, and $\mathscr{G}(1/2) = 38.2$.  $e_*$ is the rms random eccentricity.  \citet{Greenzweig92} considered a planetesimal accreting on a circular orbit, but in our case, we should use the rms relative random eccentricity which can be shown from Eq. \eqref{rayleighE} to be $e_* = 2 \sigma_e$.  Additionally, we must divide by the orbital period as Eq. \eqref{GLrate} is in units where the orbital period is unity:
\begin{equation}
R_{ij} = \frac{\Sigma_j A_{\rm g}v_K}{4\pi^3 a_p m_j}\left[\mathscr{F}(I_*) + \frac{v_{\rm esc}^2}{v_K^2} \frac{\mathscr{G}(I_*)}{e_*^2}\right].
\label{RGLprecursor}
\end{equation}

We approximate the transition to the shear-dominated regime by replacing $e_*$ in Eq. \eqref{RGLprecursor} with $e_* + c_1 e_H$, and then multiplying the entire right hand side by $1 + c_2 (e_H/e_*)$, where $c_1$ and $c_2$ are interpolation constants.  This gives us a formula 
\begin{equation}
R_{ij} = \frac{\Sigma_j A_{\rm g}v_K}{4\pi^3 a_p m_j}\left[\mathscr{F}(I_*) + \frac{v_{\rm esc}^2}{v_K^2} \frac{\mathscr{G}(I_*)}{(e_* + c_1e_H)^2}\right]\left(1 + c_2\frac{ e_H}{e_*}\right).
\label{RGLNext}
\end{equation}
We estimate $c_1$ and $c_2$ by comparing the rates given by Eq. \eqref{RGLNext} with those shown in Fig. 5 of \citet{Greenzweig92}. 
From these data 
we estimate $c_1 = 4.75$, $c_2 = 22.6$, which we use in our work.  

We expect a transition between the two dynamical regimes to occur when $e_{\rm ij} = \langle e_r \rangle = \sqrt{\pi} \sigma'_e$, where  $\sigma'_e = \sigma_e + c_1e_H/2$ (the factor of 2 arising from the fact that $e_* = 2 \sigma_e$).  The collision rates in the two regimes are not exactly equal at the transition point.  Let $R_1$ be the collision rate in the low secular velocity limit given by Eq. \eqref{RGLNext}, and let $R_2$ be the collision rate when the secular velocity dominates; $R_2$ is given by
\begin{equation}
R_2 = \frac{E(\sqrt{3}/2)\Sigma_j A_g v_K e_{ij}}{2 \pi^{3/2}\sigma_ia_pm_p}  \left[1+ \left(\frac{v_{\rm esc}}{v_{ ij}}\right)^2\right].
\label{eq:R2}
\end{equation}
 Assuming $\sigma_i/\sigma_e = 1/2$, then at $e_{ij} = \sqrt{\pi} \sigma_e$, ignoring shear, $R_2/R_1 = 2.8$ in the limit that $v_{\rm esc}/(\sigma_ev_K) \ll 1$, and $1.6/\lambda^2$ in the limit that $v_{\rm esc}/(\sigma_ev_K) \gg 1$.  To make the collision rate continuous, we approximate it as 
\begin{equation}
R_{ij}(v_{ij}) = fR_1 + (1-f)R_2,
\end{equation}
with $f = 1/(1+[e_{ij}/(\sqrt{\pi} \sigma'_e)]^2) = 1/(1+[v_{ ij}/(\sqrt{\pi} \lambda \sigma'_e v_K)^2])$.  $v_{ij} = \lambda e_{ij} v_K$ is the relative secular velocity between bodies in mass-bins $i$ and $j$ for that time step. 

The above prescription needs to be modified in the case of collisions between two bodies within the same mass bin, where the focusing factor can vary significantly depending on the location within the mass bin of the two colliding partners.  For example, two bodies with masses right at the center of the mass bin will have a higher focusing factor than one body at the bottom end of the mass bin and one body at the top end; in the latter case, the approach velocity will be higher.  

Let $v_{{\rm max},i}$ be the relative (secularly excited) velocity between the heaviest body in the $i$-th mass bin and the lightest.  Let $x_1$ = $(m_1 - m_{\rm bottom})/(m_{\rm top} - m_{\rm bottom})$ be the position of body 1 in the mass bin, where $m_{\rm top}$ and $m_{\rm bottom}$ are the masses of the heaviest and lightest bodies in the bin. We define $x_2$ analogously for the second body.  The relative secularly excited velocity $v_{12}$ between particles 1 and 2 is then equal to $v_{\rm max} |x_1 - x_2|$ and has a distribution (assuming $x_1$ and $x_2$ to be uniformly distributed in $[0,1]$)
\begin{equation}
\rho(v_{12}) = \frac{2}{v_{\rm max}} \left(1 - \frac{v_{12}}{v_{\rm max}}\right).
\end{equation}
Therefore, $\langle R_{ii} \rangle$, the rate at which a particle in bin $i$ experiences collisions with other particles in bin $i$, is given by 
\begin{equation}
\langle R_{ii} \rangle = \int_0^{v_{\rm max}} \rho(v_{12}) R_{ ii}(v_{12})dv_{12}.
\end{equation}


\subsection{Radial inspiral}
\label{sect:radialInspiral}


Using the results of \citet{RS15a, RS15b} we obtain the radial drift velocity of particles as a function of their size and semimajor axis (see Sect. \ref{inspiralEquations} for details).  Let $\Delta r_i$ be the radial width of annulus $i$, and let the drift rate of a particle in annulus $i$ and mass bin $j$ be $v_{i,j}$.  We then let the change in the number of particles in annulus $i$ and mass bin $j$ in a time interval $\Delta t_{\rm insp}$ be given by
\begin{equation}
\label{inspiralEq}
\Delta n_{i,j} = -\frac{n_{i,j} v_{i,j} \Delta t_{\rm insp}}{\Delta r_i} + \frac{n_{i+1,j} v_{i+1, j} \Delta t_{\rm insp}}{\Delta r_{i+1}},
\end{equation}

We use $N_{\rm ann} = 60$ annuli.  These were spaced so as to minimize changes in $e_c$ or $a$ within one bin.  The exact bin boundaries depend on the run of $e_c(a)$ which varies between our different disk models.  For the fiducial model, $a$ changed by less than 16\% between adjacent bins and $e_c$ by less than 41\% in the cases where $e_c > 10^{-4}$. 


\subsection{Time step determination}
\label{sect:timestep}


In this section we describe how the time steps for the coagulation process and the radial inspiral are determined.


\subsubsection{Coagulation-fragmentation time step}
\label{sect:coag_timestep}

The time step for the coagulation-fragmentation process must be short enough that the distribution does not  change appreciably during the length of a time step.  In principle, this means that we would like 
\begin{equation}
\frac{|n_i(t) - n_i(t+\Delta t_{\rm coag})|}{n_i(t)} < \epsilon
\end{equation}
for all $i$, where $\epsilon$ is some small number.  In practice, this is complicated by the fact that mass is added to bins in discrete quantities, and that some bins have a small number of particles.  In a bin with a small number of particles, even adding one additional particle will make a large fractional change to the amount of mass in that bin.  Additionally, collisions are a random process, so a fluctuation can change the mass in a bin by more than expected.  To address these concerns, we instead pick the largest time step $\Delta t_{\rm coag}$ such that for all $i$
\begin{align}
\label{timestepeq}
&\left|\left \langle \frac{n_i(t) -n_i(t + \Delta t_{\rm coag)}}{n_i(t)}\right \rangle\right| < \epsilon_1 \quad {\rm or} 
\nonumber\\
&\left|\left \langle \frac{m_i (n_i(t) -n_i(t + \Delta t_{\rm coag}))}{\Sigma_{i=1}^{N_{\rm bins}} m_i n_i} \right \rangle\right| < \epsilon_2,
\end{align}
where the angle brackets refer to averaging over the distribution of encounter velocity in Eq. \eqref{eq:vr}.  To save computational time, we approximate the average over the encounter velocity distribution by evaluating the collisional outcomes using the median encounter velocity.  Here, as before, a time step is acceptable if for each bin, either the fractional decrease in mass is less than $\epsilon_1$, or the decrease in mass is less than $\epsilon_2$ times the total mass in the system. Time step $\Delta t_{\rm coag}$ is recalculated every time we update the mass distribution, separately for each annulus, and the minimum value among all the annuli is used for each time step.  It is not obvious from first principles which values of $\epsilon_1$ and $\epsilon_2$ are appropriate. But rows 10 and 11 of Table 2 (which explores sensitivity of our calculations to the numerical parameters used; see Appendix \ref{modelParams}) show that variation of these values by a factor of two (or more) from our fiducial values of 0.05 and $10^{-6}$ have a negligible effect on $d_{\rm min}$. 
\par
In the simulation used to create Fig. \ref{f001}, we used a lower value of $\epsilon_2$ so that the distributions would remain smooth, even in regions of mass-space containing a negligible amount of mass.  The success of the coagulation process, and the shape of the distribution at sizes greater than $d_{\rm init}$ in those simulations is not affected by this change. 


\subsubsection{Radial inspiral time step}
\label{sect:drift_timestep}

The code uses a different time step for the radial drift. Mass transport due to radial drift is recalculated using Eq. (\ref{inspiralEq}) every interval of time $\Delta t_{\rm insp}$ such that 
\begin{equation}
\Delta t_{\rm insp}\times \max\left(\frac{v_{\rm drift}}{\Delta r_i}\right) = \epsilon_{\rm drift},
\end{equation}
which guarantees that the inspiraling particles do not cross the full radial bin width in time $\Delta t_{\rm insp}$ ($v_{\rm drift}$ is the inspiral speed). While we see from Eq. \eqref{inspiralEq} that $\epsilon_{\rm drift}$ must not be larger than 1, it is apparent from the results of Sect. \ref{inspiralTests} below that there is little advantage to picking $\epsilon_{\rm drift}$ less than 1.  Therefore, we use $\epsilon_{\rm drift} = 1$.


\section{Tests of the code}
\label{analTests}


In this section we describe the tests we did of the code against known analytic solutions in certain limiting cases. We independently verify three code functionalities -- coagulation, fragmentation, and radial inspiral -- and find that the code does a reasonable job of reproducing known solutions to the accuracy warranted given the imperfect knowledge of the physical processes involved.


\subsection{Tests against known solutions of the coagulation equation}
\label{coagTests}


In this section we test the coagulation module of our code. We describe the results of simulations of the coagulation alone, without fragmentation, and with simplified collision rates and initial conditions that have analytic solutions we can compare with.  There are three known analytic solutions to the Smoluchowski coagulation equation  
\begin{equation}
\frac{dn_k}{dt} = \frac{1}{2} \sum_{i+j = k} A_{ij} n_i n_j - n_k \sum_{i = 1}^{\infty} A_{ik} n_i,
\label{eq:Smol}
\end{equation}
obtained for kernels $A_{ij} = 1$ \citep{Smoluchowski16}, $A_{ij} = i+j$ and $A_{ij} = ij$ \citep{Trubnikov71}, with the initial condition of $N$ particles of a single mass bin.  Our code successfully reproduces these solutions with small deviations as discussed below.

\begin{figure}
\centering
\includegraphics[width=.5\textwidth]{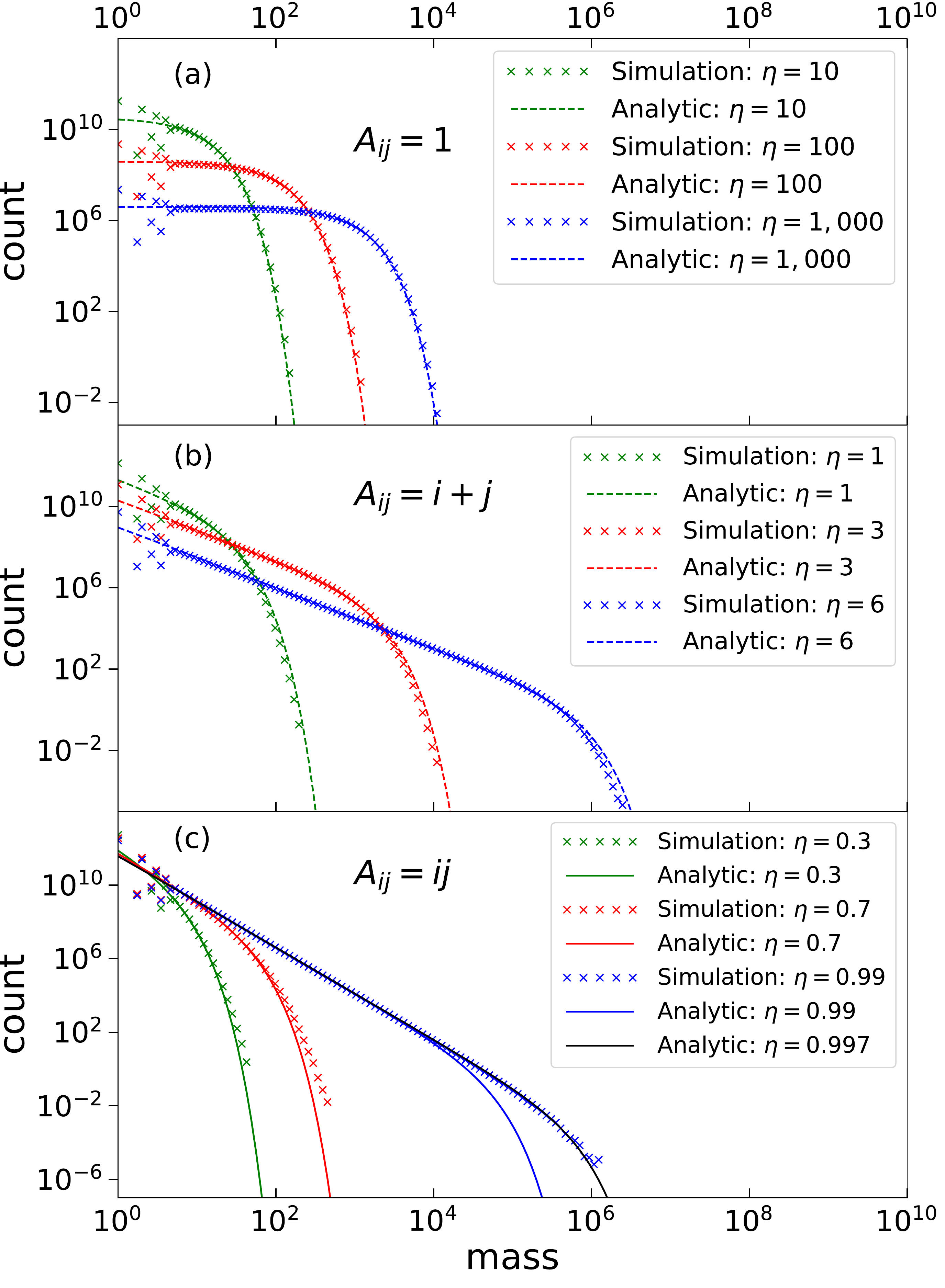}
\caption{Test of our coagulation module: comparison of simulation and analytic results for different coagulation kernels starting with $N=10^{12}$ particles of unit mass. Here $\eta = Nt$ is a time coordinate, rescaled by the number of particles initially in the system.  Each panel is made assuming a different coagulation kernel $A_{\rm ij}$, labeled in the panels and discussed in Sect. \ref{coagTests}. }
\label{coagulationFig}
\vspace{-.05cm}
\end{figure} 

The top panel of Fig. \ref{coagulationFig} shows the number of particles per unit mass as a function of mass for three different times for the kernel $A_{ij} = 1$.  In this figure we have used a mass bin spacing $M_{i+1}/M_i = 1.15$.  Initially there were $N = 10^{12}$ particles all of mass 1.  Because the distribution evolves more rapidly if there are more particles, we express time in terms of $\eta = Nt$.  In the limit of large $N$, the shape of the distribution depends only on $\eta$, rather than on $N$ or $t$ independently. The large scatter at low masses is due to the logarithmic spacing of the mass bins and the initial condition that all the mass is in bodies with unit mass.  Since we are assuming perfect sticking in collisions for this exercise, only mass bins containing integer masses will contain a nonzero number of particles, leading to an uneven distribution at low masses.  Other than this discrepancy at low mass, there is nearly perfect agreement between our code and the analytic solution.

The middle panel of Fig. \ref{coagulationFig} shows the number of particles per unit mass as a function of mass for three different values of $\eta$ for the kernel $A_{ij} = i+j$.  The distribution evolves more rapidly than in the case of the constant kernel because of the higher collision rates.  In this case, in addition to the scatter at low masses, we see that the numerical solution is slightly behind the analytic solution at the high-mass end of the distribution.

The bottom panel of Fig. \ref{coagulationFig} shows the number of particles per unit mass as a function of mass for three different values of $\eta$ for the kernel $A_{ij} = ij$.  Here the system evolves in a more complicated manner than the previous two cases \citep{Trubnikov71}.  For $\eta < 1$, orderly growth from smaller to larger particles progresses as in the previous cases.  However, for $\eta > 1$, one body becomes much larger than the others, and eventually consumes all the mass in the system.  Our simulation is not well-equipped to handle this runaway growth phase, so we restrict ourselves to reproducing the analytic solutions for $\eta < 1$.

The discrepancy between the numerical and analytic solutions at $\eta = 0.99$ looks somewhat alarming.  This is because the distribution function is evolving quite rapidly at this time.  We also plotted the analytic solution for $\eta = 0.997$, and we see that it matches the numerical solution well, implying that this difference only corresponds to a time lag of just 0.3\% in the numerical solution, which is not critical for our calculations.

\begin{figure}
\centering
\includegraphics[width=.5\textwidth]{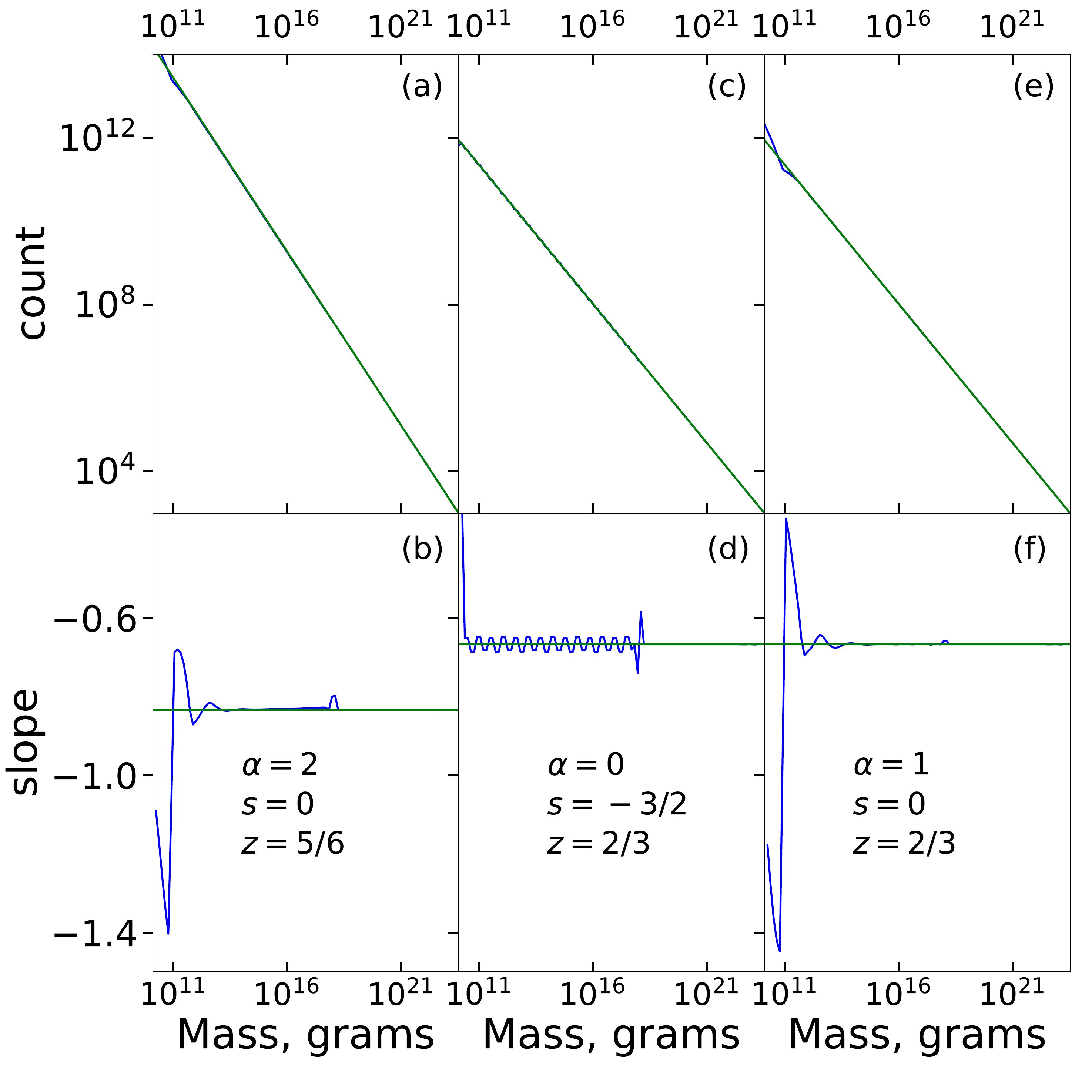}
\caption{Test of our fragmentation module: comparison of our simulation results (blue) with the theoretical predictions of  \citet{OBrien03} (green; they fall on top of the simulation outputs in the top panels).  The upper panel shows the number of particles per mass bin as a function of mass.  The bottom panel shows the power-law slope of this mass distribution. The rightmost 40\% of the mass bins are held fixed, so as not to introduce errors due to the finite extent of our simulation in mass space and time.  The values of $\alpha$, $s$, and $z$ are labeled on the panels.  See Sect. \ref{fragTests} for details.}
\label{FragmentationTest1}
\vspace{-.05cm}
\end{figure} 


\subsection{Tests of the fragmentation module}
\label{fragTests}


Next we checked the ability of our code to properly handle fragmentation. Unlike the coagulation problem, there are no known analytical solutions describing time-dependent behavior in fragmenting systems. However, there are some known results for the steady-state fragmentation cascades to which collisionally evolving systems tend to converge.  In this section we use these results to verify that the fragmentation module of our code is operating correctly.

\citet{OBrien03} present a model of collisional fragmentation that leads to a steady state size distribution given by a power-law.  Their model has one free parameter $s$, which parametrizes the strength of particles of size $d_p$ in catastrophic disruption events as
\begin{equation}
\label{qstar}
Q^*_{\rm RD} = Q_0 d_p^s.
\end{equation}
They show that the distribution of debris sizes $D_f$ in the steady-state collisional cascade is given by 
\begin{equation}
\label{eqNs}
\frac{dn_f(D_f)}{dD_f} = B D_f^{-y},
\end{equation}
where $n_f$ is the number of objects of size $D_f$ (per unit $D_f$), and $B$ is an overall scaling constant. The power-law index of the size distribution of the steady-state collisional cascade is given by 
\begin{equation}
y = \frac{21+s}{6+s}.
\end{equation}

\citet{OBrien03} have assumed that the collisional cross section between two bodies is proportional to the square of the size of the target body (i.e., the geometric cross section).  If we relax this assumption, as was done in \citet{Tanaka96}, and instead let the collision cross section be proportional to $d_p^\alpha$, one can show that $y$ gets modified to
\begin{equation}
\label{yeq}
y = \frac{15 + 3 \alpha + s}{6 + s}.
\end{equation}

Equation \eqref{yeq} was derived assuming a steady-state mass distribution with no upper or lower cut-off.  Since we are only able to simulate a finite range in masses, we must have a way of supplying mass to the system to achieve a steady state; otherwise, all mass will eventually be contained in particles smaller than the lower cutoff size for the simulation.  We do this by fixing the counts in the top 40\% of the mass bins, and letting the collisional cascade provide the mass to the mass bins corresponding to smaller particles.  By doing this we continuously resupply mass to the large mass bins.

Figure \ref{FragmentationTest1} shows a comparison of the steady-state result of the simulation against the expected power law distribution from Eq. \eqref{yeq} for three different values of $s$ and $\alpha$.  The simulations were run with bin spacing $M_{\rm i+1}/M_i = 1.15$.  In each case, the top panel shows the number of bodies in each bin as a function of bin mass.  The blue line shows the particle numbers from the simulation, while the green line shows the expected number counts based on Eq. \eqref{yeq}.  The bottom panels show the logarithmic slope $d\ln n_f/d\ln m$ as a function of mass.

\begin{figure}
\centering
\includegraphics[width=.5\textwidth]{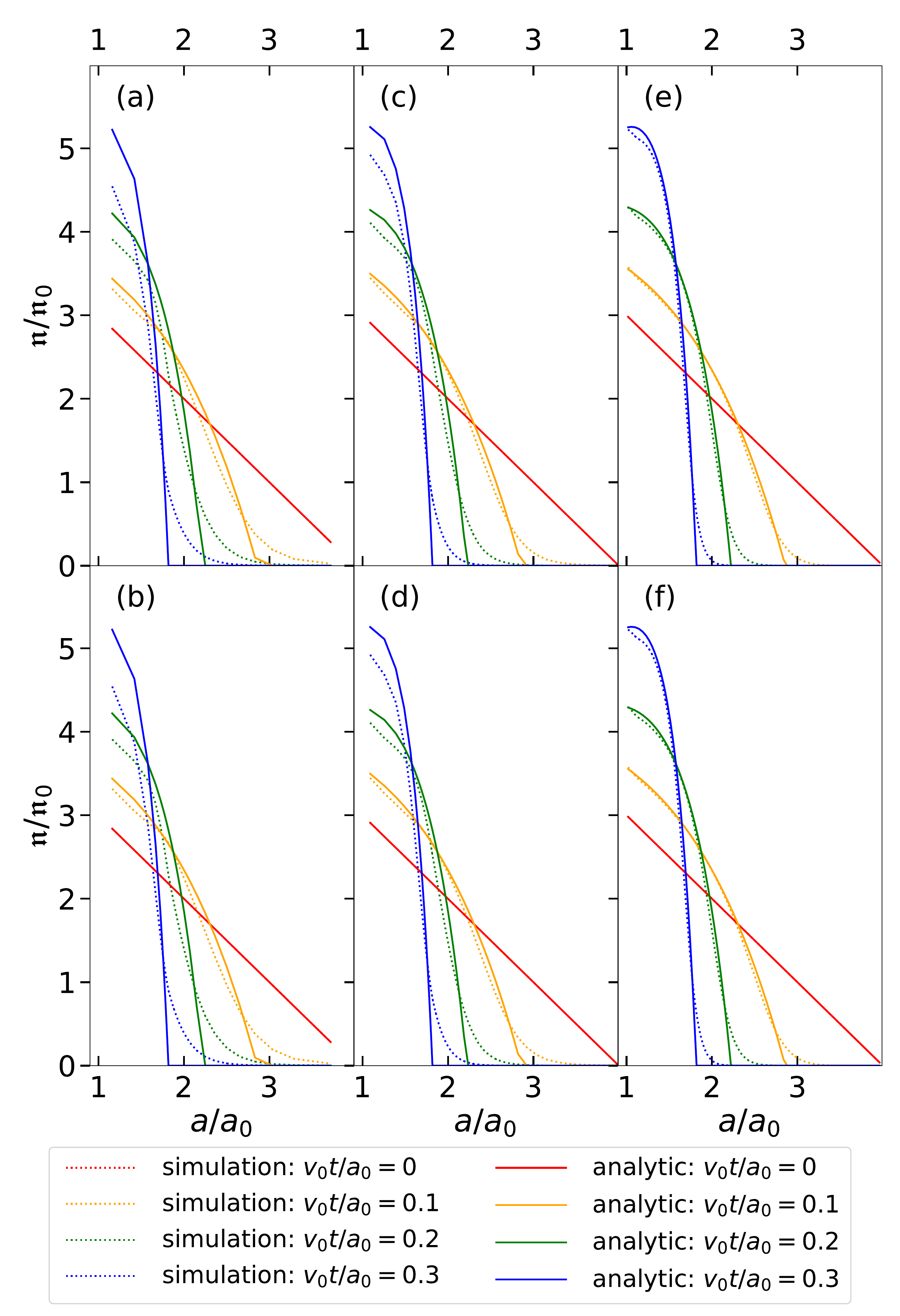}
\caption{Test of the radial inspiral module: a comparison of the evolution of the surface density profile due to radial inspiral in the simulation (dotted lines) vs. the analytic result (solid lines) at different moments of time for the inspiral rate given by Eq. \eqref{simpleInspiralRate}. Panels (a), (c), and (e) correspond to the cases with $N_{\rm ann}=30$, 60, and 240 radial annuli between 1 and 4 AU, and $\epsilon_{\rm drift} = 1.0$. The bottom panels, (b), (d), and (f), have the same $N_{\rm ann}$ values, but $\epsilon_{\rm drift} = 0.3$. }
\label{inspiral1}
\vspace{-.05cm}
\end{figure}

Panels (a) and (b) show the case where $s = 0$ and $\alpha = 2$ \citep{Dohnanyi69}.  In this case Eq. \eqref{yeq} predicts $y$ = 7/2.  Assuming constant bulk density, so that $m \propto d_p^3$, we can show that 
\begin{equation}
\frac{dn_f}{d m} = B^\prime m^{-\frac{y+2}{3}},
\end{equation}
where $B^\prime$ is the appropriate normalization.  The number of particles contained in the bin with bin mass $M_i$ scales as $\Delta n(M_i) \sim M_i^{-z}$, with $z = (y-1)/3 = (3+\alpha)/(6+s)$, due to the logarithmic spacing of the mass bins. For $s = 0$ and $\alpha = 2$, $z = 5/6$.

Panels (c) and (d) are the same as (a) and (b) except that $s = -1.5$ and $\alpha = 0$, leading to a slope $d \ln n_f/d \ln m = -z = -2/3$.  The slope is reproduced faithfully except right near the lower end of the mass distribution.  This time, there is a regular small oscillation of the slope that does not go away with decreasing time step -- a behavior that often emerges in fragmentation simulations \citep{Belyaev2011}.

Finally, in panels (e) and (f) we let $s = 0$ and $\alpha = 1$.  This leads to the expectation that $z = 2/3$. 

In these three plots, we see that our simulation has nearly perfect agreement with the expected power-laws, except for understandable deviations at specific points.  Objects at the small end of the mass distribution, have nothing smaller than themselves in the simulation, so they are destroyed less frequently than the model would predict, hence the steepening of the power-law.  Then the over-abundance of small masses creates an under-abundance of objects of slightly larger masses, giving rise to the wave-like pattern that we observe.  This effect is reduced in panels (c) and (d) because in this example we tuned the $Q_0$ in Eq. \eqref{qstar} so that objects at the lower end of the simulated mass range would not be destroyed by smaller particles, even if they were in the simulation.  The wiggle at the lowest masses in this plot is due to the finite spacing of the mass bins.  


\subsection{Tests of the radial inspiral module}
\label{inspiralTests}


In this section we describe our test of the code module handling the size-dependent radial inspiral. We made this test for a simple inspiral rate given by 

\begin{equation}
\label{simpleInspiralRate}
v_{\rm drift} = v_0\left(\frac{a}{a_0}\right)^2.
\end{equation}
Let $\mathfrak{n}(a, t)$ be the number of particles per unit semimajor axis at time $t$.  Under the influence of inspiral, $\mathfrak{n}$ solves the equation
\begin{equation}
    \frac{\partial \mathfrak{n}(a, t)}{\partial t} - \frac{\partial}{\partial a} \left[v_{\rm drift}(a) \mathfrak{n}(a, t)\right] = 0.
    \label{eq:advectionEq}
\end{equation}
We note that the minus sign in Eq. \eqref{eq:advectionEq} arises from the fact that we adopt a sign convention where positive $v_0$ corresponds to inward motion, toward smaller $a$.

Adopting the initial planetesimal distribution in the disk in the form 
\begin{equation}
\mathfrak{n}(a, t = 0) = \mathfrak{n}_0\left(4 - \frac{a}{a_0}\right),
\end{equation}
for $0 < a < 4 a_0$, this leads to an evolution of the number density given by 

\begin{equation}
\label{eq:drift-soln}
\mathfrak{n}(a, t) = \left\{
        \begin{array}{ll}
            \mathfrak{n}_0s^{-2}\left(4 - a/(a_0s)\right)  & a < 4a_0/(1 + 4v_0t/a_0), \\
            0 & a>4a_0/(1 + 4v_0t/a_0),
        \end{array}
    \right.
\end{equation}
where $s = 1-atv_0/a_0^2$.

Figure \ref{inspiral1} displays the evolution of the planetesimal radial distribution obtained by our code, in comparison with the analytical solution (\ref{eq:drift-soln}). Panels (a), (c), and (e) correspond to the cases with $N_{\rm ann}=30$, 60, and 240 radial annuli between 1 and 4 AU.  The annuli are located at the same values of $a$ as in Simulations 5, 1, and 7, respectively, and we take $\epsilon_{\rm drift} = 1.0$. The corresponding bottom panels (b), (d), and (f) are for $\epsilon_{\rm drift} = 0.3$.  

We see that generally our numerical solutions stay within several percent of the analytic solutions, except the locations where the analytic solution tends to zero.  The simulation does not do a good job of resolving the sharp outer edge of the distribution because particles are effectively redistributed through each annulus every time-step, leading to substantial numerical diffusion. The time step (i.e., the choice of $\epsilon_{\rm drift}$) makes little difference to the results, but the agreement is a strong function of the number of radial annuli $N_{\rm ann}$ used in our calculations (see Fig. \ref{inspiral1} and Table \ref{modelParametersOne}).  


\section{Sensitivity to numerical parameters}
\label{modelParams}

\begin{table*}[t]
\caption{{\large \hspace{4cm} \textbf{Variation of numerical parameters.}}}
\vspace{6mm}
\centering
\begin{tabular}{c c c c c c c c c c c}
\hline\hline
Sim \# & $M_{k+1}$/$M_k$$^1$ & $N_{\rm ann}$ &min size$^2$ (m)  & $\epsilon_1$ $^3$ & $\epsilon_2$ $^4$ & $\xi$ $^5$  & $b$ $^6$  & $d_{\rm min}$ (km) $^7$ & $d_{\rm min}^{\rm rubble}$ $^8$   \\ [0.5ex] 
\hline 
\hline
\hline
1  &    1.37   &  60  &  10   & 0.05 & $10^{-6}$  &  -1.0   &  $10^{-2}$  &   1.6 & 3.4   \\[1ex]
\hline
\hline
\hline
1A  &     {\bf 1.88}  &  60  &  10  & 0.05 & $10^{-6}$  &  -1.0   &  $10^{-2}$  &  1.4 & 3.0 \\[1ex]
1B  &    {\bf 1.17}  &  60  &  10  & 0.05 & $10^{-6}$ &    -1.0   &  $10^{-2}$   & 1.7 & 3.6  \\[1ex]
\hline
1C  &     1.37  &  {\bf 30}  &  10  &  0.05 & $10^{-6}$   & -1.0   &  $10^{-2}$  &   1.7 & 3.4  \\[1ex]
1D  &     1.37  &  {\bf 120}  &  10  & 0.05 & $10^{-6}$&  -1.0   &  $10^{-2}$  &  1.5  & 3.3 \\[1ex]
\hline
1E  &    1.37  &  60  &  {\bf 100}  & 0.05 & $10^{-6}$&   -1.0   &  $10^{-2}$  & 1.4 & 3.4  \\[1ex]
1F  &     1.37  &  60  &  {\bf 1.0} & 0.05 & $10^{-6}$   &   -1.0   &  $10^{-2}$  & 1.5  & 3.3  \\[1ex]
\hline
1G  &     1.37  &  60  &  10 & {\bf 0.01} & ${\bf 10^{-7}}$  &   -1.0   &  $10^{-2}$  & 1.6 & 3.4   \\[1ex]
1H  &     1.37  &  60  &  10  & {\bf 0.1} & ${\bf 10^{-5}}$ &   -1.0   &  $10^{-2}$  &  1.6  & 3.4 \\[1ex]
\hline
1I  &     1.37  &  60  &  10  & 0.05 & $10^{-6}$  &   {\bf -1.5}   &  $10^{-2}$  &  1.6 & 3.5   \\[1ex]
1J  &     1.37  &  60  &  10  & 0.05 & $10^{-6}$   &   {\bf 0.0}   &  $10^{-2}$  &  1.6 & 3.3    \\[1ex]
\hline
1K  &     1.37  &  60  &  10  & 0.05 & $10^{-6}$ &   -1.0   &  ${\bf 10^{-1}}$  &  1.6 & 3.5   \\[1ex]
1L  &     1.37  &  60  &  10  & 0.05 & $10^{-6}$ & -1.0   &  ${\bf 10^{-4}}$  &  1.6 & 3.3  \\[1ex]
\hline
\label{tbl:num-pars}
\end{tabular}
\vspace{20mm}
\begin{flushleft}
1. Ratio of the mass between successive mass bins. \\
2. Radius of the smallest body considered in the simulation. \\
3. Parameter which goes into determining the coagulation time step; see Sect. \ref{sect:coag_timestep}. \\
4. Parameter which goes into determining the coagulation time step; see Sect. \ref{sect:coag_timestep}. \\
5. Slope of the fragment mass distribution; see Eq. \eqref{tequation}. \\
6. Smallest possible ratio of the mass of the largest rubble particle to the total mass in the collision; see Eq. \eqref{cutoffMass}. \\
7. Smallest initial planetesimal size which results in a 300 km body forming within 1 Myr, assuming solid planetesimals.\\
8. Smallest initial planetesimal size which results in a 300 km body forming within 1 Myr, assuming rubble-pile planetesimal composition.\\
\end{flushleft}
\vspace{80mm}
\label{modelParametersOne}
\end{table*}
Our code employs a number of numerical parameters to set various time steps, properly resolve mass and inspiral-driven evolution, and so on. We varied these numerical code inputs to see the effect on the calculation outcomes. Table \ref{modelParametersOne} shows the results of this exercise, in which we use $d_{\rm min}$ and $d_{\rm min}^{\rm rubble}$ as our sensitivity metrics.  

Simulation 1 uses the fiducial set of numerical parameters that we employ in all our production runs. As in Table \ref{physicalParameters}, in all other runs the parameter that is different from Simulation 1 is highlighted in bold in each row.  We see that $d_{\rm min}$ does not depend strongly on the numerical parameters, which justifies our particular choice of their values.  

\end{document}